\documentclass[aps,prb,twocolumn,amsmath,amssymb,showpacs,superscriptaddress]{revtex4}

\usepackage{graphicx}
\usepackage{dcolumn}
\usepackage{bm}

\begin{document}

\title{Surface polaritons on left-handed cylinders: \\
A complex angular momentum analysis}

\author{St\'ephane Ancey}
\email{ancey@univ-corse.fr}
\affiliation{ UMR CNRS 6134 SPE, Equipe Ondes et Acoustique, \\
Universit\'e de Corse, Facult\'e des Sciences, Bo{\^\i}te Postale
52, 20250 Corte, France}

\author{Yves D\'ecanini}
\email{decanini@univ-corse.fr}
\affiliation{ UMR CNRS 6134 SPE,
Equipe Physique Semi-Classique (et) de la Mati\`ere Condens\'ee,
\\ Universit\'e de Corse, Facult\'e des Sciences, Bo{\^\i}te
Postale 52, 20250 Corte, France}

\author{Antoine Folacci}
\email{folacci@univ-corse.fr}
\affiliation{ UMR CNRS 6134 SPE,
Equipe Physique Semi-Classique (et) de la Mati\`ere Condens\'ee,
\\ Universit\'e de Corse, Facult\'e des Sciences, Bo{\^\i}te
Postale 52, 20250 Corte, France}

\author{Paul Gabrielli}
\email{gabrieli@univ-corse.fr}
\affiliation{ UMR CNRS 6134 SPE,
Equipe Ondes et Acoustique,
\\ Universit\'e de Corse, Facult\'e des Sciences, Bo{\^\i}te
Postale 52, 20250 Corte, France}

\date{\today}

\begin{abstract}

We consider the scattering of electromagnetic waves by a
left-handed cylinder -- i.e., by a cylinder fabricated from a
left-handed material -- in the framework of complex angular
momentum techniques. We discuss both the TE  and TM theories. We
emphasize more particularly the resonant aspects of the problem
linked to the existence of surface polaritons. We prove that the
long-lived resonant modes can be classified into distinct
families, each family being generated by one surface polariton
propagating close to the cylinder surface and we physically
describe all the surface polaritons by providing, for each one,
its dispersion relation and its damping. This can be realized by
noting that each surface polariton corresponds to a particular
Regge pole of the $S$ matrix of the cylinder. Moreover, for both
polarizations, we find that there exists a particular surface
polariton which corresponds, in the large-radius limit, to the
surface polariton which is supported by the plane interface. There
exists also an infinite family of surface polaritons of
whispering-gallery type which have no analogs in the plane
interface case and which are specific to left-handed materials.

\end{abstract}

\pacs{78.20.Ci, 41.20.Jb, 73.20.Mf, 42.25.Fx}

\maketitle

\section{Introduction}

In an article published in 1967 (see Ref.~\onlinecite{Veselago}
for the English translation), Veselago speculated upon the
alteration of electromagnetic and optical phenomena in
hypothetical materials for which the electric permittivity and the
magnetic permeability (and therefore the refractive index
\cite{Veselago,Smith2000}) were simultaneously negative in a
certain range of frequencies. He predicted more particularly the
existence, in such materials, of anomalous effects such as a
reversed Doppler shift, reversed Cerenkov radiation, negative
radiation pressure, and inverse Snell-Descartes law. Since in such
substances the electric field, the magnetic field and the wave
vector of a plane wave form a left-handed system, so that the
Poynting vector and the wave vector have opposite direction, he
referred to them as left-handed media and here we shall use this
terminology, even if today other authors prefer to call this kind
of materials ``negative index media", ``double negative media" or
``Veselago's media". Four years ago, following insights from
Pendry and coworkers \cite{Pendry96,Pendry98,Pendry99}, Schultz,
Smith and colleagues \cite{Smith2000a,Smith2001a,Smith2001b} have
been able to build, for the first time, an artificial left-handed
medium by combining arrays of wires and split-ring resonators and
to experimentally test ``left-handed electromagnetism" in the
microwave frequency range. Since then, several other groups have
successfully fabricated left-handed media and experimentally
studied the alteration of electromagnetic phenomena (see, for a
comprehensive list of references, the popular article by Pendry
and Smith \cite{PendrySmith2004}). Now, the possibility to create
left-handed materials working in the optical domain is seriously
foreseen and, very recently, negative refraction has been even
observed at infrared wavelengths in the context of photonic
crystal physics \cite{BerrierETAL2004}. Of course, the unusual and
remarkable properties of left-handed media could revolutionize
optoelectronics and communications. Many technological
applications are already considered including superlenses,
bandpass filters, beam guiders and light-emitting devices. As a
consequence, this recent new field of physics has attracted the
interest of many researchers and is rapidly evolving and the
corresponding scientific literature is exploding.

In this article, we shall focus our attention on a particular
problem of left-handed electromagnetism: namely, the scattering of
an electromagnetic wave by a ``left-handed cylinder" -- i.e., by
an homogeneous cylinder fabricated from a left-handed material.
This problem has been already considered by Kuzmiak and Maradudin
\cite{KuzmiakMaradudin02} but we shall reexamine it in the
framework of complex angular momentum (CAM) techniques
\cite{New82,Nus92}. In fact, we are above all interested in the
resonant aspects of scattering linked to the existence of surface
polaritons (SP's) and, by using CAM techniques, we shall be able
to provide a theoretical description of SP's propagating close to
the left-handed cylinder surface as well as a physical explanation
for the excitation of the associated resonant modes.

Let us recall here that SP's are electromagnetic surface waves
propagating close to the interface separating two different media
with an amplitude that decays in an exponential fashion
perpendicularly to the interface and into both media. SP's
supported by metal-dielectric or semiconducting-dielectric
interfaces have been extensively studied during the last 40 years
because of the fundamental role they play in the context of the
interaction of electromagnetic radiation with matter but also
because of their numerous practical applications in physics,
chemistry and biology (for reviews on this subject we refer to
Refs.~\onlinecite{MillsB74,Agranovich82,Raether88,Sernelius01}).
Recently, activity has focused particularly on the role of SP's in
photonic crystal physics (see, for example,
Refs.~\onlinecite{McGurnMaradudin93,KusmiakMP1994,SigalasCHS1995,
ZhangHLXM1996,
KusmiakMG1997,KusmiakM1997,Moroz2000,Sakoda2001,Sakoda2001b,
MorenoEH2002,OchiaiSD2002}) as well as on their role in the
enhanced transmission of light through periodic arrays of holes in
a metal film \cite{Ebbesen98} (see also
Ref.~\onlinecite{Ebbesen2004} and references therein). Of course,
the scientific literature dealing with SP's localized at the
interface separating a left-handed medium and a conventional one
is much less developed and it is still too early to judge their
importance in left-handed electromagnetism. However, it is already
obvious that it will not be negligible: indeed, it seems that SP's
play a central role in the superlensing phenomenon
\cite{Pendry00,Feise01,Haldane02,Rao03} and in the giant
Goos-H{\"a}nchen effect recently discovered
\cite{ShadrivovEtAl03}.

The theory of SP's supported by left-handed media has been studied
in Refs.~\onlinecite{RuppinPLA00,Darmanyanetal03,ShadrivovEtAl04}
in the case of a plane interface. In that configuration, we can
consider that the properties of SP's are now completely known.
They have been obtained from rather elementary calculations
involving homogeneous and inhomogeneous plane waves. The existence
of SP's in the presence of a curved interface has been noted in
the article by Kuzmiak and Maradudin \cite{KuzmiakMaradudin02}
previously cited and in an article by Ruppin \cite{RuppinSSC00}
dealing with scattering from a sphere. But it should be noted that
neither of these two articles provide a clear physical description
of SP's. This is mainly due to the fact that, unlike the case for
a plane interface, in the case of a curved interface the wave
equation cannot be solved in terms of elementary functions. As a
consequence, the description of SP's involves transcendental
equations which obscures any analysis. Fortunately, it is possible
to circumvent these difficulties by using CAM techniques (here, we
refer to the Introduction of Ref.~\onlinecite{AnceyDFG2004} for a
short bibliography on this topic) in connection with modern
aspects of
asymptotics\cite{Dingle73,Berry89,BerryHowls90,SegurTL91}. It is
exactly what we did in a recent article \cite{AnceyDFG2004}
dealing with scattering of electromagnetic waves from metallic and
semiconducting cylinders -- i.e., in the presence of dispersive
materials. This allowed us to provide a clear physical explanation
for the excitation mechanism of the resonant surface polariton
modes (RSPM's) as well as a simple mathematical description of the
SP's that generate them. In the present paper, we shall extend
this analysis to left-handed cylinders but before we embark on
this work we shall briefly discuss the limitations of our
approach. We first recall that the CAM method is an asymptotic
approach and that formally it is only valid for high frequencies
or more precisely when the radius $a$ of the cylinder is large
compared to the wavelength $\lambda = 2\pi c / \omega$ of the
electromagnetic field. In fact, this limitation must not be taken
too seriously. Indeed, in practical applications CAM techniques
provide good results even if $\lambda \approx a$. Moreover, in the
present problem we cannot assume that $a$ is very large compared
to $\lambda$. Indeed, in that case, the wavelength of the
electromagnetic field could become comparable to the size $a'$ of
the unit cell of the left-handed material and the cylinder could
not be treated as homogeneous. In other words, the analysis we
shall provide in the next sections is formally valid for $a' \ll
\lambda \le a$.

Our paper is organized as follows. In Sec. II, we introduce our
notations and we construct the $S$ matrix of the system. We
consider the TE theory ($H$ polarization) as well as the TM theory
($E$ polarization). We then discuss the resonant aspects of our
problem for both theories. In Sec. III, by working in the CAM
plane, we qualitatively describe the SP's supported by the
left-handed cylinder. Then, in Sec. IV, by using CAM techniques,
we establish the connection between these SP's and the associated
RSPM's. In Sec. V, by using asymptotic techniques, we describe
semiclassically the different SP's and we provide analytic
expressions for their dispersion relations and their damping. We
show more particularly the existence of SP's of whispering-gallery
type. Finally, in Sec. VI, we conclude our article by emphasizing
the main results of our work.

\section{Exact $S$ matrices and scattering resonances
for the TE and the TM theories}

From now on, we consider the scattering of electromagnetic waves
by a cylinder with circular cross section and radius $a$ having an
effective frequency-dependent permittivity $\epsilon (\omega)$ and
an effective frequency-dependent permeability $\mu (\omega)$. Here
and in the following, we implicitly assume the time dependence
$\exp(-i\omega t)$ for electric and magnetic fields. We consider
that the cylinder is embedded in a host medium of infinite extent
having the electromagnetic properties of the vacuum. We introduce
the usual cylindrical polar coordinate system $(\rho ,\theta ,z)$.
It is chosen so that the cylinder and surrounding medium,
respectively, occupy the regions corresponding to the range $0 \le
\rho < a$ (region II) and to the range $\rho > a$ (region I).
Furthermore, in order to describe wave propagation we also
introduce the wave number
\begin{equation}
k(\omega)=\frac{\omega }{c}
\end{equation}
where $c$ denotes the velocity of light in vacuum, and the
refractive index of the cylinder
\begin{equation}
n(\omega)=\sqrt{ \epsilon (\omega) \mu (\omega)}.
\end{equation}

As far as the electric permittivity $\epsilon (\omega)$ and the
magnetic permeability $\mu (\omega)$ of the cylinder are
concerned, we assume they are, respectively, given by
\begin{equation}
\epsilon (\omega) = 1- \frac{\omega_p^2}{\omega ^2} \label{PetP1}
\end{equation}
and
\begin{equation}
\mu (\omega) =  1 - \frac{F \omega^2}{\omega ^2 - \omega_0^2}
=(1-F) \left( \frac{\omega^2 - \omega_b^2 }{\omega ^2 -
\omega_0^2} \right)  \label{PetP2}
\end{equation}
where $0<F<1$ and $\omega_b=\omega_0/\sqrt{1-F}$. Here we have
considered that the cylinder is fabricated from a metamaterial. Of
course, the parameters $\omega_p$, $\omega_0$ and $F$ depend on
its structure. But we would like to be as general as possible in
our analysis and therefore we do not restrict our study to any
particular metamaterial (see, however, the discussion in the last
paragraph of our conclusion). As a consequence, we do not
attribute any ``microscopic" interpretation to the parameters
$\omega_p$, $\omega_0$ and $F$ in terms of the internal structure
of the metamaterial considered. Similarly,  we do not precise the
frequency range where $\omega_p$ and $\omega_0$ lie. We only
assume that  $\omega_0 < \omega_b < \omega_p$. We then have
$\epsilon (\omega) <0$ in the frequency range $\omega \in \left]0,
\omega_p \right[$ and $\mu (\omega) <0$ in the frequency range
$\omega \in \left]\omega_0, \omega_b \right[$. Thus, the electric
permittivity, the magnetic permeability and the refractive index
are simultaneously negative in the region $\omega_0 < \omega <
\omega_b$. In that region the metamaterial presents left-handed
behavior. As far as the numerical aspects of our work are
concerned, we shall work with $F=0.4$ and with the reduced
frequencies $\omega_0a/c=5.52$, $\omega_ba/c \approx 7.127$ and
$\omega_pa/c=11.04$. Even though we restrict ourselves to that
particular configuration, the results we shall obtain numerically
are in fact very general and they permit us to correctly
illustrate the theory.

Here and from now on, we choose to treat our problem in a
two-dimensional setting, ignoring the $z$ coordinate. We briefly
recall the equations governing the TE theory where the magnetic
field $\mathbf{H}$ is parallel to the cylinder axis ($H$
polarization) and the TM theory where the electric field
$\mathbf{E}$ is parallel to the cylinder axis ($E$ polarization).
From Maxwell's equations it is easy to show that the $z$
components of the magnetic and electric fields satisfy the
Helmholtz equation
\begin{subequations}\label{HEquHzEz}
\begin{eqnarray}&& \left[ \Delta_{\bf x} + n^2(\omega)\left(
\frac{\omega}{c} \right)^2  \right]
  \begin{cases}
    H_z^{\mathrm {II}} ({\bf x})
=0  &   \\
    E_z^{\mathrm {II}} ({\bf x})
=0  &
  \end{cases}
  \mathrm{for} \  0  \le \rho < a,
\nonumber \\
&&    \label{HEqu1HzEz} \\
&& \left[ \Delta_{\bf x} +\left( \frac{\omega}{c} \right)^2\right]
\begin{cases}
    H_z^{\mathrm {I}} ({\bf x})
=0  &   \\
    E_z^{\mathrm {I}} ({\bf x})
=0  &
  \end{cases}
  \mathrm{for}  \ \rho
> a,  \label{HEqu2HzEz}
\end{eqnarray}
\end{subequations}
with ${\bf x}=(\rho,\theta)$ and where the Laplacian $\Delta_{\bf
x}$ is expressed in the polar coordinate system. For the TE
theory, from the continuity of the tangential components of the
electric and magnetic fields -- i.e., of $E_\theta$ and $H_z$ --
at the interface between regions I and II, it can be shown that
the $z$ component of the magnetic field satisfies, for
$0\leq\theta<2\pi$,
\begin{subequations}\label{BCHz}
\begin{eqnarray}
&&H_z^{\mathrm I}(\rho =a,\theta)=H_z^{\mathrm {II}}(\rho
=a,\theta) , \label{BCHz1} \\
&&  \frac{\partial H_z^{\mathrm I}}{\partial \rho}(\rho
=a,\theta)=\frac{1}{\epsilon  (\omega)} \, \frac{\partial
H_z^{\mathrm {II}}}{\partial \rho}(\rho =a,\theta).  \label{BCHz2}
\end{eqnarray}
\end{subequations}
For the TM theory, the continuity of the tangential components of
the electric and magnetic fields -- i.e., of $E_z$ and $H_\theta$
-- at the interface between regions I and II permits us to show
that the $z$ component of the electric field satisfies, for
$0\leq\theta<2\pi$,
\begin{subequations}\label{BCEz}
\begin{eqnarray}
&&E_z^{\mathrm I}(\rho =a,\theta)=E_z^{\mathrm {II}}(\rho
=a,\theta) , \label{BCEz1} \\
&&  \frac{\partial E_z^{\mathrm I}}{\partial \rho}(\rho
=a,\theta)=\frac{1}{\mu (\omega)} \, \frac{\partial E_z^{\mathrm
{II}}}{\partial \rho}(\rho =a,\theta).  \label{BCEz2}
\end{eqnarray}
\end{subequations}

We can now construct the $S$ matrix for the cylinder for both
polarizations. Because of the cylindrical symmetry of the
scatterer, the $S$ matrix is diagonal and its elements $S_{\ell
\ell'}(\omega)$ are given by $S_{\ell \ell'}(\omega)=S_\ell
(\omega)\ \delta _{\ell \ell'}$. It should be recalled that the
$S$ matrix is of fundamental importance because it contains all
the information about the scattering process. Its components
appear in the Green functions of the problem, in the scattered
field when a plane wave excites the cylinder as well as in both
the scattering amplitude and the total scattering cross section.
For the TE and the TM theories, we shall denote, respectively, by
$S^{H}_\ell (\omega)$  and $S^{E}_\ell (\omega)$ the $S$ matrix
diagonal elements. For a given angular momentum index $\ell \in
\mathbb{Z}$, the coefficients $S^{H}_\ell$ and $S^{E}_\ell$ are
respectively obtained  from the partial wave ${(H_z)}_\ell$ and
${(E_z)}_\ell$ solutions of the following problem \cite{Mott65}:
\begin{description}
  \item (i) ${(H_z)}_\ell$ and ${(E_z)}_\ell$ satisfy the Helmholtz equation (\ref{HEquHzEz}),
  \item (ii) ${(H_z)}_\ell$ and ${(E_z)}_\ell$, respectively, satisfy the boundary
  conditions (\ref{BCHz}) and (\ref{BCEz}),
  \item (iii) at large distance, ${(H_z)}_\ell$ and ${(E_z)}_\ell$ respectively present
  the asymptotic behaviors
\begin{eqnarray}
&&{(H_z)}_\ell (\rho,\theta) \underset{\rho \to +\infty}{\sim}
\frac{1}{\sqrt{2\pi k \rho}}
\left(e^{-i(k \rho -\ell\pi/2-\pi/4)} \right. \nonumber \\
&& \qquad\qquad\qquad +   \left. S^{H}_\ell (\omega) e^{i(k\rho
-\ell\pi/2-\pi/4)}\right)e^{i\ell\theta},  \nonumber
\\ &&{(E_z)}_\ell (\rho,\theta) \underset{\rho \to +\infty}{\sim}
\frac{1}{\sqrt{2\pi k \rho}}
\left(e^{-i(k \rho -\ell\pi/2-\pi/4)} \right. \nonumber \\
&& \qquad\qquad\qquad +   \left. S^{E}_\ell (\omega) e^{i(k\rho
-\ell\pi/2-\pi/4)}\right)e^{i\ell\theta}. \nonumber
\end{eqnarray}
\end{description}
Outside the cylinder (region I), the solution of
Eq.~(\ref{HEquHzEz}) is expressible in terms of Bessel functions
(see Ref.~\onlinecite{AS65}) as a linear combination of
$J_{\ell}(\omega \rho/c)e^{i\ell\theta}$ and $H^{(1)}_{\ell
}(\omega \rho/c)e^{i\ell\theta}$. Inside the cylinder (region II),
it is proportional to $J_\ell(n(\omega) \, \omega
\rho/c)e^{i\ell\theta}$. As a consequence, the partial waves
${(H_z)}_\ell$  and ${(E_z)}_\ell$ solutions of (i) and (ii) can
be obtained exactly. Then, by using the standard asymptotic
behavior of Hankel functions $H_\ell^{(1)}(x)$ and
$H_\ell^{(2)}(x)$ for $x \to \infty$ (see Ref.~\onlinecite{AS65}),
we find from (iii) the expressions of the diagonal elements
$S^{H}_\ell$  and $S^{E}_\ell$ of the $S$ matrix for the TE and
the TM theories. We have
\begin{equation}\label{ExprMSHE}
S^{H}_\ell(\omega)=1-2\frac{C^{H}_\ell(\omega)}{D^{H}_\ell(\omega)}
~~\mathrm{and}~~
S^{E}_\ell(\omega)=1-2\frac{C^{E}_\ell(\omega)}{D^{E}_\ell(\omega)}
\end{equation}
where $C^{H}_\ell(\omega)$, $D^{H}_\ell(\omega)$,
$C^{E}_\ell(\omega)$ and $D^{E}_\ell(\omega)$ are $2\times 2$
determinants which are explicitly given by
\begin{subequations}\label{DDH}
\begin{eqnarray}
&C^{H}_\ell(\omega)=\sqrt{\epsilon (\omega)/ \mu (\omega)}
J'_\ell (\omega a/c) J_\ell(n(\omega) \, \omega a/c) \nonumber \\
& \quad -J_\ell(\omega a/c)J'_\ell(n(\omega) \, \omega a/c),  \label{DDHa}\\
&D^{H}_\ell(\omega)=\sqrt{\epsilon (\omega)/ \mu (\omega)}
H_\ell^{(1)'} (\omega a/c) J_\ell(n(\omega) \, \omega a/c) \nonumber \\
& \quad - H_\ell^{(1)} (\omega a/c)J'_\ell(n(\omega) \, \omega
a/c), \label{DDHb}
\end{eqnarray}
\end{subequations}
and
\begin{subequations}\label{DDE}
\begin{eqnarray}
&C^{E}_\ell(\omega)=\sqrt{\mu (\omega)/ \epsilon (\omega)}
J'_\ell (\omega a/c) J_\ell(n(\omega) \, \omega a/c) \nonumber \\
& \quad -J_\ell(\omega a/c)J'_\ell(n(\omega) \, \omega a/c),  \label{DDEa}\\
&D^{E}_\ell(\omega)=\sqrt{\mu (\omega)/ \epsilon (\omega)}
H_\ell^{(1)'} (\omega a/c) J_\ell(n(\omega) \, \omega a/c) \nonumber \\
& \quad - H_\ell^{(1)} (\omega a/c)J'_\ell(n(\omega) \, \omega
a/c). \label{DDEb}
\end{eqnarray}
\end{subequations}
For both polarizations, the unitarity of the $S$ matrix
\cite{New82}, which expresses the energy conservation, and the
reciprocity property \cite{New82}, which is associated with
time-reversal invariance, can be easily verified by using
elementary properties of Bessel functions.

\begin{figure}
\includegraphics[height=7cm,width=8.6cm]{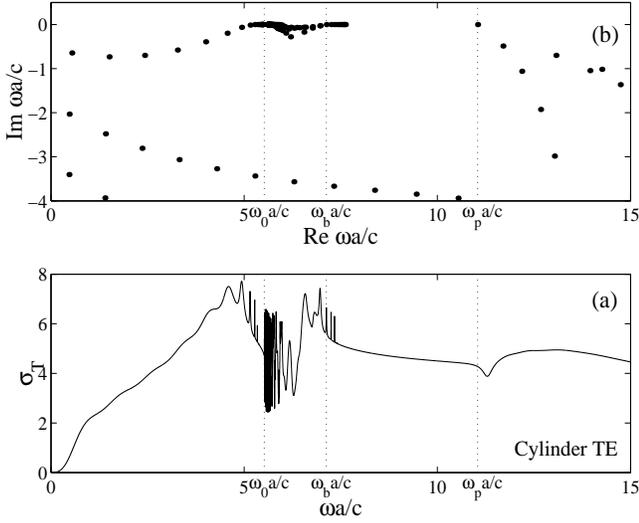}
\caption{\label{fig:general_TE} TE theory. a) Total cross section
$\sigma^H_T$. b) Scattering resonances in the complex $\omega a/c$
plane.}
\end{figure}
\begin{figure}
\includegraphics[height=7cm,width=8.6cm]{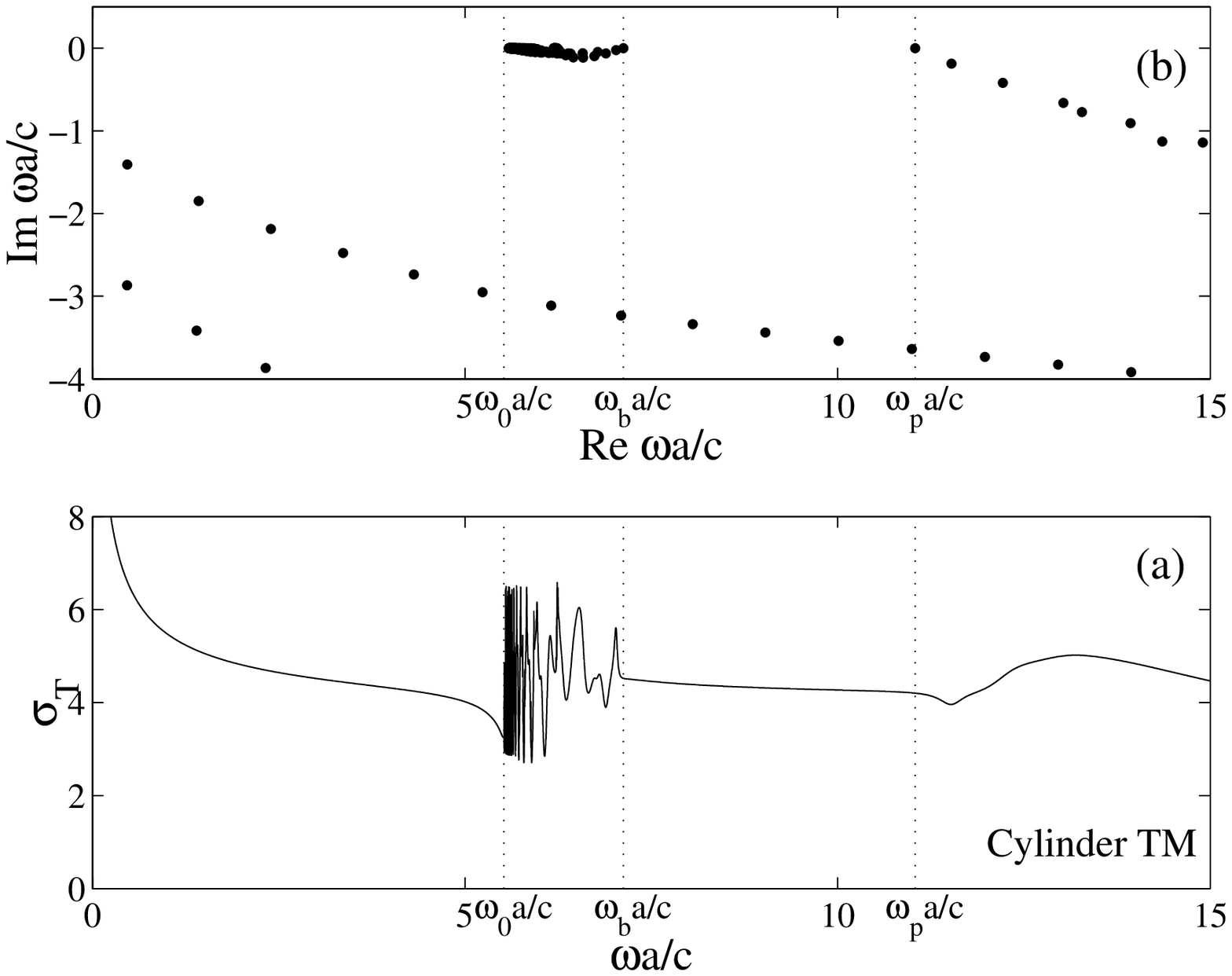}
\caption{\label{fig:general_TM} TM theory. a) Total cross section
$\sigma^E_T$. b) Scattering resonances in the complex  $\omega
a/c$ plane.}
\end{figure}

As far as the scattering amplitude $f(\omega, \theta)$ and the
total scattering cross section per unit length of the cylinder
$\sigma _T(\omega)$ are concerned, they are, respectively, given
by
\begin{equation}\label{ampli}
f(\omega, \theta)=\sqrt{\frac{1}{2i\pi k(\omega)}}
\sum_{\ell=0}^{+\infty} \gamma_\ell \left[S_\ell(\omega) - 1
\right] \cos ({\ell\theta}).
\end{equation}
where $\gamma_\ell$ is the Neumann factor ($\gamma_0=1$ and for
$\ell \not= 0$,  $\gamma_\ell=2$) and
\begin{equation}\label{CS}
\sigma _T(\omega)=\sqrt{\frac{8\pi}{k (\omega)}} \mathrm{Im}
\left( e^{-i\pi/4}f (\omega, \theta =0)\right).
\end{equation}
It should be noted that the definition (\ref{CS}) assumes the
unitarity of the $S$ matrix. It is obtained by using the optical
theorem \cite{New82}. From the expressions
(\ref{ExprMSHE})-(\ref{DDE}) of the $S$ matrices and by using
expressions (\ref{ampli}) and (\ref{CS}), we obtain the total
cross sections $\sigma^H _T(\omega)$ and $\sigma^E _T(\omega)$ for
the TE  and  TM theories. We have
\begin{equation}\label{CSH}
\sigma^H _T(\omega)=\frac{4}{k (\omega)} \sum_{\ell=0}^{+\infty}
\gamma_\ell {\left| \frac{C^{H}_\ell(\omega)}{D^{H}_\ell(\omega)}
\right|}^2
\end{equation}
and
\begin{equation}\label{CSE}
\sigma^E _T(\omega)=\frac{4}{k (\omega)} \sum_{\ell=0}^{+\infty}
\gamma_\ell {\left| \frac{C^{E}_\ell(\omega)}{D^{E}_\ell(\omega)}
\right|}^2.
\end{equation}
Here, it should be noted that our expressions for the total cross
sections $\sigma^H _T(\omega)$ and $\sigma^E _T(\omega)$ are
different from the expressions given by Kuzmiak and Maradudin
\cite{KuzmiakMaradudin02}. It seems to us that these authors have
incorrectly applied the boundary conditions for the electric and
magnetic fields at the cylinder surface.

In Figs.~\ref{fig:general_TE}a and \ref{fig:general_TM}a, we
display the total cross sections for the TE and the TM theories.
They are both plotted as functions of the reduced frequency
$\omega a /c$. In the two figures, rapid variations of sharp
characteristic shapes can be observed. For the $H$ polarization,
this strongly fluctuating behavior is localized within and
slightly around the frequency range where the cylinder presents
left-handed behavior, while for the $E$ polarization, it is
totally localized within that frequency range. For both
polarizations, such strongly fluctuating behavior is due to the
scattering resonances associated with the long-lived resonant
modes of the cylinder -- i.e., the long-lived resonant states of
the photon-cylinder system. These resonances are the poles of the
$S$ matrix lying in the fourth quadrant of the complex $\omega$
plane near the real $\omega$ axis. Resonances are determined by
solving
\begin{equation}\label{detH}
D^H_\ell(\omega)=0 \quad \mathrm{for} \quad \ell \in \mathbb{N}
\end{equation}
for the TE theory and
\begin{equation}\label{detE}
D^E_\ell(\omega)=0 \quad \mathrm{for} \quad \ell \in \mathbb{N}
\end{equation}
for the TM theory. In Figs.~\ref{fig:general_TE}b and
\ref{fig:general_TM}b, resonances are exhibited for both theories.
For certain frequencies, we can clearly observe a one-to-one
correspondence between the peaks of $\sigma _T(\omega)$ and the
resonances near the real $\omega a/c$ axis but in general the
situation seems very confused. This is due to the profusion of
long-lived resonant modes in and around the frequency range where
the cylinder presents left-handed behavior. Furthermore, by
zooming in on the distribution of resonances in regions close to
the real axis of the complex $\omega$ plane, we have also observed
accumulations of resonances for large values of $\ell$:

\qquad -- For the TE theory, there exists an accumulation of
resonances which converges to the limiting frequency $\omega_s$
satisfying
\begin{equation}\label{TEaccSPinf1}
\epsilon (\omega_s) + 1 =0
\end{equation}
and given by
\begin{equation}\label{TEaccSPinf2}
\omega_s =  \frac{\omega_p}{ \sqrt{2}}.
\end{equation}
We have for the corresponding numerical reduced frequency
$\omega_sa/c \approx 7.806$.

\qquad -- For the TM theory, there exists an accumulation of
resonances at the limiting frequency $\omega_f$ satisfying
\begin{equation}\label{TMaccSPinf1}
\mu (\omega_f) + 1 =0
\end{equation}
and given by
\begin{equation}\label{TMaccSPinf2}
\omega_f = \omega_0 \sqrt{\frac{2}{2-F}}.
\end{equation}
We have for the corresponding numerical reduced frequency
$\omega_fa/c \approx 6.172$.

\qquad -- For both theories, there exists an accumulation of
resonances at the pole $\omega_0$ of $\mu (\omega)$ and which
corresponds more precisely to
\begin{equation}\label{TMaccWGSP}
\mu (\omega_0) = - \infty.
\end{equation}

In summary, as far as the spectrum of resonances is concerned, the
left-handed cylinder is a physical system much richer than the
metallic or the semiconducting cylinder (see Figs.~1 and 2 of
Ref.~\onlinecite{AnceyDFG2004} and the discussion at the end of
Sec. II of that reference) and this is certainly very interesting
for practical applications of left-handed electromagnetism. With
this aim in view, it is necessary to understand the resonance
spectrum of the left-handed cylinder from a physical point of view
-- i.e., to decode the underlying physics. That is what we shall
do in the next sections. More precisely, we shall prove that this
resonance spectrum is generated by  SP's orbiting around the
cylinder and we shall provide a numerical and a theoretical
description of these surface waves.

It should be noted that the spectrum of resonances lies beyond the
frequency range where $n(\omega) <0$ (long-wave and short-wave
scattering). The corresponding resonant modes are associated with
bulk polaritons. Their complex resonances appear in Figs.~1b and
2b. Their imaginary parts are much larger in modulus than those
associated with the resonant modes generated by SP's. As a
consequence, they have a shorter lifetime, they do not play a
significant role in the scattering process (see Fig.1a and Fig.2a
beyond the frequency range  where $n(\omega) < 0$) and so they are
much less interesting with in mind practical applications. For all
these reasons, we focus our interest only on the resonant modes
associated with surface polaritons.

\section{Regge poles and surface polaritons}

\begin{figure}
\includegraphics[height=5.0cm,width=8.6cm]{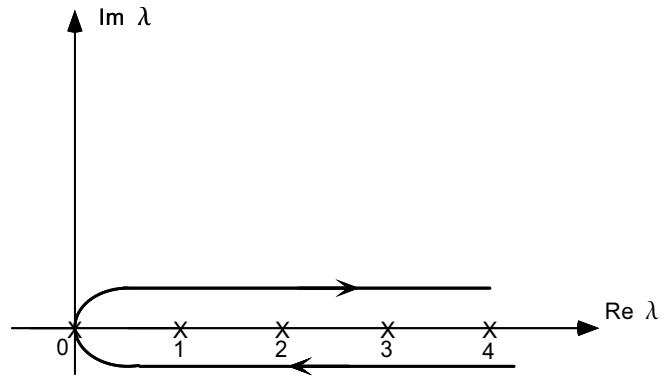}
\caption{\label{fig:watson} The Watson integration contour. }
\end{figure}

Using the CAM method, we shall first provide a physical picture of
the scattering of electromagnetic waves by the left-handed
cylinder in term of diffraction by surface waves. By means of a
Watson transformation \cite{Watson18} applied to the scattering
amplitude (\ref{ampli}), we can write
\begin{equation}\label{ampliII}
f(\omega, \theta)=\sqrt{ \frac{i} {2\pi k(\omega)}} ~ {\mathcal P}
\int_{\cal C} \frac{\left(S_{\lambda} (\omega) - 1 \right)}{\sin
\pi \lambda} \cos \left[\lambda (\pi -\theta)\right] d\lambda .
\end{equation}
Here, in order to simplify the notation, we have not specified the
polarization. In Eq.~(\ref{ampliII}), $\mathcal{C}$ is the
integration contour in the complex $\lambda$ plane (CAM plane)
illustrated in Fig.~\ref{fig:watson} and which encircles the real
axis in the clockwise sense. ${\mathcal P}$ which stands for
Cauchy's principal value at the origin is used in order to
reproduce the Neumann factor.  The Watson transformation has
permitted us to replace the ordinary angular momentum $\ell $ by
the complex angular momentum $\lambda$. $S_{\lambda} (\omega)$ is
now an analytic extension of $S_\ell (\omega )$ into the complex
$\lambda$ plane which is regular in the vicinity of the positive
real $\lambda$ axis. Using Cauchy's theorem and by noting that
inside the contour $\mathcal{C}$ the only singularities of the
integrand in (\ref{ampliII}) are the integers, we can easily
recover (\ref{ampli}) from (\ref{ampliII}).

\begin{figure}
\includegraphics[height=5.4cm,width=8.6cm]{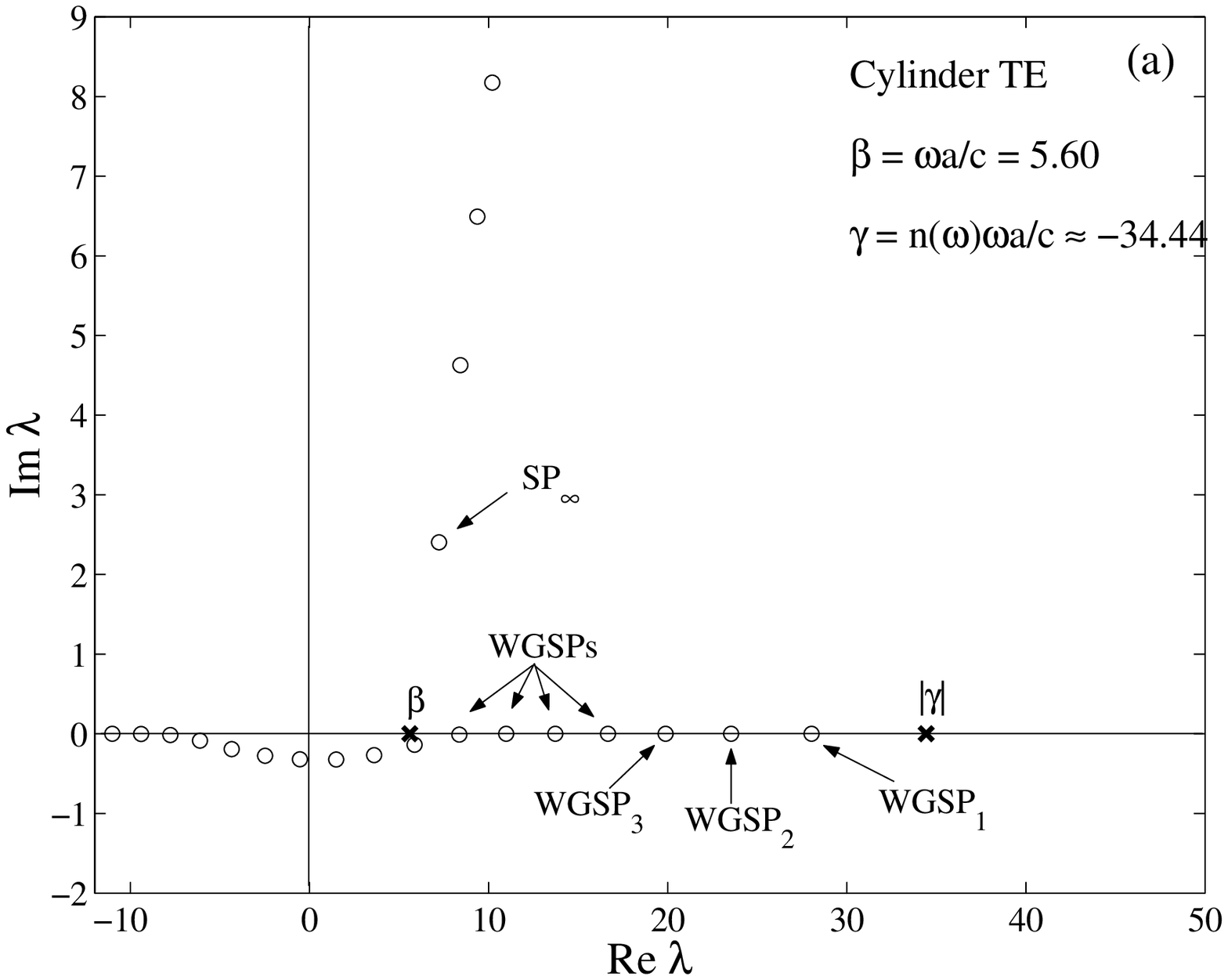}
\includegraphics[height=5.4cm,width=8.6cm]{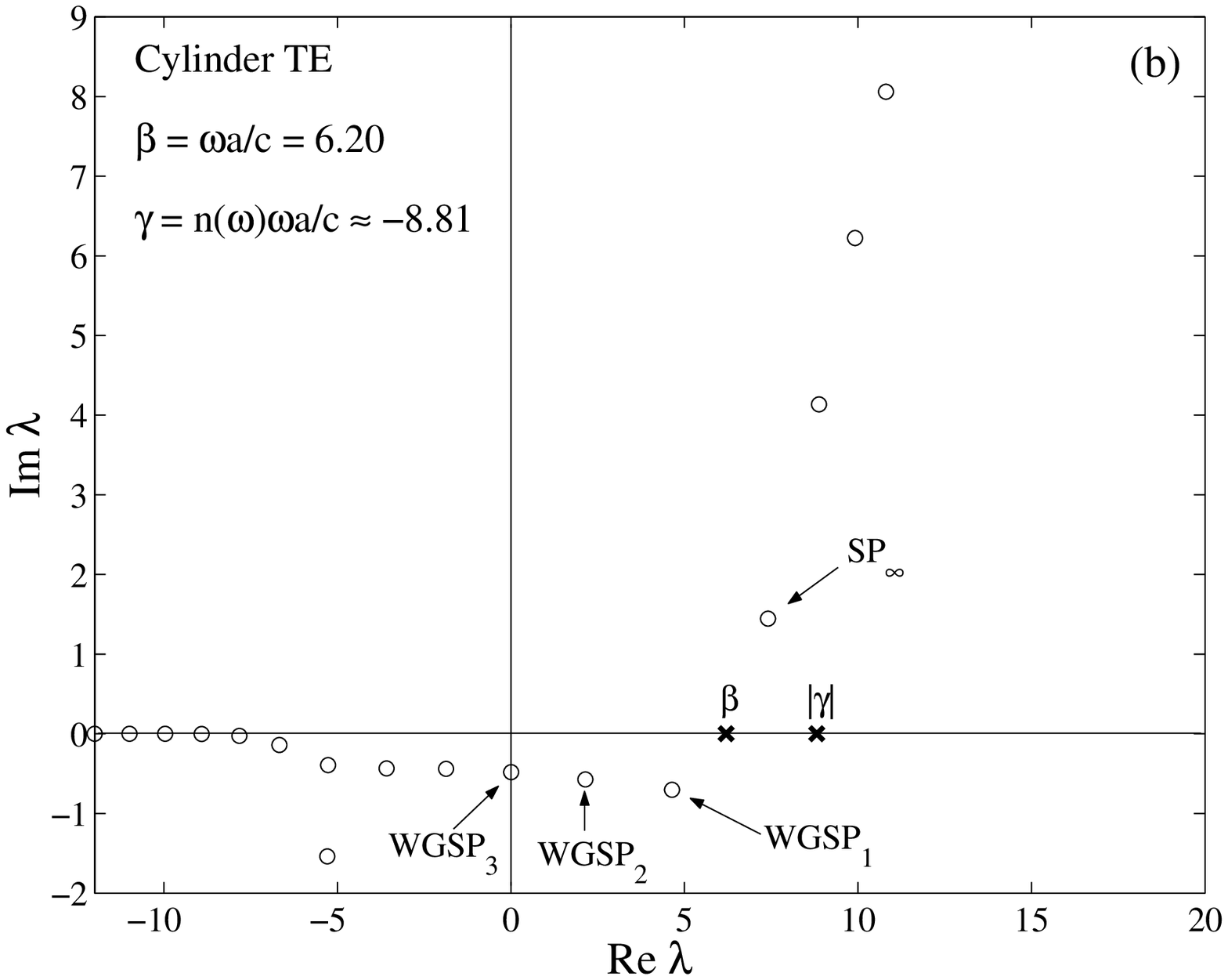}
\includegraphics[height=5.4cm,width=8.6cm]{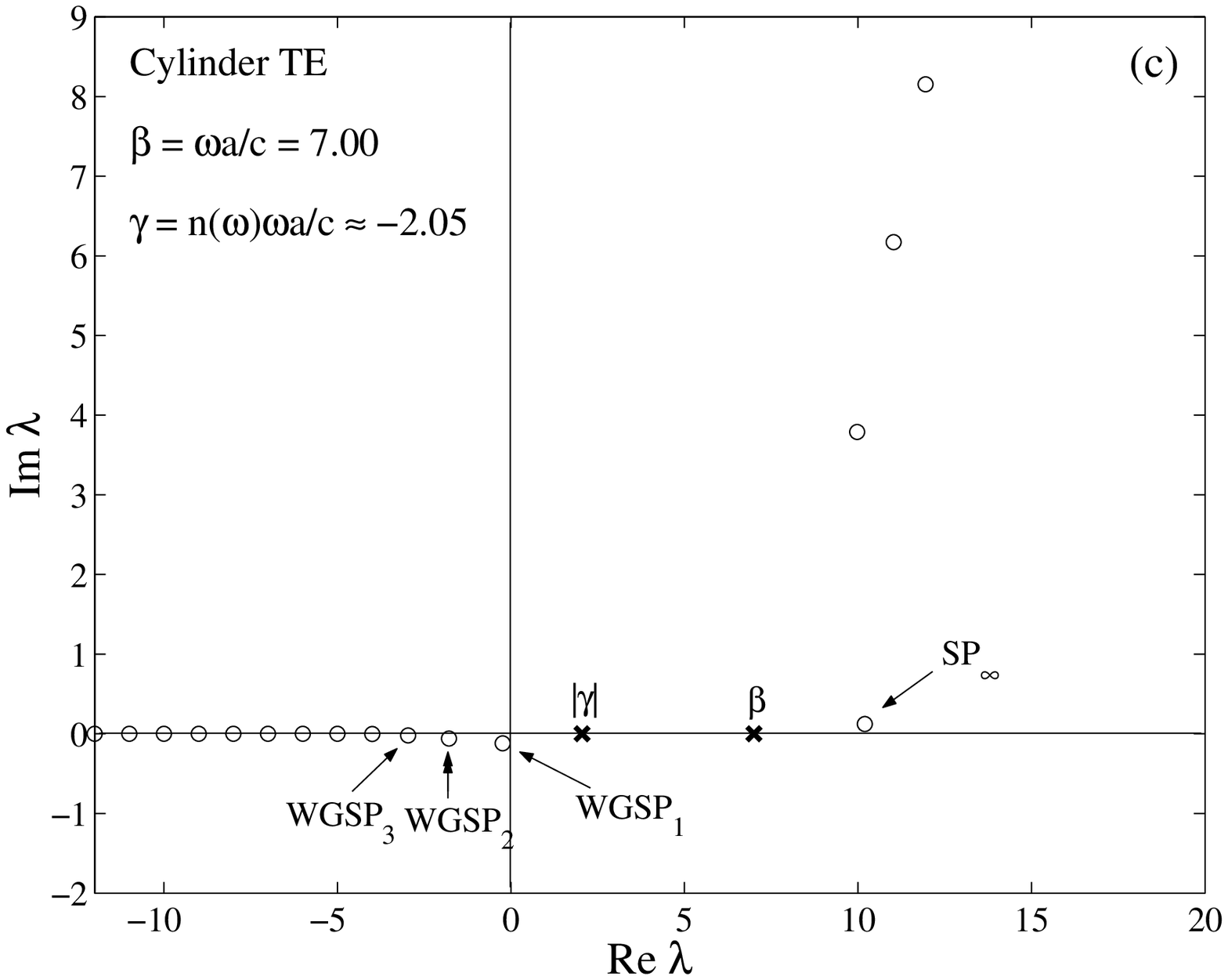}
\caption{\label{fig:RP1} Regge poles in the CAM plane for the TE
theory. a) The distribution corresponds to $\omega a/c= 5.6$. b)
The distribution corresponds to $\omega a/c= 6.2$. c) The
distribution corresponds to $\omega a/c= 7.0$.}
\end{figure}
\begin{figure}
\includegraphics[height=5.4cm,width=8.6cm]{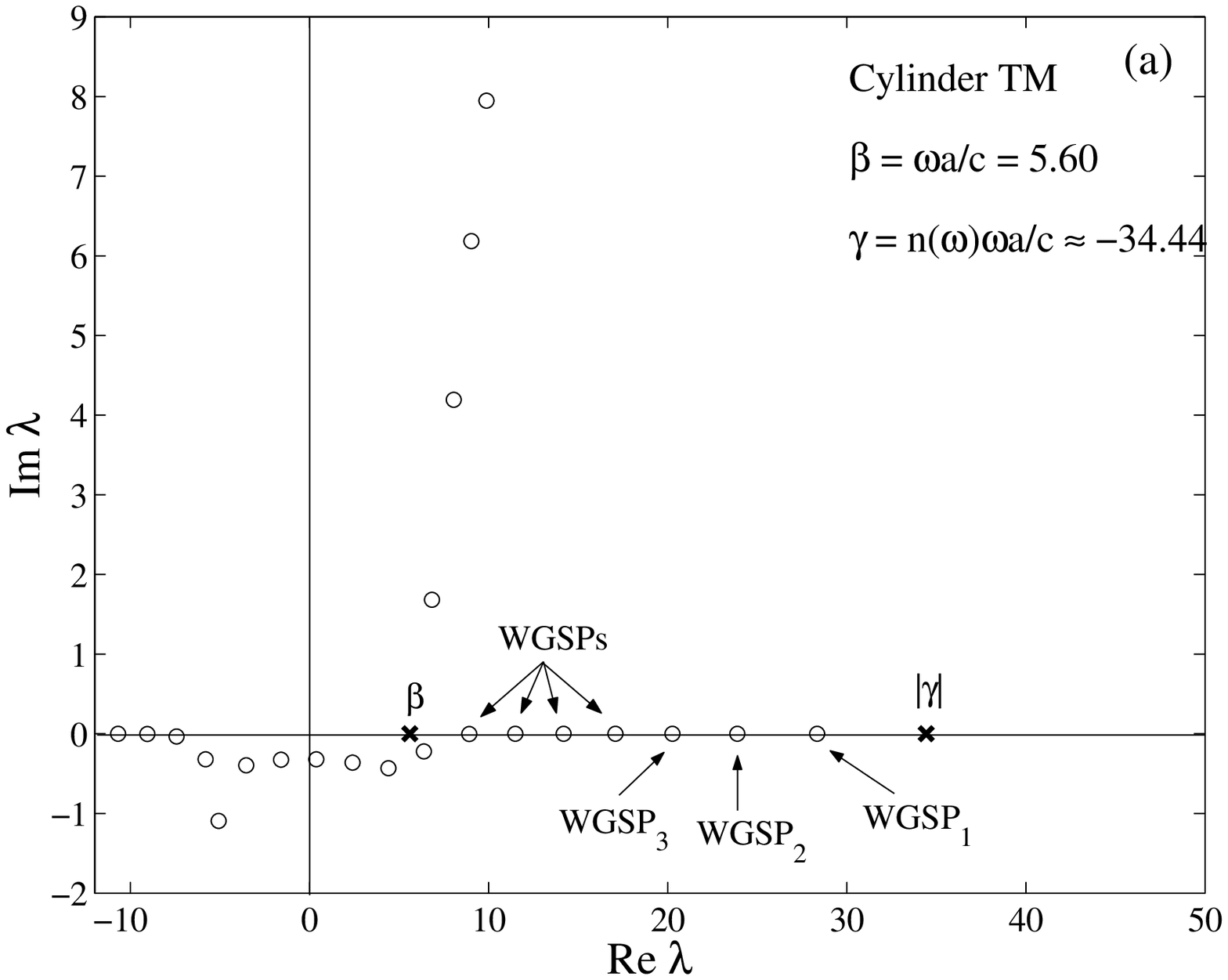}
\includegraphics[height=5.4cm,width=8.6cm]{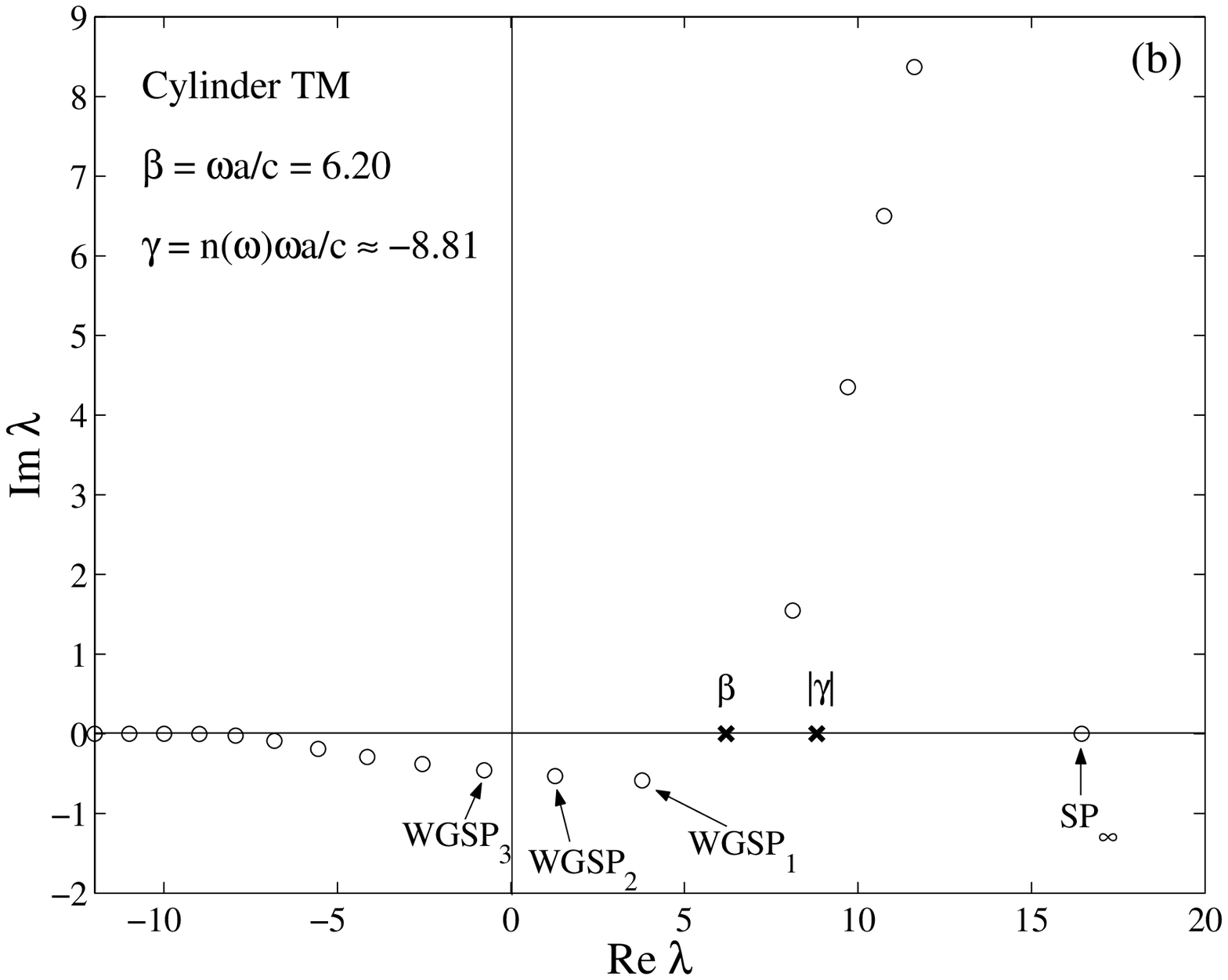}
\includegraphics[height=5.4cm,width=8.6cm]{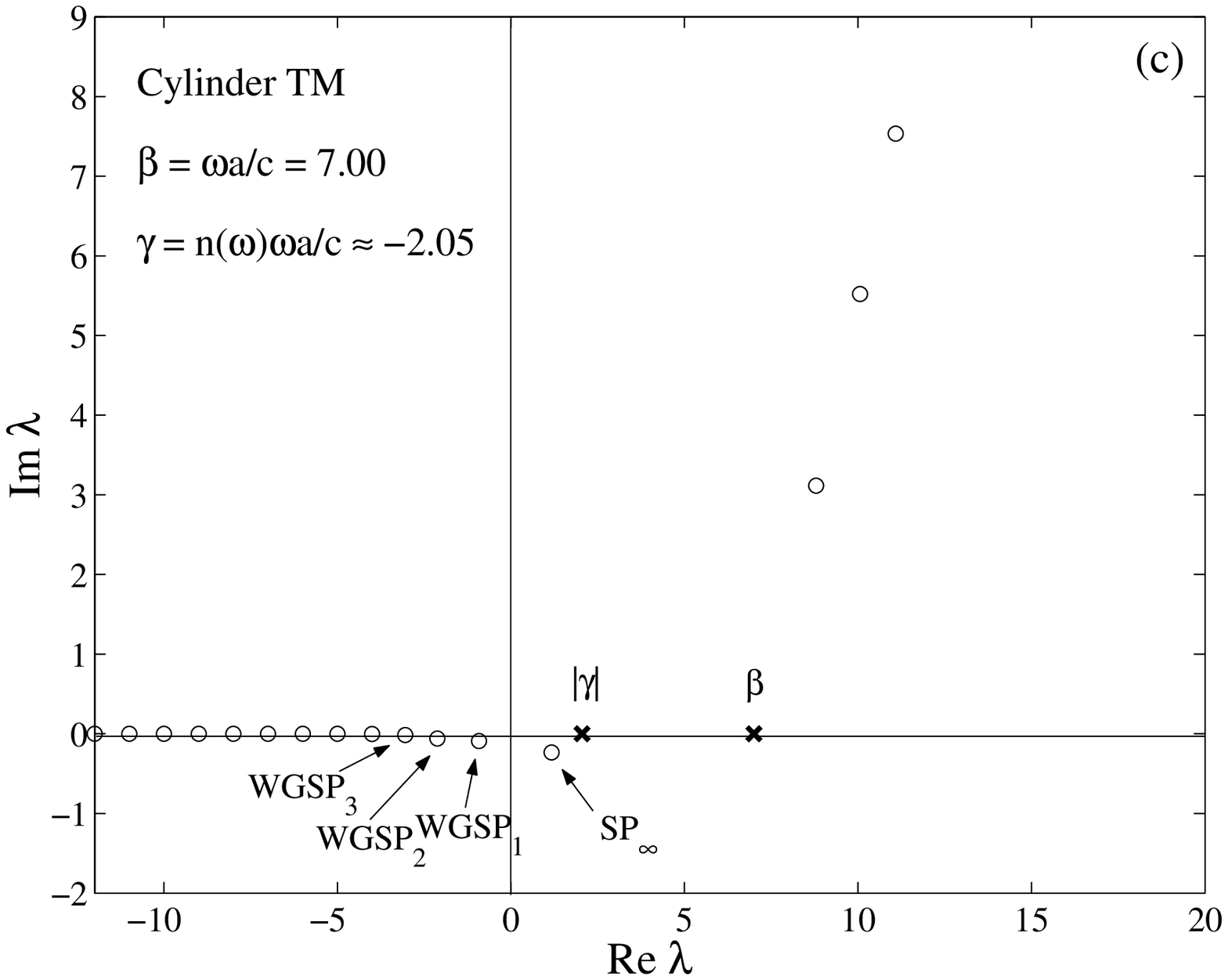}
\caption{\label{fig:RP2} Regge poles in the CAM plane for the TM
theory. a) The distribution corresponds to $\omega a/c= 5.6$. b)
The distribution corresponds to $\omega a/c= 6.2$. c) The
distribution corresponds to $\omega a/c= 7.0$.}
\end{figure}

We can then deform the path of integration in (\ref{ampliII})
taking into account the possible singularities. The only
singularities that are encountered are the poles of the S matrix
lying in the CAM plane. They are known as Regge poles
\cite{New82,Nus92} and are determined by solving
\begin{equation}\label{RPH}
D^H_\lambda( \omega)=0 \quad \mathrm{for} \quad \omega > 0
\end{equation}
for the TE theory and
\begin{equation}\label{RPE}
D^E_\lambda( \omega)=0 \quad \mathrm{for} \quad \omega > 0
\end{equation}
for the TM theory. Figures \ref{fig:RP1} and \ref{fig:RP2} exhibit
the distribution of Regge poles for both theories for three
different reduced frequencies lying in the frequency region where
$n(\omega) <0$. At first sight, these Regge pole distributions are
more complicated than the distributions associated with the
metallic and the semiconducting cylinders studied in
Ref.~\onlinecite{AnceyDFG2004}. However, we have identified and
indicated some particular Regge poles which, as we shall see
below, are associated with surface waves orbiting around the
cylinder and which explain its resonant behavior. For both
polarizations:

\qquad -- One of these Regge poles is associated with the SP
denoted ${\mathrm{SP}_\infty}$ which, as we shall show in Sec. V,
corresponds, in the large-radius limit (i.e., for $a \to \infty$),
to a SP which is supported by the plane interface and which has
been theoretically described in
Refs.~\onlinecite{RuppinPLA00,Darmanyanetal03,ShadrivovEtAl04}.

\qquad -- The other Regge poles are associated with an infinite
family of SP's of whispering-gallery type denoted by
${\mathrm{WGSP}_n}$ with $n \in \mathbb{N}$ and which have no
analogs in the plane interface case (see Sec. V).

\noindent For the TE theory, the Regge pole
$\lambda_{\mathrm{SP}_\infty}$ lies in the first quadrant of the
CAM plane while for the TM theory it lies in the fourth quadrant
of that plane. For both theories, the Regge poles
$\lambda_{\mathrm{WGSP}_n}$ lie in the fourth quadrant of the CAM
plane but as $\omega \to \omega_b$ they migrate to the third
quadrant of that plane where they do not play any role.

By Cauchy's theorem we can now extract from (\ref{ampliII}) the
contribution of a residue series over Regge poles.  For each Regge
pole $\lambda_\mathrm{SP}(\omega)$ (here $\mathrm{SP}$ stands for
$\mathrm{SP}_\infty$  as well for $\mathrm{WGSP}_n$), we capture
an associated contribution given by
\begin{equation}\label{ampliIIIa}
f_\mathrm{SP}(\omega, \theta)=\sqrt{\frac{2\pi }{i k (\omega)}}
\frac{r_\mathrm{SP}(\omega)}{\sin \left[\pi
\lambda_\mathrm{SP}(\omega)\right]} \cos
\left[\lambda_\mathrm{SP}(\omega) (\pi -\theta)\right]
\end{equation}
where $r_\mathrm{SP}(\omega)=\mathrm{residue}
\left(S_\lambda(\omega)\right)_{\lambda = \lambda
_\mathrm{SP}(\omega)}$.

By using
\begin{equation}
\frac{1}{\sin \pi \lambda }=-2i \sum_{m=0}^{+\infty} e^{
+i\pi(2m+1)\lambda} \nonumber
\end{equation}
which is true if $\mathrm{Im} \ \lambda > 0$, we can write
\begin{eqnarray}\label{ampliIIIb1}
&   &   f_\mathrm{SP}(\omega, \theta)=-\sqrt{\frac{2i\pi }{k
(\omega)}}~
r_\mathrm{SP}(\omega) \nonumber \\
&  & \qquad  \qquad  \times \sum_{m=0}^{+\infty}  \left(
e^{+i\lambda_\mathrm{SP} (\omega
)(\theta +2m\pi) } \right. +  \nonumber \\
&  & \qquad \qquad \qquad \qquad  \left.
e^{+i\lambda_\mathrm{SP}(\omega )(2\pi - \theta +2m\pi)} \right)
\quad
\end{eqnarray}
when the Regge pole $\lambda _\mathrm{SP}$ lies in the first
quadrant of the CAM plane. Expression (\ref{ampliIIIb1}) is
therefore associated with the surface wave ${\mathrm{SP}_\infty}$
of the TE theory. By using
\begin{equation}
\frac{1}{\sin \pi \lambda }=2i \sum_{m=0}^{+\infty} e^{
-i\pi(2m+1)\lambda} \nonumber
\end{equation}
which is true if $\mathrm{Im} \ \lambda < 0$, we can write
\begin{eqnarray}\label{ampliIIIb2}
&   &   f_\mathrm{SP}(\omega, \theta)=+\sqrt{\frac{2i\pi }{k
(\omega)}}~
r_\mathrm{SP}(\omega) \nonumber \\
&  & \qquad  \qquad  \times \sum_{m=0}^{+\infty}  \left(
e^{-i\lambda_\mathrm{SP} (\omega
)(\theta +2m\pi) } \right. +  \nonumber \\
&  & \qquad \qquad \qquad \qquad  \left.
e^{-i\lambda_\mathrm{SP}(\omega )(2\pi - \theta +2m\pi)} \right)
\quad
\end{eqnarray}
when the Regge pole $\lambda _\mathrm{SP}$ lies in the fourth
quadrant of the CAM plane. Expression (\ref{ampliIIIb2}) is
therefore associated with the surface waves ${\mathrm{WGSP}_n}$
with $n\in {\mathbb N}$ of the TE theory as well as with the
surface waves ${\mathrm{SP}_\infty}$ and ${\mathrm{WGSP}_n}$ with
$n\in {\mathbb N}$ of the TM theory.

In Eqs.~(\ref{ampliIIIb1}) and (\ref{ampliIIIb2}), exponential
terms correspond to diffractive contributions. This clearly
appears by taking into account the time dependence $\exp(-i\omega
t)$. The physical interpretations slightly differ according to the
position of the Regge pole $\lambda _\mathrm{SP}$ in the complex
$\lambda$ plane:

\qquad -- When the Regge pole $\lambda_\mathrm{SP}(\omega )$ lies
in the first quadrant of the CAM plane -- i.e., when we can use
Eq.~(\ref{ampliIIIb1}) -- the term
$\exp[i\lambda_\mathrm{SP}(\omega )(\theta)]$ (resp.
$\exp[i\lambda_\mathrm{SP}(\omega )(2\pi - \theta)]$) describes
the SP propagating counterclockwise (resp. clockwise) around the
cylinder  and $\mathrm{Re} \ \lambda_\mathrm{SP}(\omega)$
represents its azimuthal propagation constant while $\mathrm{Im} \
\lambda_\mathrm{SP}(\omega)$ is its damping constant. The
corresponding exponential decay reads $\exp[-\mathrm{Im} \
\lambda_\mathrm{SP}(\omega)\theta]$ (resp. $\exp[-\mathrm{Im} \
\lambda_\mathrm{SP}(\omega)(2\pi - \theta)]$).

\qquad -- When the Regge pole $\lambda_\mathrm{SP}(\omega )$ lies
in the fourth quadrant of the CAM plane -- i.e., when we can use
Eq.~(\ref{ampliIIIb2}) -- the term
$\exp[-i\lambda_\mathrm{SP}(\omega )(\theta)]$ (resp.
$\exp[-i\lambda_\mathrm{SP}(\omega )(2\pi - \theta)]$) describes
the SP propagating clockwise (resp. counterclockwise) around the
cylinder and $\mathrm{Re} \ \lambda_\mathrm{SP}(\omega)$
represents its azimuthal propagation constant while $-\mathrm{Im}
\ \lambda_\mathrm{SP}(\omega)$ is its damping constant. The
corresponding exponential decay reads $\exp[+\mathrm{Im} \
\lambda_\mathrm{SP}(\omega)\theta]$ (resp. $\exp[+\mathrm{Im} \
\lambda_\mathrm{SP}(\omega)(2\pi - \theta)]$).

\noindent Finally, in both cases, the sum over $m$ in
(\ref{ampliIIIb1}) and (\ref{ampliIIIb2}) takes into account the
multiple circumnavigations of the surface wave around the cylinder
as well as the associated radiation damping.

From the previous discussion, it is important to keep in mind that
the function ${\mathrm{Re} \ \lambda_\mathrm{SP} (\omega)}$
provides the dispersion relation for the SP associated with the
Regge pole $\lambda_\mathrm{SP}$ and that the phase velocity $v_p$
and the group velocity $v_g$ of that SP  are therefore given by
\begin{equation}\label{VpGg}
v_p = \frac{a \omega}{\mathrm{Re} \lambda_\mathrm{SP} (\omega)}
\quad \mathrm{and} \quad v_g = \frac{d~a \omega}{d~\mathrm{Re}
\lambda_\mathrm{SP} (\omega)}.
\end{equation}
Here we have taken into account the fact that the SP is supported
by the cylinder surface at $\rho = a$ and therefore that its wave
number is given by
\begin{equation}\label{WNSP}
k_\mathrm{SP} (\omega) = \frac{\mathrm{Re} \lambda_\mathrm{SP}
(\omega)}{a}.
\end{equation}
Moreover, it is also important to note that the Regge poles
$\lambda_{\mathrm{SP}_\infty}$ and $\lambda_{\mathrm{WGSP}_n}$
with $n \in \mathbb{N}$ are always close to the real axis in the
complex $\lambda$ plane. As a consequence, they all correspond to
SP's which are slightly attenuated during their propagation and
which contribute significantly to the scattering process and to
the resonance excitation mechanism.

\begin{figure}
\includegraphics[height=8cm,width=8.6cm]{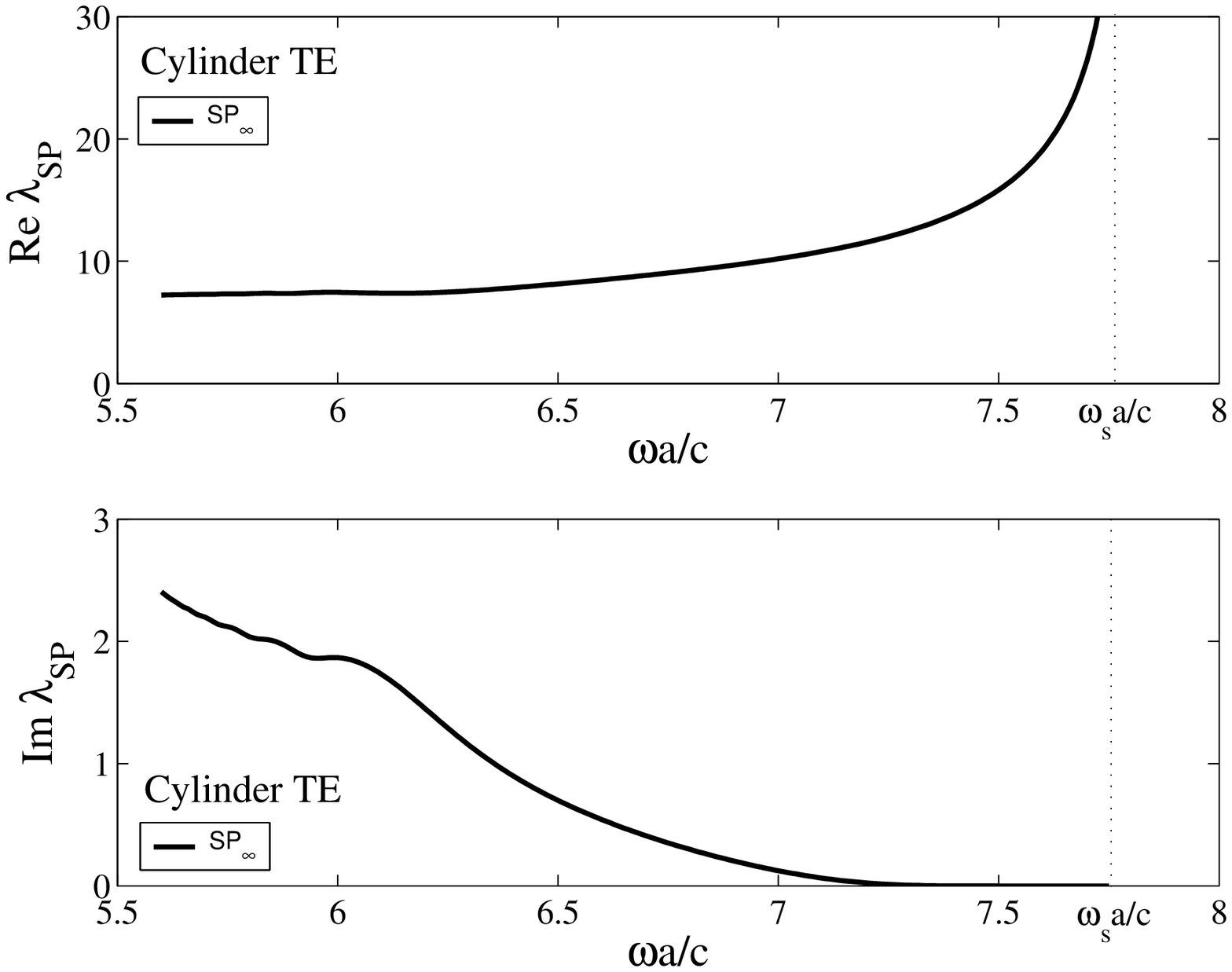}
\caption{\label{fig:RTSPiH} Regge trajectory for the Regge pole
associated with ${\mathrm{SP}_\infty}$ (TE theory). }
\includegraphics[height=8cm,width=8.6cm]{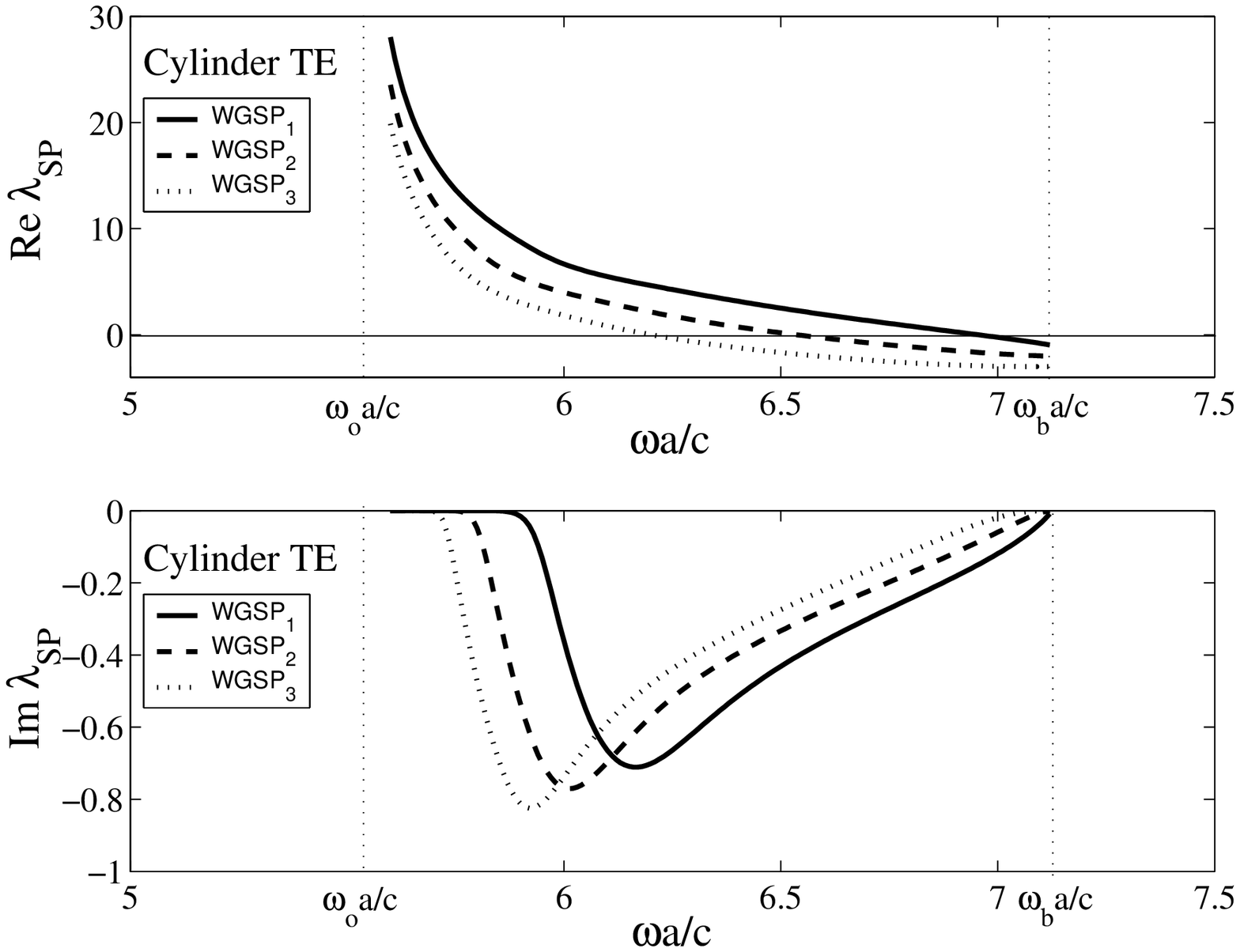}
\caption{\label{fig:RTWGSPH} Regge trajectories for the Regge
poles associated with the first three whispering-gallery SP's (TE
theory). }
\end{figure}

\begin{figure}
\includegraphics[height=8cm,width=8.6cm]{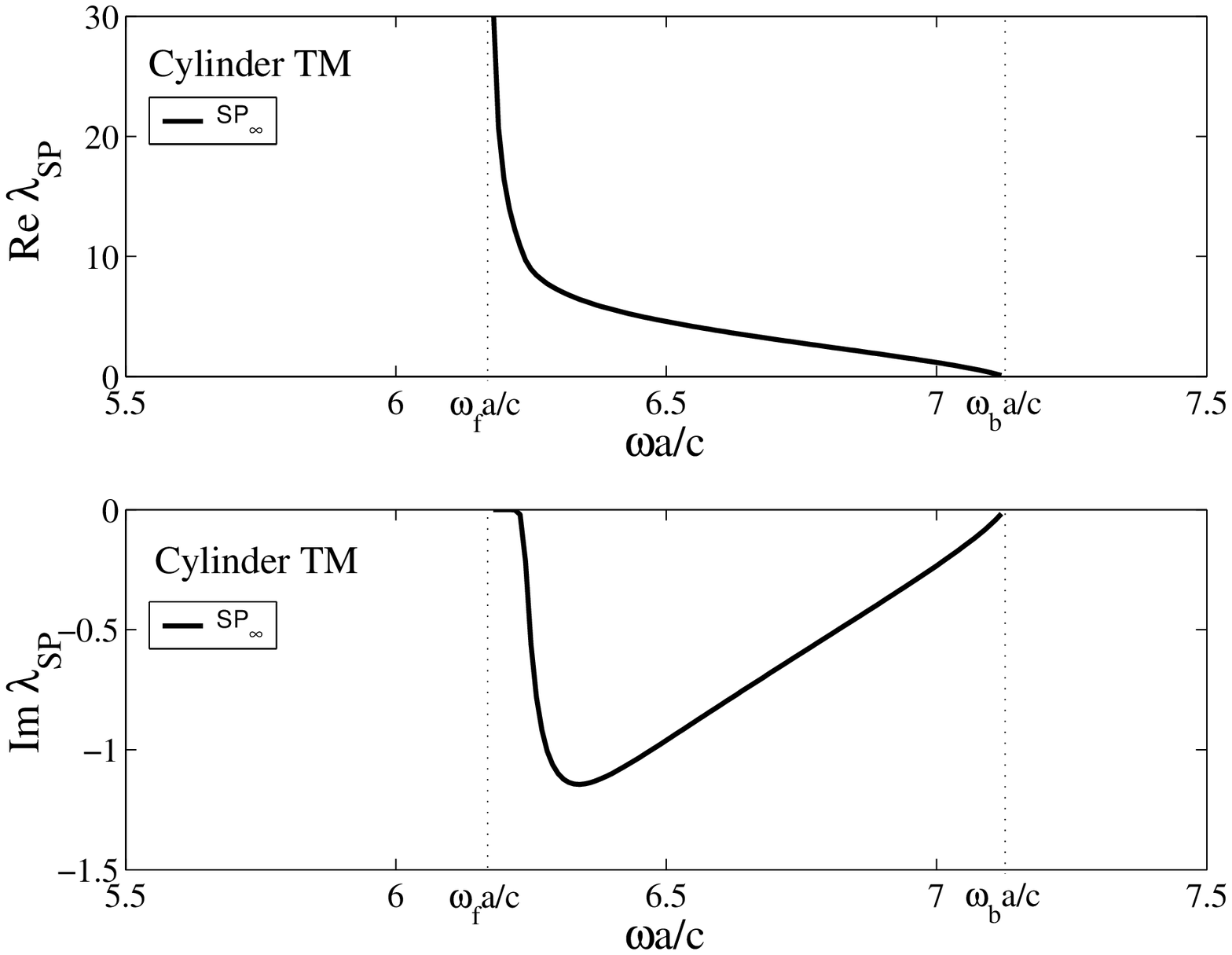}
\caption{\label{fig:RTSPiE} Regge trajectory for the Regge pole
associated with ${\mathrm{SP}_\infty}$ (TM theory). }
\includegraphics[height=8cm,width=8.6cm]{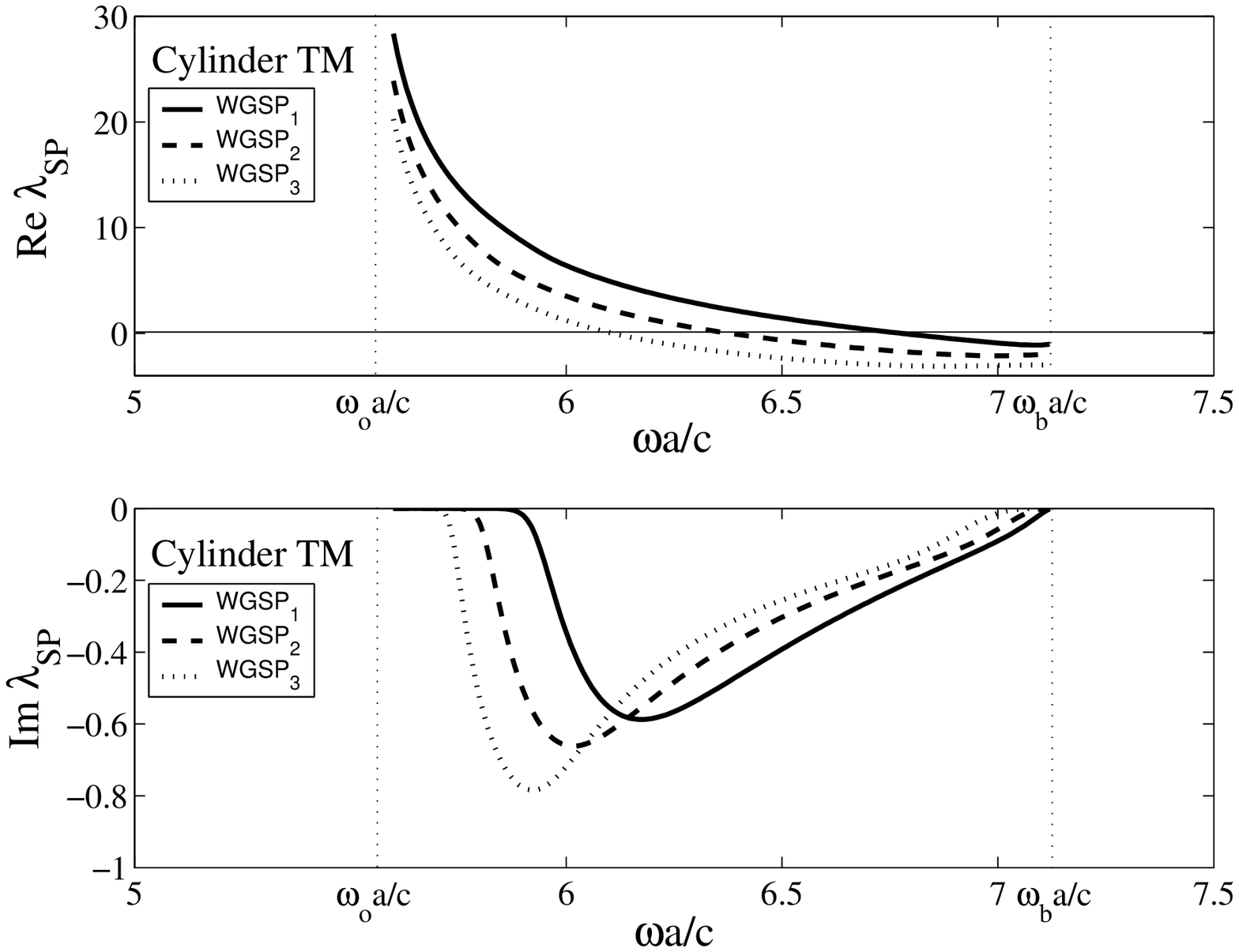}
\caption{\label{fig:RTWGSPE} Regge trajectories for the Regge
poles associated with the first three whispering-gallery SP's (TM
theory). }
\end{figure}

As $\omega $ varies, a given Regge pole
$\lambda_\mathrm{SP}(\omega )$ describes a curve in the CAM plane.
Such a curve is called a Regge trajectory \cite{New82}. From a
physical point of view, the Regge trajectory corresponding to
$\lambda_\mathrm{SP}(\omega )$ provides the dispersion relation as
well as the damping of the surface wave associated with this pole.
In Figs.~\ref{fig:RTSPiH}-\ref{fig:RTWGSPE}, we have displayed the
Regge trajectories of some SP's for the TE and TM theories. They
have been obtained by solving numerically Eqs.~(\ref{RPH}) and
(\ref{RPE}). We can observe some interesting features:

\qquad -- The dispersion curve for the surface wave
${\mathrm{SP}_\infty}$ of the TE theory is a positive and
monotonically increasing function of $\omega$. As a consequence,
the associated group and phase velocities given by
Eq.~(\ref{VpGg}) are both positive and ${\mathrm{SP}_\infty}$ has
an ordinary behavior. It should be also noted that this SP exists
in the frequency range $\omega \in ]0, \omega_s[$ and therefore in
the range $]\omega_0, \omega_b[$ where the refraction index is
negative but also outside this range. However, for low values of
$\omega$, its damping becomes very high and thus this surface wave
has a negligible role in the scattering process and in the
resonance excitation mechanism. Furthermore, it should be noted
that as $\omega \to \omega_s$ the dispersion curve increases
indefinitely. In the next section, this result will permit us to
explain the accumulation of resonances which converges to the
limiting frequency $\omega_s$ for the TE theory.

\qquad -- The dispersion curve for the surface wave
${\mathrm{SP}_\infty}$  of the TM theory is a positive and
monotonically decreasing function of $\omega$. As a consequence,
the associated phase velocity is positive while the group velocity
is negative (see Eq.~(\ref{VpGg})). ${\mathrm{SP}_\infty}$ has a
``left-handed behavior". It should be also noted that this SP only
exists in the frequency range $]\omega_f, \omega_b[$ which is
included in the frequency range $]\omega_0, \omega_b[$ where the
refraction index is negative, that its damping is always weak and
thus that this surface wave always plays a significant role in the
scattering process and in the resonance excitation mechanism.
Finally, it should be noted that as $\omega \to \omega_f$ the
dispersion curve increases indefinitely. In the next section, this
result will permit us to explain the accumulation of resonances
which converges to the limiting frequency $\omega_f$ for the TM
theory.

\qquad -- As far as the surface waves ${\mathrm{WGSP}_n}$ with $n
\in \mathbb{N}$ of the TE  and  TM theories are concerned, it
seems at first sight they present a behavior which is rather
independent of the polarization. It should be also noted that the
real part of a given Regge pole $\lambda_{\mathrm{WGSP}_n}$
vanishes for a frequency in the frequency range $]\omega_0,
\omega_b[$ and becomes negative. The Regge pole then lies in the
third quadrant of the CAM plane and is not taken into account by
the theory previously developed. So we can consider that the
surface waves ${\mathrm{WGSP}_n}$ with $n \in \mathbb{N}$ only
exist in a sub-domain of the frequency range $]\omega_0,
\omega_b[$ where the refraction index is negative. The dispersion
relations of all these surface waves are positive but
monotonically decreasing functions.  Their group velocities are
always negative while their phase velocities are positive (see
Eq.~(\ref{VpGg})). All these SP's thus have a ``left-handed
behavior". Furthermore, because the dampings of these surface
waves are always weak, they all play a significant role in the
scattering process and in the resonance excitation mechanism.
Finally, it should be noted that as $\omega \to \omega_0$, the
dispersion curves increase indefinitely. In the next section, this
result will permit us to explain the accumulation of resonances
which converges to the limiting frequency $\omega_0$ for the TE
and TM theories.

\section{From surface polaritons to resonances}

\begin{figure}
\includegraphics[height=5.3cm,width=8.6cm]{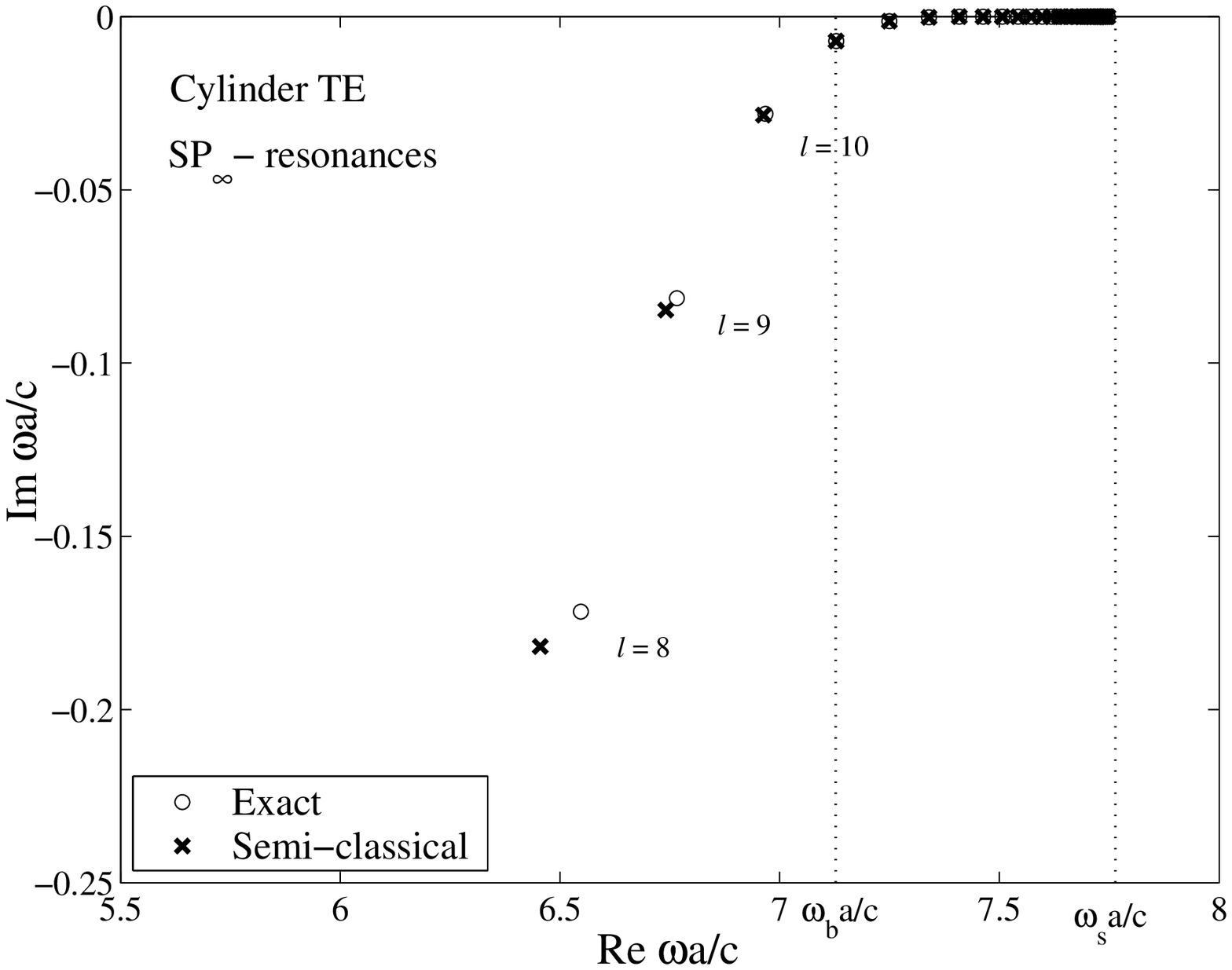}
\caption{\label{fig:ResSPinfH} Resonances generated by
${\mathrm{SP}_\infty}$ (TE theory). }
\includegraphics[height=5.3cm,width=8.6cm]{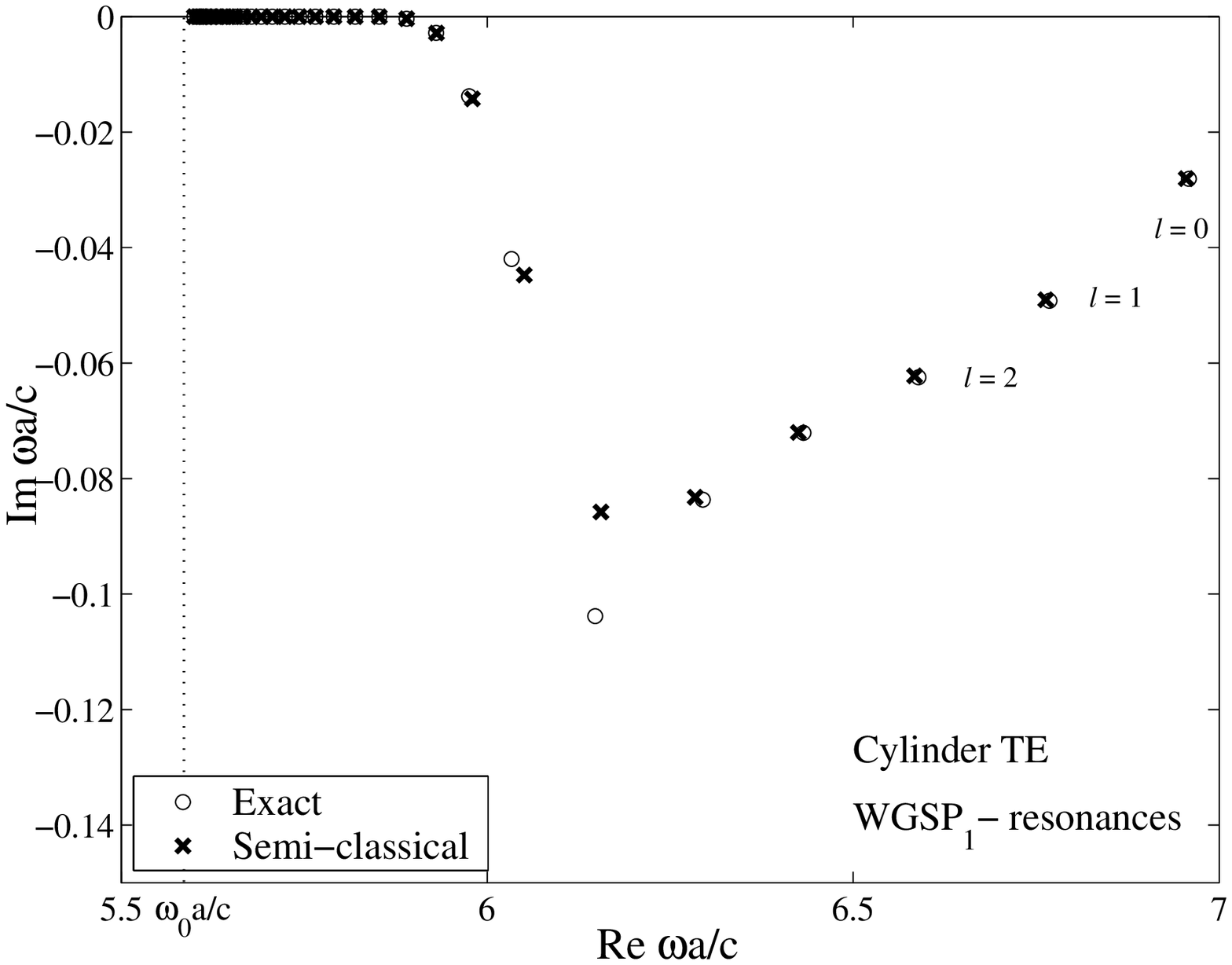}
\includegraphics[height=5.3cm,width=8.6cm]{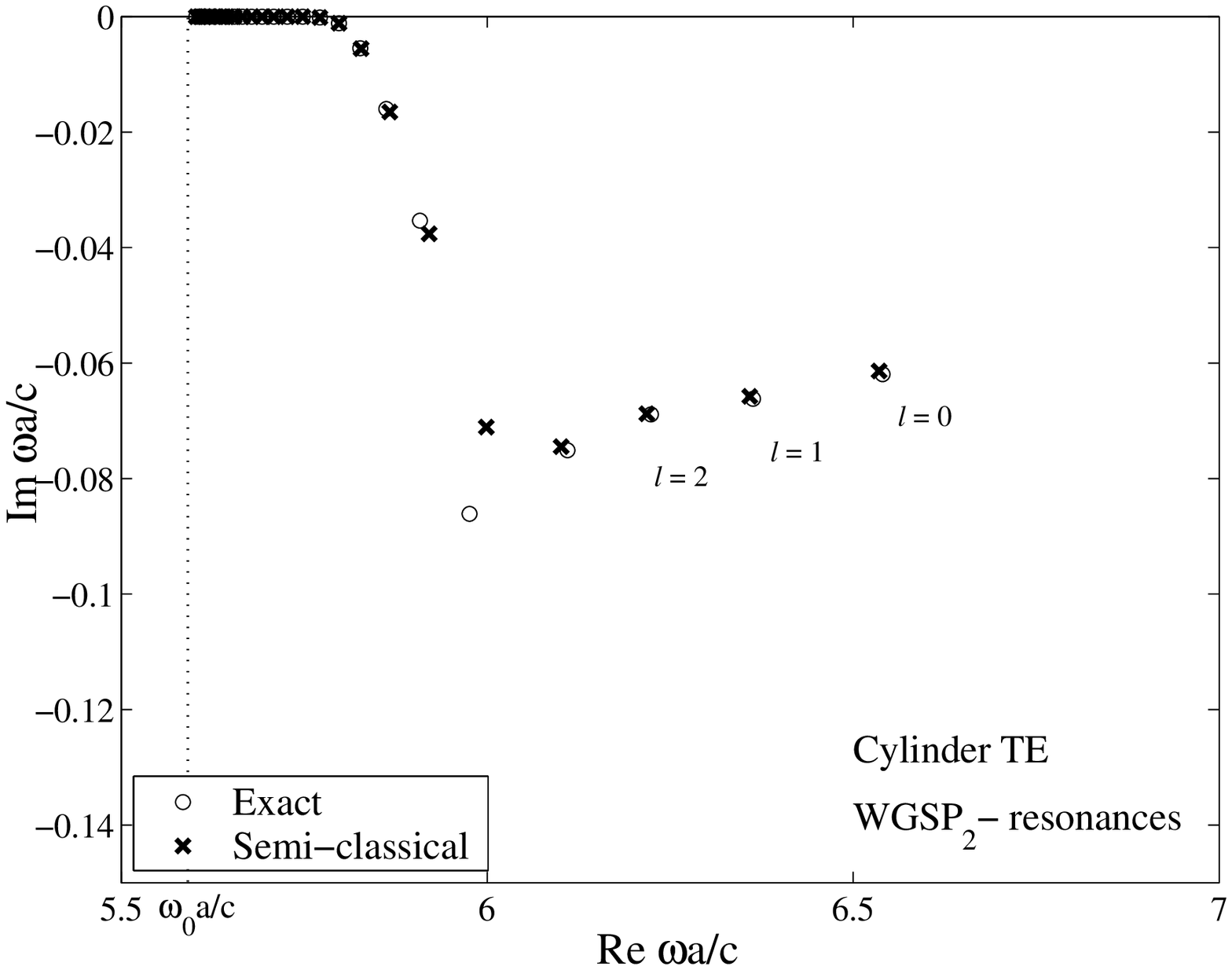}
\includegraphics[height=5.3cm,width=8.6cm]{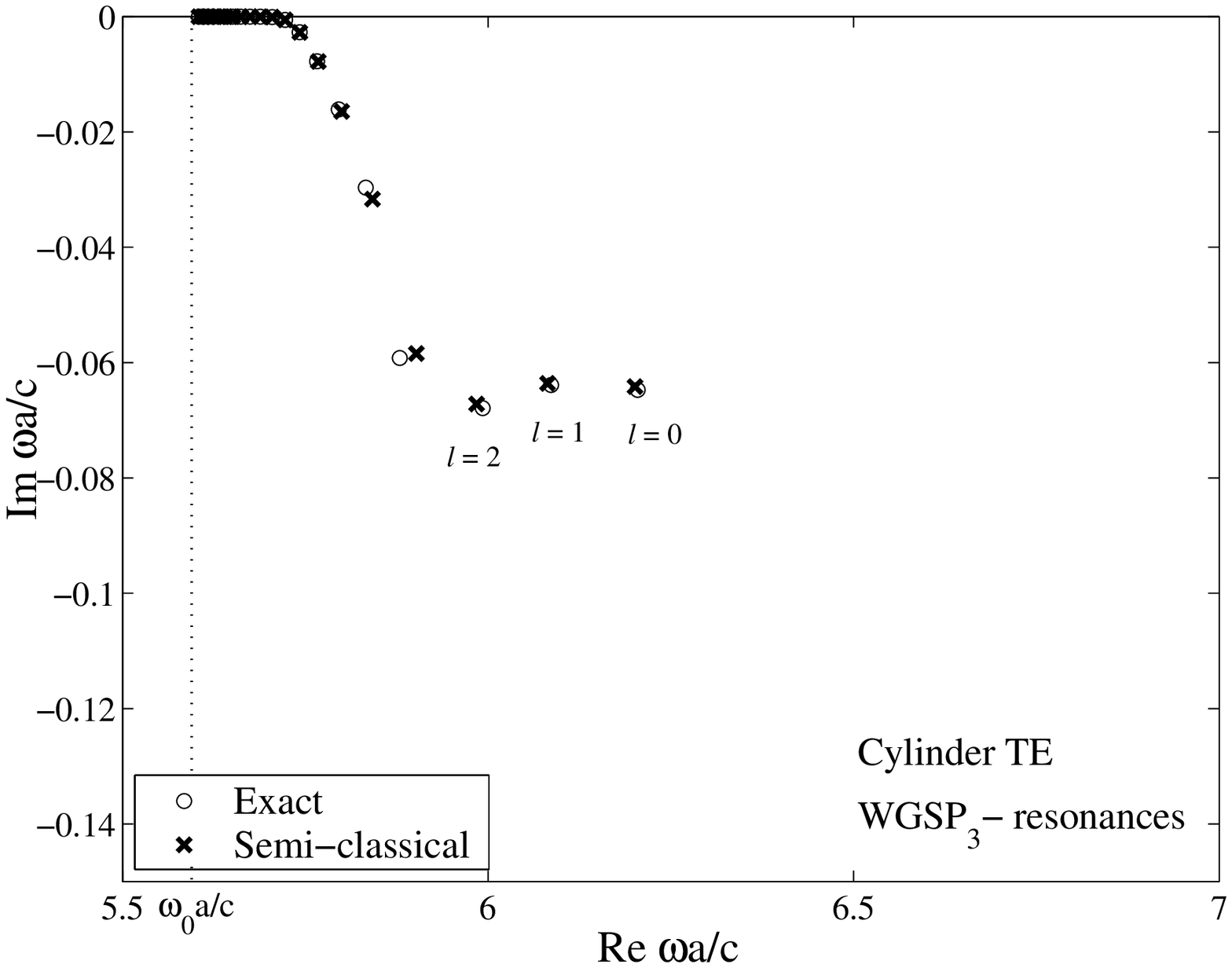}
\caption{\label{fig:ResWGSPH} Resonances generated by the first
three whispering-gallery SP's (TE theory). }
\end{figure}
\begin{figure}
\includegraphics[height=5.3cm,width=8.6cm]{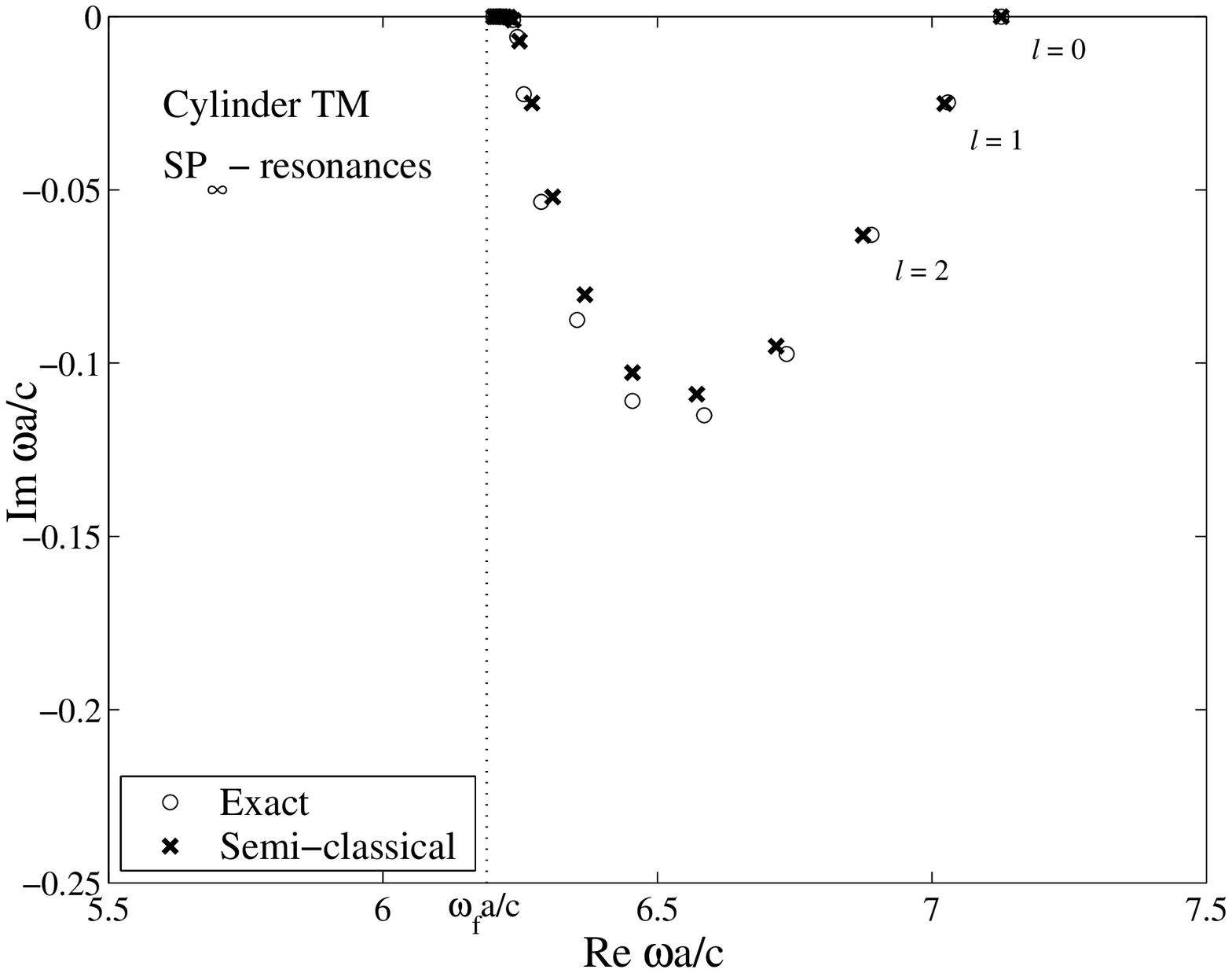}
\caption{\label{fig:ResSPinfE}  Resonances generated by
${\mathrm{SP}_\infty}$ (TM theory). }
\includegraphics[height=5.3cm,width=8.6cm]{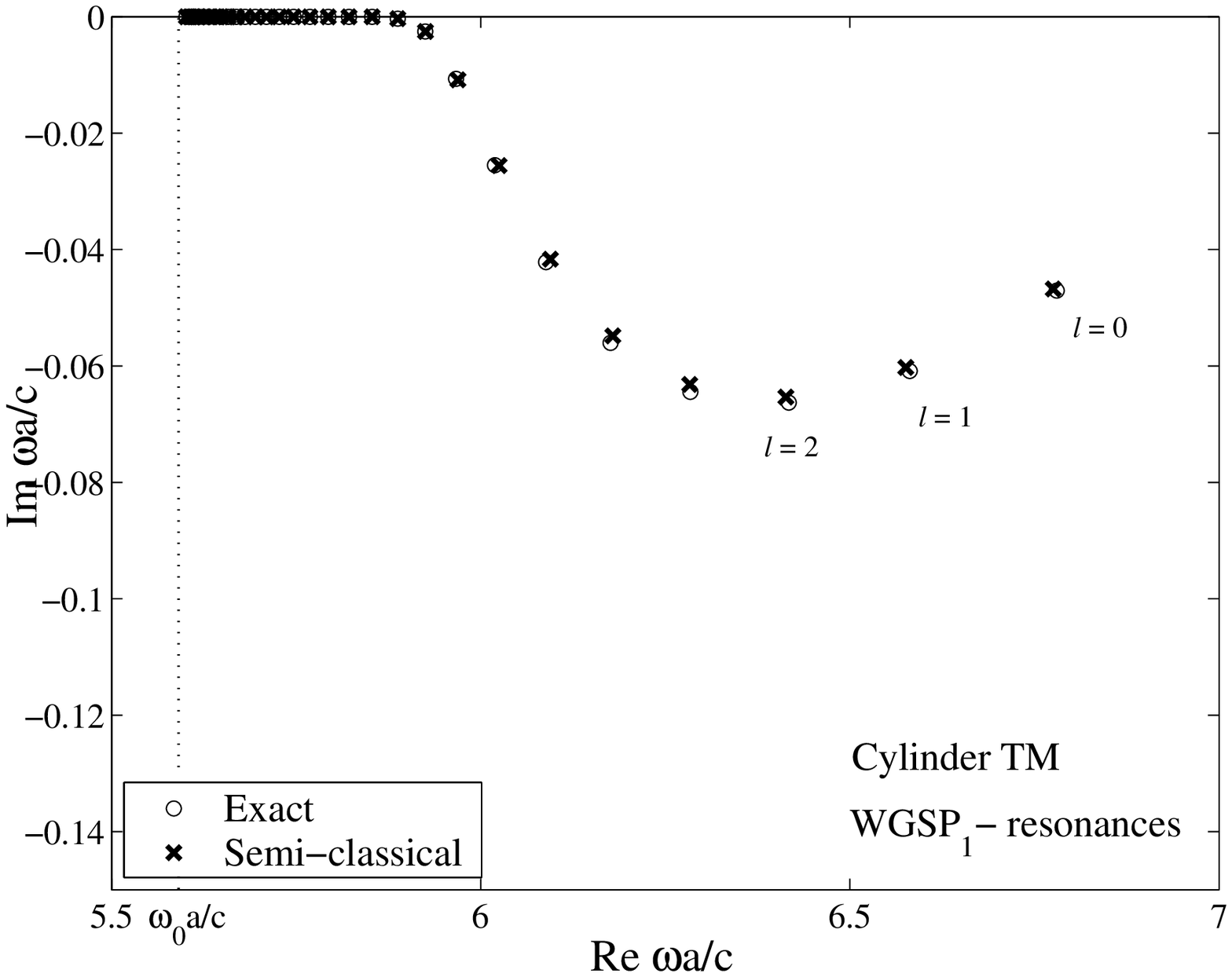}
\includegraphics[height=5.3cm,width=8.6cm]{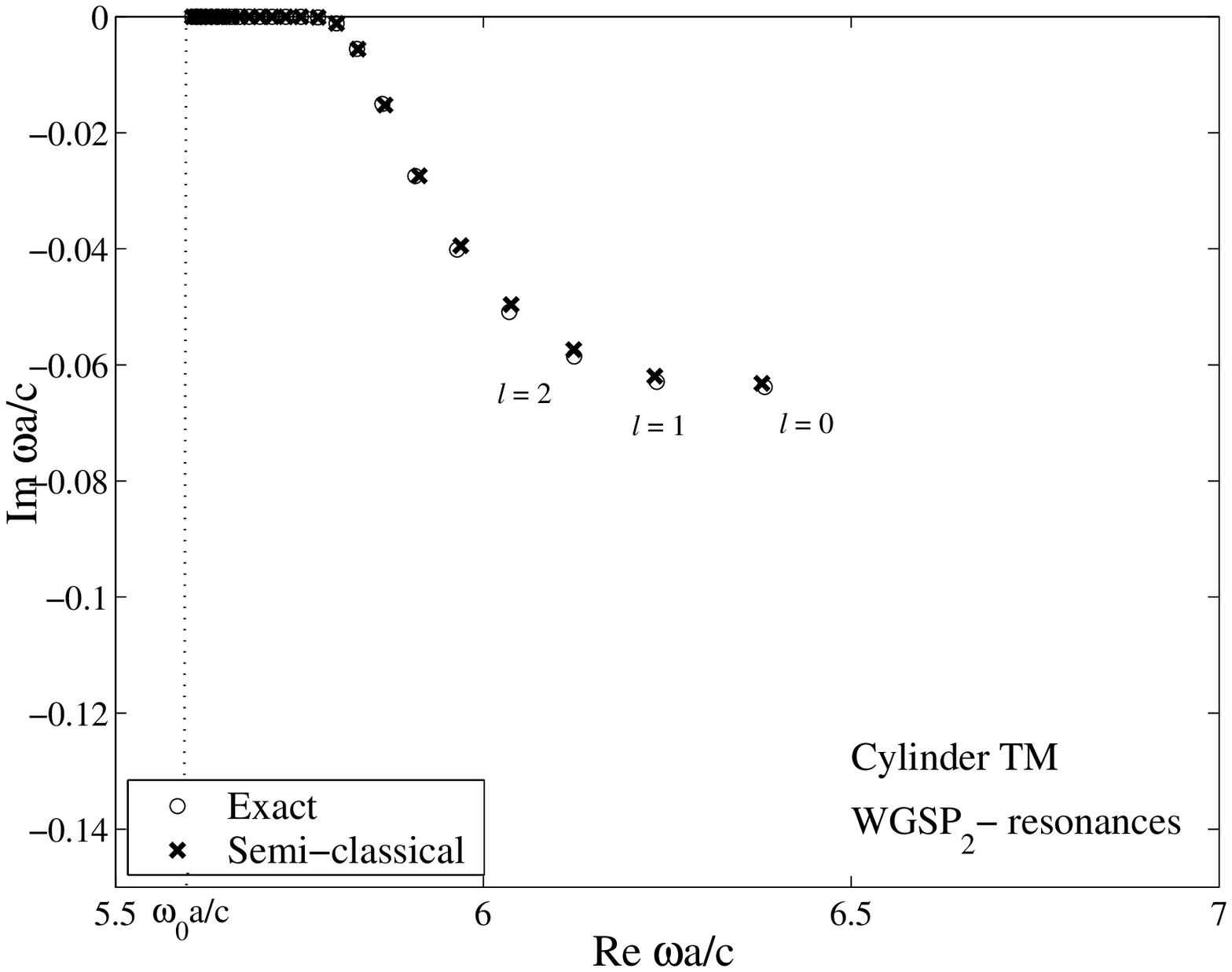}
\includegraphics[height=5.3cm,width=8.6cm]{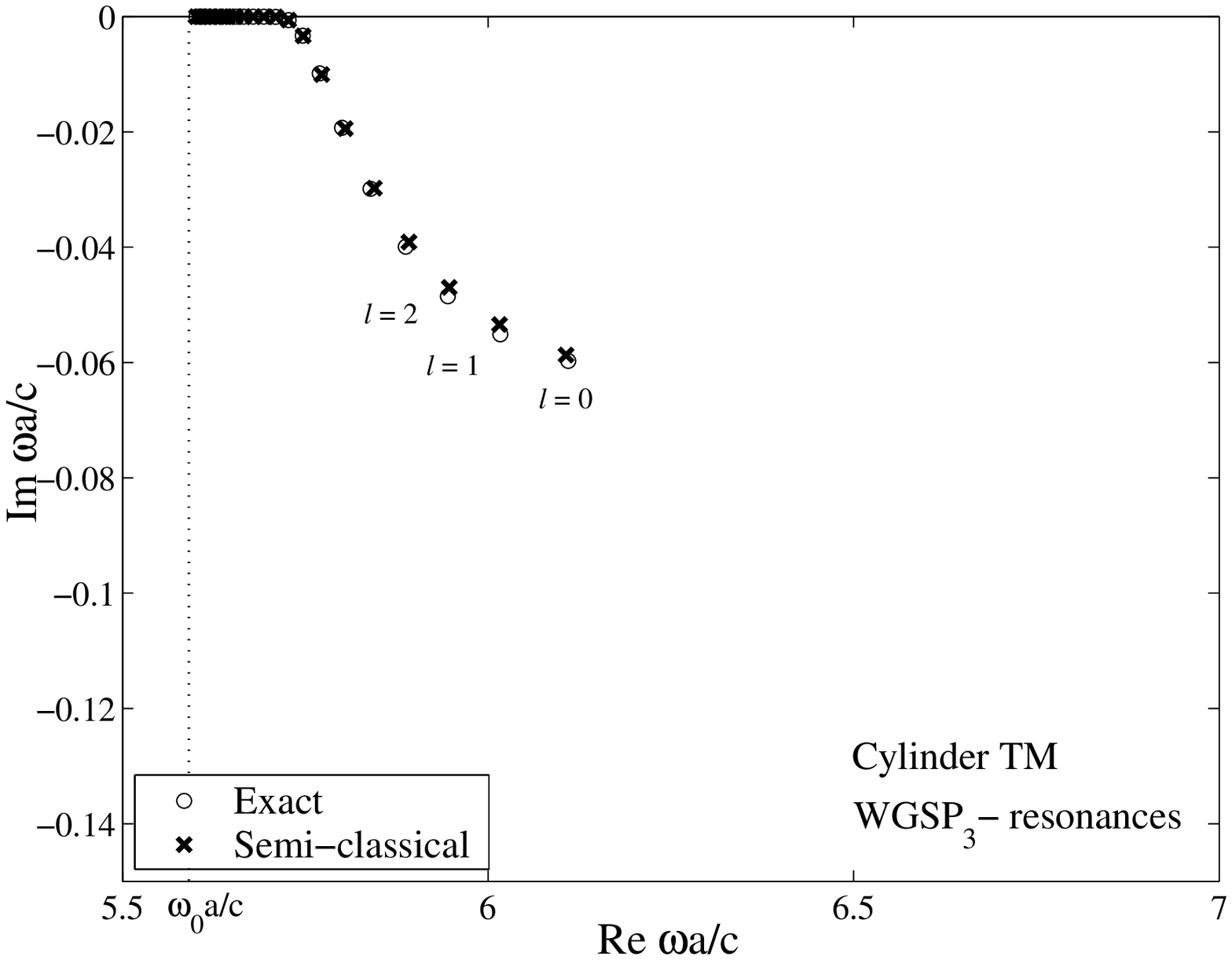}
\caption{\label{fig:ResWGSPE} Resonances generated by the first
three whispering-gallery SP's (TM theory). }
\end{figure}

From now on, we shall focus our attention on the physical
interpretation of the long-lived resonant modes whose excitation
frequencies are localized within or slightly around the frequency
range $\omega_0  < \omega  < \omega_b $ where the cylinder
presents left-handed behavior. In this section, we shall prove
that they are generated by the SP's described in Sec. III and for
this reason we have called them RSPM's.

We first consider that the solutions of (\ref{detH}) or
(\ref{detE}) are denoted by $\omega_{\ell p}=\omega^{(0)}_{\ell
p}-i\Gamma _{\ell p}/2$ where $\omega^{(0)}_{\ell p}>0$ and
$\Gamma _{\ell p}>0$, the index $p$ permitting us to distinguish
between the different roots of (\ref{detH}) or (\ref{detE}) for a
given $\ell$. We also assume that in the immediate neighborhood of
the resonance $\omega_{\ell p}$, $S^H_\ell(\omega)$ or
$S^E_\ell(\omega)$ and therefore the corresponding scattering
amplitudes obtained from (\ref{ampli}) present a Breit-Wigner type
behavior -- i.e., are proportional to
\begin{equation}\label{BW}
\frac{\Gamma _{\ell p}/2}{\omega -\omega^{(0)}_{\ell p}+i\Gamma
_{\ell p}/2}.
\end{equation}
The resonant behavior of the cylinder (or more precisely of the
cylinder-photon system) can then be understood in terms of SP
Regge trajectories. Let us consider a given SP (we do not precise
its polarization) described by the Regge pole denoted
$\lambda_\mathrm{SP}(\omega )$. When the quantity $\mathrm{Re} \
\lambda_\mathrm{SP}(\omega )$ coincides with an integer, a
resonance occurs which is produced by a constructive interference
between the different components of the surface wave, each
component corresponding to a different number of circumnavigations
of the cylinder (see Eqs.~(\ref{ampliIIIb1}) and
(\ref{ampliIIIb2})). Resonance excitation frequencies
$\omega^{(0)}_{\ell \mathrm{SP}}$ associated with that SP are
therefore obtained from the Bohr-Sommerfeld-type quantization
condition
\begin{equation}\label{sc1}
\mathrm{Re}   \lambda_\mathrm{SP} \left(\omega^{(0)}_{\ell
\mathrm{SP}} \right)= \ell  \qquad \ell =0,1,2,\dots .
\end{equation}
Thus, they can be obtained numerically from the dispersion curves.
Now, by assuming that $\omega $ is in the neighborhood of
$\omega^{(0)}_{\ell \mathrm{SP}}$, we can expand
$\lambda_\mathrm{SP} (\omega)$ in a Taylor series about
$\omega^{(0)}_{\ell \mathrm{SP}}$ and write
\begin{eqnarray} \label{TS}
& &  \lambda_\mathrm{SP}(\omega) \approx \ell  +  \left. \frac{d
\mathrm{Re} \lambda_\mathrm{SP}(\omega)}{d\omega} \right|_{\omega
=\omega^{(0)}_{\ell \mathrm{SP}}} (\omega - \omega^{(0)}_{\ell
\mathrm{SP}}  ) \nonumber \\
& & \quad +  i \mathrm{Im} \lambda_\mathrm{SP} (\omega^{(0)}_{\ell
\mathrm{SP}}) + i \left. \frac{d \mathrm{Im}
\lambda_\mathrm{SP}(\omega)}{d\omega} \right|_{\omega
=\omega^{(0)}_{\ell \mathrm{SP}}} (\omega - \omega^{(0)}_{\ell
\mathrm{SP}}  ) \nonumber \\ &  &   \quad  + \dots .
\end{eqnarray}
Then, by replacing (\ref{TS}) in the term ${\sin \left[\pi
\lambda_\mathrm{SP}(\omega )  \right]}$ of (\ref{ampliIIIa}), we
can see that $f_\mathrm{SP}(\omega, \theta)$ presents a resonant
behavior given by the Breit-Wigner formula (\ref{BW}) with
\begin{equation}\label{sc2}
\frac{\Gamma _{\ell \mathrm{SP}}}{2}= \left.  \frac{\mathrm{Im}
\lambda_\mathrm{SP} (\omega )(d \mathrm{Re} \lambda_\mathrm{SP}
(\omega ) /d\omega)}{(d  \mathrm{Re} \lambda_\mathrm{SP} (\omega )
/d\omega)^2 + (d  \mathrm{Im} \lambda_\mathrm{SP} (\omega )
/d\omega)^2 } \right|_{\omega =\omega^{(0)}_{\ell \mathrm{SP}}}.
\end{equation}
Our reasoning also predicts a shift in the excitation frequency
$\omega^{(0)}_{\ell \mathrm{SP}} $ solution of Eq.~(\ref{sc1})
which is given by
\begin{equation}\label{sc3}
\Delta \omega^{(0)}_{\ell \mathrm{SP}} = -\left. \frac{\mathrm{Im}
\lambda_\mathrm{SP} (\omega )(d \mathrm{Im} \lambda_\mathrm{SP}
(\omega ) /d\omega)}{(d \mathrm{Re} \lambda_\mathrm{SP} (\omega )
/d\omega)^2 + (d \mathrm{Im} \lambda_\mathrm{SP} (\omega )
/d\omega)^2 } \right|_{\omega =\omega^{(0)}_{\ell \mathrm{SP}}}.
\end{equation}
In fact, such a shift is always negligible and it can be
forgotten. Furthermore, it should be noted that in some frequency
range we have
\begin{equation}
 | d  \mathrm{Re} \lambda_\mathrm{SP} (\omega
) /d\omega | \gg |d  \mathrm{Im} \lambda_\mathrm{SP} (\omega )
/d\omega |
\end{equation}
so that (\ref{sc2}) then reduces to
\begin{equation}\label{sc4}
\frac{\Gamma _{\ell \mathrm{SP}}}{2}= \left.  \frac{\mathrm{Im} \
\lambda_\mathrm{SP} (\omega )}{d \ \mathrm{Re} \lambda_\mathrm{SP}
(\omega ) /d\omega } \right|_{\omega =\omega^{(0)}_{\ell
\mathrm{SP}}}.
\end{equation}

Equations (\ref{sc1}) and (\ref{sc2}) (or (\ref{sc4})) are two
semiclassical formulas which permit us to determine the location
of the resonances from the Regge trajectory of
$\lambda_\mathrm{SP}$. In
Figs.~\ref{fig:ResSPinfH}-\ref{fig:ResWGSPE} we present samples of
complex frequencies for the RSPM's associated with the surface
waves ${\mathrm{SP}_\infty}$ and ${\mathrm{WGSP}_n}$ with $n = 1$,
$2$ and $3$. They have been calculated from the semiclassical
formulas (\ref{sc1}) and (\ref{sc2}) by using the Regge
trajectories determined numerically by solving Eqs.~(\ref{RPH})
and (\ref{RPE}) (see Figs.~\ref{fig:RTSPiH}-\ref{fig:RTWGSPE}). A
comparison between the semiclassical spectra and the ``exact ones"
(calculated by solving numerically Eqs.~(\ref{detH}) or
(\ref{detE})) shows a very good agreement.  Moreover, we can also
observe some interesting features:

\qquad -- The resonance spectrum associated with the surface wave
${\mathrm{SP}_\infty}$ of the TE theory (see
Fig.~\ref{fig:ResSPinfH}) extends beyond the frequency range
$]\omega_0, \omega_b[$ where the cylinder presents left-handed
behavior because ${\mathrm{SP}_\infty}$ exists for $\omega \in ]0,
\omega_s[$. Furthermore, inserted into the semiclassical formulas
(\ref{sc1}) and (\ref{sc2}), the behavior of the Regge trajectory
of $\lambda_{\mathrm{SP}_\infty}$ near $\omega_s$ easily explains
the existence of the family of resonances close to the real axis
of the complex $\omega$ plane which converges for large $\ell$ to
the limiting frequency $\omega_s$.

\qquad -- The resonance spectrum associated with the surface wave
${\mathrm{SP}_\infty}$ of the TM theory (see
Fig.~\ref{fig:ResSPinfE}) fully lies inside the frequency range
$]\omega_0, \omega_b[$ where the cylinder presents left-handed
behavior because ${\mathrm{SP}_\infty}$ exists only in that range.
Furthermore, inserted into the semiclassical formulas (\ref{sc1})
and (\ref{sc2}), the behavior of the Regge trajectory of
$\lambda_{\mathrm{SP}_\infty}$ near $\omega_f$ explains the
existence of the family of resonances close to the real axis of
the complex $\omega$ plane which converges for large $\ell$ to the
limiting frequency $\omega_f$.

\qquad -- The resonance spectra associated with the surface waves
${\mathrm{WGSP}_n}$ with $n \in \mathbb{N}$ of the TE  and TM
theories (see Figs.~\ref{fig:ResWGSPH} and \ref{fig:ResWGSPE})
fully lie inside the frequency range $]\omega_0, \omega_b[$ where
the cylinder presents left-handed behavior because all these
surface waves exist only in that range. Furthermore, inserted into
the semiclassical formulas (\ref{sc1}) and (\ref{sc2}), the
behavior of the Regge trajectory of a given Regge pole
$\lambda_{\mathrm{WGSP}_n}$ near $\omega_0$ explains the existence
of a corresponding family of resonances close to the real axis of
the complex $\omega$ plane which converges for large $\ell $ to
the limiting frequency $\omega_0$. Since there is an infinity of
surface waves ${\mathrm{WGSP}_n}$, the accumulation of resonances
in Fig.~\ref{fig:general_TE}b (resp. Fig.~\ref{fig:general_TM}b)
is more intense at the point $\omega_0$ than at the point
$\omega_s$ (resp. $\omega_f$).

\begin{figure}
\includegraphics[height=6cm,width=7.6cm]{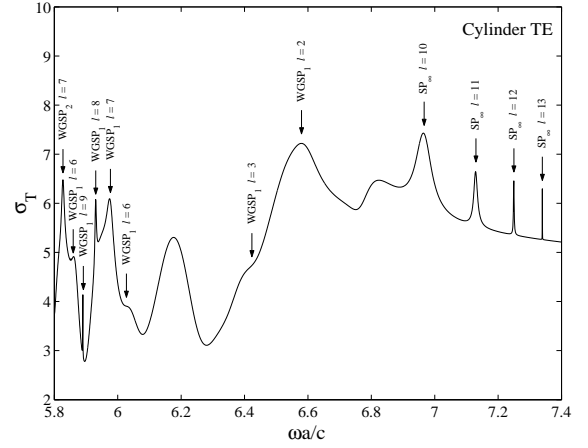}
\caption{\label{fig:ZoomTE} Zoom-in on the total cross section
$\sigma^H_T$.}
\end{figure}
\begin{figure}
\includegraphics[height=6cm,width=8.6cm]{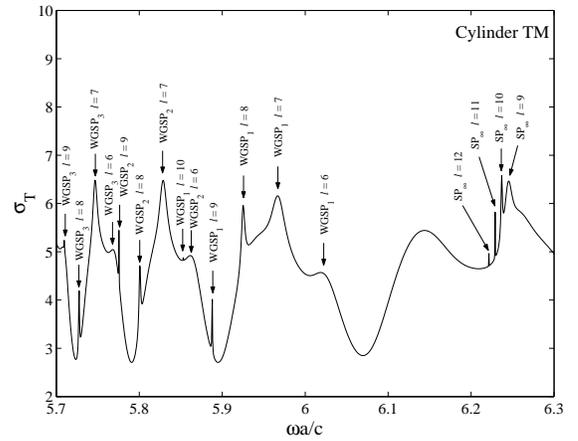}
\caption{\label{fig:ZoomTM} Zoom-in on the total cross section
$\sigma^E_T$.}
\end{figure}

In conclusion, we have established a connection between the
complex frequencies of the long-lived resonant modes (or RSPM's)
of the cylinder and the SP's noted ${\mathrm{SP}_\infty}$ and
${\mathrm{WGSP}_n}$ with $n \in \mathbb{N}$ which are supported by
its surface. In other words, in spite of the great confusion which
seems to prevail in the resonance spectrum of the left-handed
cylinder (see Sec. II), we have been able to fully classify and
physically interpret the resonances thanks to CAM techniques. We
now invite the reader to look at Figs.~\ref{fig:ZoomTE} and
\ref{fig:ZoomTM} where we have zoomed in on
Figs.~\ref{fig:general_TE}a and \ref{fig:general_TM}a. On the
total cross sections $\sigma^H _T(\omega)$ and $\sigma^E
_T(\omega)$ we have identified the peaks corresponding to
resonances and, for each one, we have specified the SP which has
generated it as well as the  associated ``quantum number" $\ell$.

\section{Asymptotics for
surface polaritons and physical description}

In order to obtain a deeper physical understanding of the SP's
orbiting around the left-handed cylinder and to justify the
terminology previously used, we must ``analytically" solve
Eqs.~(\ref{RPH}) and (\ref{RPE}) for $\lambda =
\lambda_\mathrm{SP}$ or equivalently
\begin{equation} \label{RPSP1H}
-\frac{J'_{\lambda_\mathrm{SP}} (|n(\omega)| \, \omega
a/c)}{J_{\lambda_\mathrm{SP}}(|n(\omega)| \, \omega
a/c)}=\sqrt{\frac{\epsilon (\omega)}{\mu (\omega)}}
\frac{H_{\lambda_\mathrm{SP}}^{(1)'} (\omega
a/c)}{H_{\lambda_\mathrm{SP}}^{(1)} (\omega a/c)}
\end{equation}
for the TE theory and
\begin{equation} \label{RPSP1E}
-\frac{J'_{\lambda_\mathrm{SP}}(|n(\omega)| \, \omega
a/c)}{J_{\lambda_\mathrm{SP}}(|n(\omega)| \, \omega
a/c)}=\sqrt{\frac{\mu (\omega)}{\epsilon (\omega)}}
\frac{H_{\lambda_\mathrm{SP}}^{(1)'} (\omega
a/c)}{H_{\lambda_\mathrm{SP}}^{(1)} (\omega a/c)}
\end{equation}
for the TM theory. Here, Eqs.~(\ref{RPSP1H}) and (\ref{RPSP1E})
have been obtained from Eqs.~(\ref{RPH}) and (\ref{RPE}) by using
the relations $J_\lambda (-z)=e^{i\pi \lambda }J_\lambda (z) $ and
$ J'_\lambda (-z)=-e^{i\pi \lambda }J'_\lambda (z) $ (see
Ref.~\onlinecite{AS65}). These two equations can be solved
approximatively by using asymptotic analysis. More precisely, by
considering adequate asymptotic expansions for the Bessel
functions occurring in (\ref{RPSP1H}) and (\ref{RPSP1E}), we shall
be able to provide analytic expressions for the Regge poles
$\lambda_{\mathrm{SP}_\infty}(\omega)$ and
$\lambda_{\mathrm{WGSP}_n} (\omega)$ for the two polarizations. Of
course, the main difficulty with this method concerns the choice
of the asymptotic expansions for the Bessel functions. Such a
choice strongly depends on the relative positions of the arguments
$|n(\omega)|\omega a/c$ and $\omega a/c$ with respect to the
complex order $\lambda_\mathrm{SP}$. In order to simplify the
discussion, we choose to describe theoretically the SP's in the
frequency ranges where they generate the RSPM's with the longest
lifetime (such modes are the most important from the physical
point of view). In other words, we shall seek
$\lambda_{\mathrm{SP}_\infty}(\omega)$ of the TE theory for
$\omega$ in the neighborhood of $\omega_s$,
$\lambda_{\mathrm{SP}_\infty}(\omega)$ of the TM theory for
$\omega$ in the neighborhood of $\omega_f$ and
$\lambda_{\mathrm{WGSP}_n}(\omega)$ of the TE  and TM theories for
$\omega$ in the neighborhood of $\omega_0$ . In fact, in spite of
these restrictions, we shall obtain asymptotic results valid in
large frequency ranges.

Let us first consider the Regge pole associated with
${\mathrm{SP}_\infty}$ of the TE theory. We assume $\omega$ in the
neighborhood of $\omega_s$ and then we can also assume that
$\mathrm{Re} \lambda_{\mathrm{SP}_\infty}(\omega) > \omega a/c >
|n(\omega)|\omega a/c$ and formally that $\mathrm{Re}
\lambda_{\mathrm{SP}_\infty}(\omega) \gg 1$ and $\mathrm{Re}
\lambda_{\mathrm{SP}_\infty}(\omega) \gg \mathrm{Im}
\lambda_{\mathrm{SP}_\infty}(\omega)$. The configuration there is
displayed in Fig.~\ref{fig:RP1}c. As a consequence, we can use the
Debye asymptotic expansions for $J_\lambda(z)$ and
$H_\lambda^{(1)}(z)$ valid for large orders (see Appendix A of
Ref.~\onlinecite{Nuss65} or Ref.~\onlinecite{WatsonBessel}) in the
form
\begin{eqnarray}
& &  J_{\lambda} (z) {\sim}  \frac{1}{2} A(\lambda ,z)
e^{+ \alpha(\lambda ,z)},  \label{DebyeJetH1a} \\
 & & H_{\lambda}^{(1)}(z) {\sim} -i A(\lambda ,z)
e^{-\alpha(\lambda ,z)},  \label{DebyeJetH1b}
\end{eqnarray}
where
\begin{subequations}
\begin{eqnarray}
&&A(\lambda ,z) = \left( \frac{2}{\pi} \right)^{1/2} (\lambda^2 -
z^2)^{-1/4},
\label{AsympDebyeIb} \\
&& \alpha(\lambda ,z)   =  (\lambda^2 - z^2)^{1/2} -\lambda \ln
\left( \frac{\lambda + (\lambda^2 - z^2)^{1/2}}{z} \right).  \nonumber \\
&& \label{AsympDebyeIc}
\end{eqnarray}
\end{subequations}
Thus,  we have
\begin{equation} \label{lhsRPSP1H}
\frac{J_{\lambda_{\mathrm{SP}_\infty}}^{'}( |n(\omega)| \omega a/c
)}{J_{\lambda_{\mathrm{SP}_\infty}}( |n(\omega)| \omega a/c )}
\sim \frac{\left[ \lambda_{\mathrm{SP}_\infty}^2 -
(|n(\omega)|\omega a/c)^2 \right]^{1/2}}{|n(\omega)|(\omega a/c)}
\end{equation}
and
\begin{equation} \label{rhsRPSP1H}
\frac{H_{\lambda_{\mathrm{SP}_\infty}}^{(1)'}(\omega a/c
)}{H_{\lambda_{\mathrm{SP}_\infty}}^{(1)}( \omega a/c )} \sim -
\frac{\left[ \lambda_{\mathrm{SP}_\infty}^2 - (\omega a/c)^2
\right]^{1/2}}{(\omega a/c)}.
\end{equation}
Now, by inserting  (\ref{lhsRPSP1H}) and (\ref{rhsRPSP1H}) into
Eq.~(\ref{RPSP1H}), we obtain an equation which can be easily
solved and we deduce
\begin{equation} \label{ReRPSPinfTE}
\lambda_{\mathrm{SP}_\infty} (\omega ) \sim  \left( \frac{\omega
a}{c} \right) \sqrt{ \frac{|\epsilon(\omega )| (|\epsilon(\omega
)| +\mu (\omega))}{\epsilon ^2 (\omega )-1}} .
\end{equation}

\begin{figure}
\includegraphics[height=5.4cm,width=8.3cm]{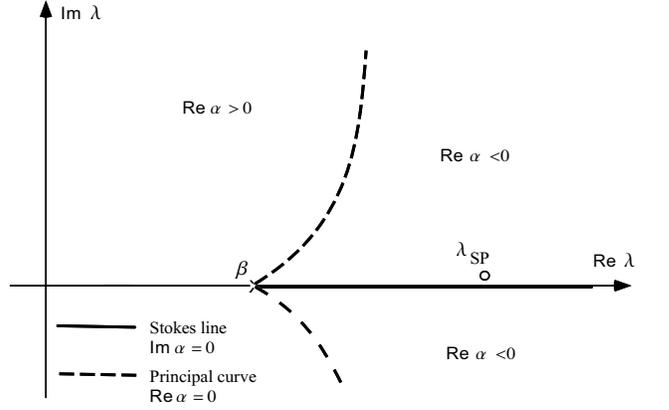}
\caption{\label{fig:StokesH1} The relative positions, in the
$\lambda$ complex plane, of the reduced frequency $\beta= \omega
a/c$ and the Regge pole $\lambda_{\mathrm{SP}_\infty}$ of the TE
theory.}
\end{figure}

We have obtained an asymptotic expansion for
$\lambda_{\mathrm{SP}_\infty} (\omega )$ or, more exactly, for the
real part of that function. Indeed, it should be noted that the
right-hand side of (\ref{ReRPSPinfTE}) is purely real. The
perturbative method previously used did not permit us to extract
the small imaginary part of $\lambda_{\mathrm{SP}_\infty} (\omega
)$. Even by taking into account higher orders in the asymptotic
expansions (\ref{DebyeJetH1a}) and (\ref{DebyeJetH1b}), we are not
able to provide the imaginary part of
$\lambda_{\mathrm{SP}_\infty} (\omega )$ because it corresponds to
an exponentially small contribution which lies beyond all orders
in perturbation theory and which can be captured only by carefully
taking into account the Stokes phenomenon
\cite{StokesPh1,StokesPh2}. We have already encountered such a
problem in Ref.~\onlinecite{AnceyDFG2004} and, {\it mutatis
mutandis}, we shall solve it in the same way. Instead of
(\ref{DebyeJetH1b}), we must use the Debye asymptotic expansion of
$H_{\lambda}^{(1)}(z)$ in the form
\begin{eqnarray} \label{DebyeJetH1bCORR}
&  & H_{\lambda}^{(1)}(z) {\sim} -i A(\lambda ,z)
e^{-\alpha(\lambda ,z)}(1 + \dots ) \nonumber \\
& & \qquad + S\left[ \alpha(\lambda ,z) \right]A(\lambda ,z)(1 +
\dots ) e^{\alpha(\lambda ,z)}.
\end{eqnarray}
In the right-hand side of (\ref{DebyeJetH1bCORR}), the first term
is the usual Debye asymptotic expansion truncated near its least
term. The second one is obtained by decoding the divergent tail of
that asymptotic expansion. This can be done (see
Refs.~\onlinecite{Berry89,BerryHowls90}) by Borel summation after
exploiting a resurgence formula discovered by Dingle
\cite{Dingle73}. In the region of the $\lambda$ complex plane
where the Regge pole $\lambda_{\mathrm{SP}_\infty}$ lies (see
Fig.~\ref{fig:StokesH1}), we have $\mathrm{Re}~\alpha < 0$. As a
consequence, the first term of the right-hand side of
(\ref{DebyeJetH1bCORR}) is the dominant contribution while the
second one is a subdominant term which can be forgotten when
$|\lambda| \to \infty$. That is what we did previously by using
(\ref{DebyeJetH1b}). The Stokes multiplier function $S\left[
\alpha(\lambda ,z) \right]$ is a complicated function involving
the exponential integral function $E_1$. It goes continuously from
$0$ to $1$ at the crossing of the Stokes line $\mathrm{Im}~\alpha
= 0$ emerging from the turning point $z=\beta$ (see
Fig.~\ref{fig:StokesH1}). Below the Stokes line, it rapidly
vanishes. On the Stokes line it is equal to $1/2$ and above the
Stokes line it rapidly becomes equal to $1$. It thus describes the
rapid but continuous birth of the subdominant contribution near
the Stokes line\cite{Berry89}. From (\ref{DebyeJetH1bCORR}) we can
now write
\begin{eqnarray} \label{rhsRPSP1Hcoor}
&  & \frac{H_{\lambda_{\mathrm{SP}_\infty}}^{(1)'}( \omega a/c
)}{H_{\lambda_{\mathrm{SP}_\infty}}^{(1)}( \omega a/c )} \sim -
\frac{\left[ \lambda_{\mathrm{SP}_\infty}^2 -
(\omega a/c)^2 \right]^{1/2}}{(\omega a/c)}  \nonumber \\
& &  \qquad  \times \left( 1- 2i S \left[
\alpha(\lambda_{\mathrm{SP}_\infty} , \omega a/c) \right] e^{2
\alpha(\lambda_{\mathrm{SP}_\infty}, \omega a/c)}
\right)  \nonumber \\
&  &
\end{eqnarray}
instead of (\ref{rhsRPSP1H}). Because we can assume that
$\lambda_{\mathrm{SP}_\infty} (\omega)$ is very close to the
Stokes line emerging from the turning point $\omega a/c$, we can
consider that $S  \left[
\alpha(\lambda_{\mathrm{SP}_\infty}(\omega) , \omega a/c)
\right]=1/2$. Then, by inserting (\ref{lhsRPSP1H}) and
(\ref{rhsRPSP1Hcoor}) into Eq.~(\ref{RPSP1H}), we obtain
\begin{subequations}
\begin{eqnarray} \label{RPSPiHas}
&  & \mathrm{Re}   \lambda_{\mathrm{SP}_\infty} (\omega ) \sim
\left( \frac{\omega a}{c}  \right) \sqrt{ \frac{|\epsilon(\omega
)| (|\epsilon(\omega )| +\mu (\omega))}{\epsilon ^2 (\omega )-1}},
\nonumber\\
\label{reRPSPiHas}  \\
&  & \mathrm{Im}  \lambda_{\mathrm{SP}_\infty}(\omega) \sim
\left(\frac{\epsilon ^2 (\omega ) }{ \epsilon ^2
(\omega )-1} \right)  \nonumber \\
&  &  \qquad \qquad \times \frac{ \left(\mathrm{Re}
\lambda_{\mathrm{SP}_\infty} (\omega )\right)^2-\left( {\omega
a/c} \right)^2 }{\mathrm{Re} \lambda_{\mathrm{SP}_\infty} (\omega
)}  \nonumber \\
&  & \qquad  \qquad \times \exp[2 \alpha(\mathrm{Re}
\lambda_{\mathrm{SP}_\infty} (\omega ) ,\omega a/c)].
\label{imRPSPiHas}
\end{eqnarray}
\end{subequations}

Equations (\ref{reRPSPiHas}) and (\ref{imRPSPiHas}) provide
analytic expressions for the dispersion relation and the damping
of the surface polariton ${\mathrm{SP}_\infty}$ of the TE theory.
The following important features must be noted:

\qquad -- The wave number $k_{\mathrm{SP}_\infty} (\omega)$
associated with this SP is obtained from (\ref{reRPSPiHas}) and
(\ref{WNSP}) and is given by
\begin{equation}\label{WNSPinfH}
k_{\mathrm{SP}_\infty} (\omega) \sim \left( \frac{\omega}{c}
\right) \sqrt{ \frac{|\epsilon(\omega )| (|\epsilon(\omega )| +\mu
(\omega))}{\epsilon ^2 (\omega )-1}}.
\end{equation}
This expression is the usual dispersion relation found in
Refs.~\onlinecite{RuppinPLA00,Darmanyanetal03,ShadrivovEtAl04} for
the $p$-polarized SP - i.e., the SP for which the magnetic field
$\mathbf{H}$ is normal to the incidence plane - supported by the
flat interface. Here, we have recovered exactly the same
dispersion relation because we have limited the perturbative
resolution of Eq.~(\ref{RPSP1H}) to the lowest order. By taking
into account higher orders in the asymptotic expansions
(\ref{DebyeJetH1a}) and (\ref{DebyeJetH1bCORR}), we could obtain
corrections for (\ref{WNSPinfH}) which vanish for $a \to \infty$
-- i.e., in the flat interface limit. The surface wave of the TE
theory denoted by ${\mathrm{SP}_\infty}$ is therefore the
counterpart of the $p$-polarized SP supported by the flat
interface.

\qquad -- The imaginary part (\ref{imRPSPiHas}) of
$\lambda_{\mathrm{SP}_\infty}$ vanishes for $a \to \infty$ --
i.e., in the flat interface limit. Then, $\mathrm{SP}_\infty$ has
no damping like the $p$-polarized SP supported by the flat
interface. This result strengthens our previous claim.

\qquad -- The function $\mathrm{Re} \lambda_{\mathrm{SP}_\infty}
(\omega )$ given by (\ref{reRPSPiHas}) has a simple pole when
$\epsilon(\omega)+1=0$ -- i.e., for $\omega = \omega_s$.
Furthermore, the imaginary part (\ref{imRPSPiHas}) of
$\lambda_{\mathrm{SP}_\infty}$ vanishes for $\omega =\omega_s$.
These two results justify all our previous remarks concerning the
accumulation of resonances which converges to the limiting
frequency $\omega_s$.

\begin{figure}
\includegraphics[height=8cm,width=8.6cm]{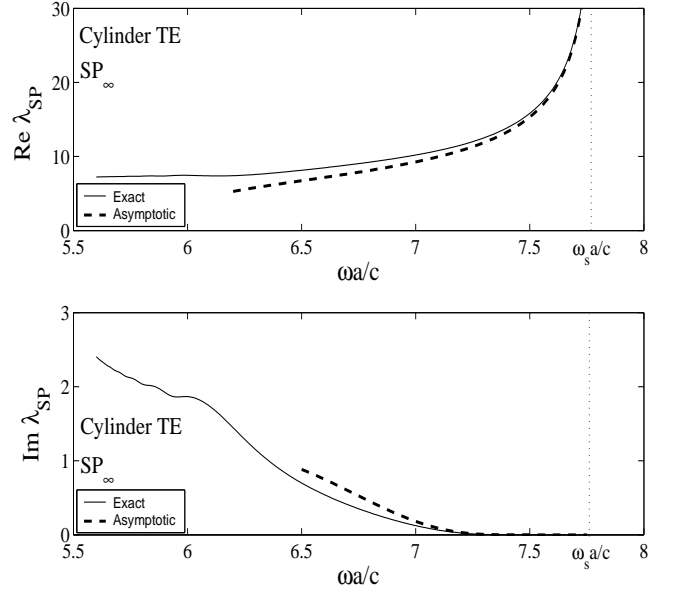}
\caption{\label{fig:RPSinfasTE} Regge trajectory for the Regge
pole associated with ${\mathrm{SP}_\infty}$ (TE theory).
Comparison between exact and asymptotic theories.}
\end{figure}

\qquad -- We have numerically tested formulas (\ref{reRPSPiHas})
and (\ref{imRPSPiHas}) (see Fig.~\ref{fig:RPSinfasTE}). They
provide rather good approximations for $\mathrm {Re}
\lambda_{\mathrm{SP}_\infty}(\omega)$ in a large frequency range
and for $\mathrm {Im} \lambda_{\mathrm{SP}_\infty}(\omega)$ in the
neighborhood  of $\omega_s$. It would be possible to improve
(\ref{imRPSPiHas}) by taking into account the variation of the
Stokes multiplier and to obtain a formula valid in a larger
frequency domain but then the expression of $\mathrm {Im}
\lambda_{\mathrm{SP}_\infty}(\omega)$ would become much more
complicated.

\begin{figure}
\includegraphics[height=8cm,width=8.6cm]{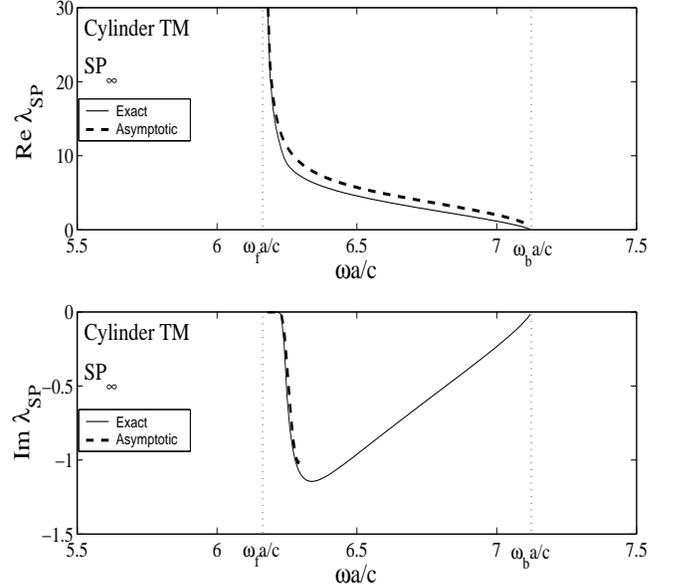}
\caption{\label{fig:RPSinfasTM} Regge trajectory for the Regge
pole associated with ${\mathrm{SP}_\infty}$ (TM theory).
Comparison between exact and asymptotic theories.}
\end{figure}

Let us now consider the Regge pole associated with
${\mathrm{SP}_\infty}$ of the TM theory. We must solve
Eq.~(\ref{RPSP1E}) which only differs from Eq.~(\ref{RPSP1H}) by
the factor in front of the Bessel functions. We assume $\omega$ in
the neighborhood of $\omega_f$ and then we can also assume that
$\mathrm{Re} \lambda_{\mathrm{SP}_\infty}(\omega)
> |n(\omega)|\omega a/c > \omega a/c$ and formally that
$\mathrm{Re} \lambda_{\mathrm{SP}_\infty}(\omega) \gg 1$ and
$\mathrm{Re} \lambda_{\mathrm{SP}_\infty}(\omega) \gg \mathrm{Im}
\lambda_{\mathrm{SP}_\infty}(\omega)$. The configuration there is
displayed in Fig.~\ref{fig:RP2}b. As a consequence, we can use
again the Debye asymptotic expansions for $J_\lambda(z)$ and
$H_\lambda^{(1)}(z)$ given in (\ref{DebyeJetH1a}) and
(\ref{DebyeJetH1bCORR}) and the resolution of Eq.~(\ref{RPSP1E})
can be modelled on that of Eq.~(\ref{RPSP1H}). Formulas
(\ref{lhsRPSP1H}) and (\ref{rhsRPSP1Hcoor}) are still valid and
the Stokes multiplier in (\ref{rhsRPSP1Hcoor}) can be taken equal
to $1/2$. Indeed, in this case $\lambda_{\mathrm{SP}_\infty}
(\omega)$ is below the Stokes line emerging from the turning point
$\omega a/c$ but it remains very close to this Stokes line. Then,
by inserting (\ref{lhsRPSP1H}) and (\ref{rhsRPSP1Hcoor}) into
Eq.~(\ref{RPSP1E}), we obtain
\begin{subequations}
\begin{eqnarray} \label{RPSPiEas}
&  & \mathrm{Re}   \lambda_{\mathrm{SP}_\infty} (\omega ) \sim
\left( \frac{\omega a}{c}  \right) \sqrt{ \frac{|\mu(\omega )|
(|\mu(\omega )| +\epsilon (\omega))}{\mu ^2 (\omega )-1}},
\nonumber\\
\label{reRPSPiEas}  \\
&  & \mathrm{Im}  \lambda_{\mathrm{SP}_\infty}(\omega) \sim
\left(\frac{\mu ^2 (\omega ) }{ \mu ^2
(\omega )-1} \right)  \nonumber \\
&  &  \qquad \qquad \times \frac{ \left(\mathrm{Re}
\lambda_{\mathrm{SP}_\infty} (\omega )\right)^2-\left( {\omega
a/c} \right)^2 }{\mathrm{Re} \lambda_{\mathrm{SP}_\infty} (\omega
)}  \nonumber \\
&  & \qquad  \qquad \times \exp[2 \alpha(\mathrm{Re}
\lambda_{\mathrm{SP}_\infty} (\omega ) ,\omega a/c)].
\label{imRPSPiEas}
\end{eqnarray}
\end{subequations}
Equations (\ref{reRPSPiEas}) and (\ref{imRPSPiEas}) provide
analytic expressions for the dispersion relation and the damping
of the surface polariton ${\mathrm{SP}_\infty}$ of the TM theory.
The following important features must be also noted:

\qquad -- The wave number $k_{\mathrm{SP}_\infty} (\omega)$
associated with this SP is obtained from (\ref{reRPSPiEas}) and
(\ref{WNSP}) and is given by
\begin{equation}\label{WNSPinfE}
k_{\mathrm{SP}_\infty} (\omega) \sim \left( \frac{\omega}{c}
\right) \sqrt{ \frac{|\mu(\omega )| (|\mu(\omega )| +\epsilon
(\omega))}{\mu ^2 (\omega )-1}}.
\end{equation}
This expression is the usual dispersion relation found in
Refs.~\onlinecite{RuppinPLA00,Darmanyanetal03,ShadrivovEtAl04} for
the $s$-polarized SP - i.e., the SP for which the electric field
$\mathbf{E}$ is normal to the incidence plane - supported by the
flat interface. The surface wave of the TM theory denoted by
${\mathrm{SP}_\infty}$ is therefore the counterpart of the
$s$-polarized SP supported by the flat interface.

\qquad -- The imaginary part (\ref{imRPSPiEas}) of
$\lambda_{\mathrm{SP}_\infty}$ vanishes for $a \to \infty$ -- i.e.
in the flat interface limit. Then, $\mathrm{SP}_\infty$ has no
damping like the $s$-polarized SP supported by the flat interface.
This result strengthens our previous claim.

\qquad -- The function $\mathrm{Re} \lambda_{\mathrm{SP}_\infty}
(\omega )$ given by (\ref{reRPSPiEas}) has a simple pole when
$\mu(\omega)+1=0$ -- i.e., for $\omega = \omega_f$. Furthermore,
the imaginary part (\ref{imRPSPiEas}) of
$\lambda_{\mathrm{SP}_\infty}$ vanishes for $\omega =\omega_f$.
These two results justify all our previous remarks concerning the
accumulation of resonances which converges to the limiting
frequency $\omega_f$.

\qquad -- We have numerically tested formulas (\ref{reRPSPiEas})
and (\ref{imRPSPiEas}) (see Fig.~\ref{fig:RPSinfasTM}). They
provide good approximations for $\mathrm {Re}
\lambda_{\mathrm{SP}_\infty}(\omega)$ in the frequency range
$]\omega_f,\omega_b[$ and for $\mathrm {Im}
\lambda_{\mathrm{SP}_\infty}(\omega)$ in the neighborhood  of
$\omega_f$. Of course, it would be possible to improve
(\ref{imRPSPiEas}) by taking into account the variation of the
Stokes multiplier and to obtain a formula for the imaginary part
of $\lambda_{\mathrm{SP}_\infty}(\omega)$ valid in a larger
frequency domain.

\begin{figure}
\includegraphics[height=5.4cm,width=8.3cm]{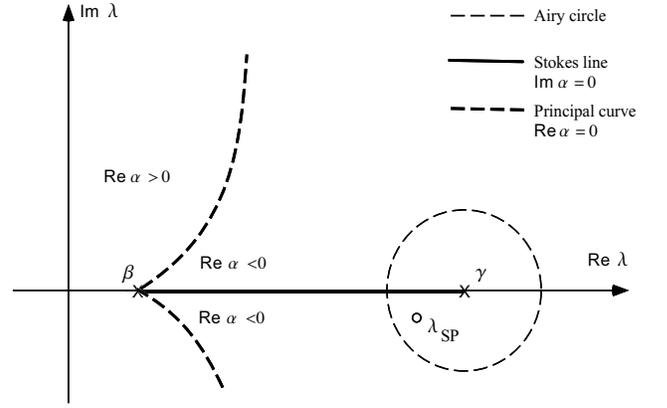}
\caption{\label{fig:StokesH2} The relative positions, in the
$\lambda$ complex plane, of the reduced frequencies $\beta= \omega
a/c$ and $\gamma =|n (\omega)|\omega a /c$ and of the Regge pole
$\lambda_{\mathrm{WGSP}_1}$ of the TE and TM theories.}
\end{figure}

Let us finally consider the Regge poles associated with the
surface waves ${\mathrm{WGSP}_n}$ for the TE  and TM theories. We
must now solve Eq.~(\ref{RPSP1H}) and Eq.~(\ref{RPSP1E}) for
$\lambda_\mathrm{SP}=\lambda_{\mathrm{WGSP}_n}$ by assuming
$\omega$ in the neighborhood of $\omega_0$. We are in the
configurations described in Fig.~\ref{fig:RP1}a for the TE theory
and Fig.~\ref{fig:RP2}a for the TM theory (see also
Fig.~\ref{fig:StokesH2}). As far as the asymptotic expansion for
$H_\lambda (z)$ is concerned, we can still use (\ref{DebyeJetH1b})
as well as its modification (\ref{DebyeJetH1bCORR}) because we can
assume that $\mathrm{Re} \lambda_{\mathrm{WGSP}_n}(\omega) >
\omega a/c $ and formally that $\mathrm{Re}
\lambda_{\mathrm{WGSP}_n}(\omega) \gg 1$ and $\mathrm{Re}
\lambda_{\mathrm{WGSP}_n}(\omega) \gg \mathrm{Im}
\lambda_{\mathrm{WGSP}_n}(\omega)$. We then obtain
\begin{eqnarray} \label{rhsRPWGSPcoor}
&  & \frac{H_{\lambda_{\mathrm{WGSP}_n}}^{(1)'}( \omega a/c
)}{H_{\lambda_{\mathrm{WGSP}_n}}^{(1)}( \omega a/c )} \sim -
\frac{\left[ \lambda_{\mathrm{WGSP}_n}^2 -
(\omega a/c)^2 \right]^{1/2}}{(\omega a/c)}  \nonumber \\
& &  \qquad  \times \left( 1- 2i S \left[
\alpha(\lambda_{\mathrm{WGSP}_n} , \omega a/c) \right] e^{2
\alpha(\lambda_{\mathrm{WGSP}_n}, \omega a/c)}
\right).  \nonumber \\
&  &
\end{eqnarray}
As far as the asymptotic expansion for $J_{\lambda} (z)$ is
concerned, the situation is a little bit more complicated.
$\lambda_{\mathrm{WGSP}_n}$ is in the immediate neighborhood of
$|n(\omega)|\omega a/c$ or more precisely in the Airy circle
centered on $|n(\omega)|\omega a/c$ (see Fig.~\ref{fig:StokesH2}).
This is true for the Regge pole corresponding to $n=1$ but it is
also formally true for all the other Regge poles with $n\not= 1$
when $\omega \to \omega_0$. Then, the Debye asymptotic expansion
(\ref{DebyeJetH1b}) for $J_\lambda(z)$ fails because $\lambda
\approx z$ and we must use the uniform asymptotic expansion (see
Appendix A of Ref.~\onlinecite{Nuss65} or
Ref.~\onlinecite{WatsonBessel})
\begin{equation}
J_{\lambda} (z) {\sim}  \left( 2/z \right)^{1/3} Ai \left[ \left(
2/z \right)^{1/3}(\lambda -z) \right]   \label{AiryBesselJ}
\end{equation}
where $Ai (z)$ denotes the Airy function\cite{AS65}. Thus we have
\begin{eqnarray} \label{lhsRPWGSP}
 &  & \frac{J_{\lambda_{\mathrm{WGSP}_n}}^{'}( |n(\omega)| \omega a/c
)}{J_{\lambda_{\mathrm{WGSP}_n}}( |n(\omega)| \omega a/c )} \sim
\left(2/(|n(\omega)| \omega a/c )\right)^{1/3}  \nonumber \\ &  &
\quad \times \frac{{Ai}^{'} \left[ \left(2/(|n(\omega)| \omega a/c
)\right)^{1/3}(\lambda_{\mathrm{WGSP}_n} - |n(\omega)| \omega a/c
) \right]}{Ai \left[ \left(2/(|n(\omega)| \omega a/c
)\right)^{1/3}(\lambda_{\mathrm{WGSP}_n} - |n(\omega)| \omega a/c
) \right]}.  \nonumber  \\
\end{eqnarray}
We can now insert (\ref{lhsRPWGSP}) and (\ref{rhsRPWGSPcoor}) into
Eqs.~(\ref{RPSP1H}) and (\ref{RPSP1E}). In order to solve the
resulting equations, we first note that their rhs tend to infinity
when $\omega \to \omega_0$. Hence, the argument of the Airy
function is very close to the zeros $x_n$ of that function (let us
recall here that the first three ones are $x_1\approx -2.3381...$,
$x_2\approx -4.0879...$ and $x_3\approx -5.5205...$) and we then
have
\begin{equation} \label{1orderRPWGSP}
\lambda_{\mathrm{WGSP}_n} (\omega ) \sim  |n (\omega)|  \omega a/c
+ \left( \frac{|n (\omega)| \omega a /c}{2} \right)^{1/3} x_n .
\end{equation}
Of course, Eq.~(\ref{1orderRPWGSP}) provides an approximation
which can be easily improved by taking into account the
subdominant contribution in (\ref{rhsRPWGSPcoor}). We assume that
the Stokes multiplier is equal to $1/2$. This is true because the
Regge pole $\lambda_{\mathrm{WGSP}_n} (\omega)$ is below but very
close to the Stokes line emerging from the turning point $\omega
a/c$. We then obtain
\begin{subequations}\label{RPWGSPHetEas}
\begin{eqnarray}
&  & \mathrm{Re}   \lambda_{\mathrm{WGSP}_n} (\omega ) \sim |n
(\omega)|  \omega a/c + \left( \frac{|n (\omega)| \omega a /c}{2}
\right)^{1/3} x_n
\nonumber\\
\label{reRPWGSPHetEas}
\end{eqnarray}
for the TE  and TM theories,
\begin{eqnarray}
 &  & \mathrm{Im}
\lambda_{\mathrm{WGSP}_n}(\omega) \sim
\sqrt{\frac{\epsilon(\omega)}{\mu(\omega)}} \left( \frac{|n
(\omega)| ~\omega a
/c}{2} \right)^{2/3}  \nonumber \\
& & \qquad \qquad  \times  \frac{\left[ \left(\mathrm{Re} \
\lambda_{\mathrm{WGSP}_n} (\omega )\right)^2-
\left( {\omega a/c} \right)^2 \right]^{1/2}}{\left({\omega a/c}\right) x_n} \nonumber \\
& & \qquad \qquad  \times \exp \left[ 2 \alpha(\mathrm{Re} \
\lambda_{\mathrm{WGSP}_n}(\omega ), \omega a/c) \right]
\label{imRPWGSPHas}
\end{eqnarray}
for the TE theory and
\begin{eqnarray}
 &  & \mathrm{Im}
\lambda_{\mathrm{WGSP}_n}(\omega) \sim
\sqrt{\frac{\mu(\omega)}{\epsilon(\omega)}} \left( \frac{|n
(\omega)| ~\omega a
/c}{2} \right)^{2/3}  \nonumber \\
& & \qquad \qquad  \times  \frac{\left[ \left(\mathrm{Re} \
\lambda_{\mathrm{WGSP}_n} (\omega )\right)^2-
\left( {\omega a/c} \right)^2 \right]^{1/2}}{\left({\omega a/c}\right) x_n} \nonumber \\
& & \qquad \qquad \times \exp \left[ 2 \alpha(\mathrm{Re} \
\lambda_{\mathrm{WGSP}_n}(\omega ), \omega a/c) \right]
\label{imRPWGSPEas}
\end{eqnarray}
for the TM theory .
\end{subequations}

\begin{figure}
\includegraphics[height=8cm,width=8.6cm]{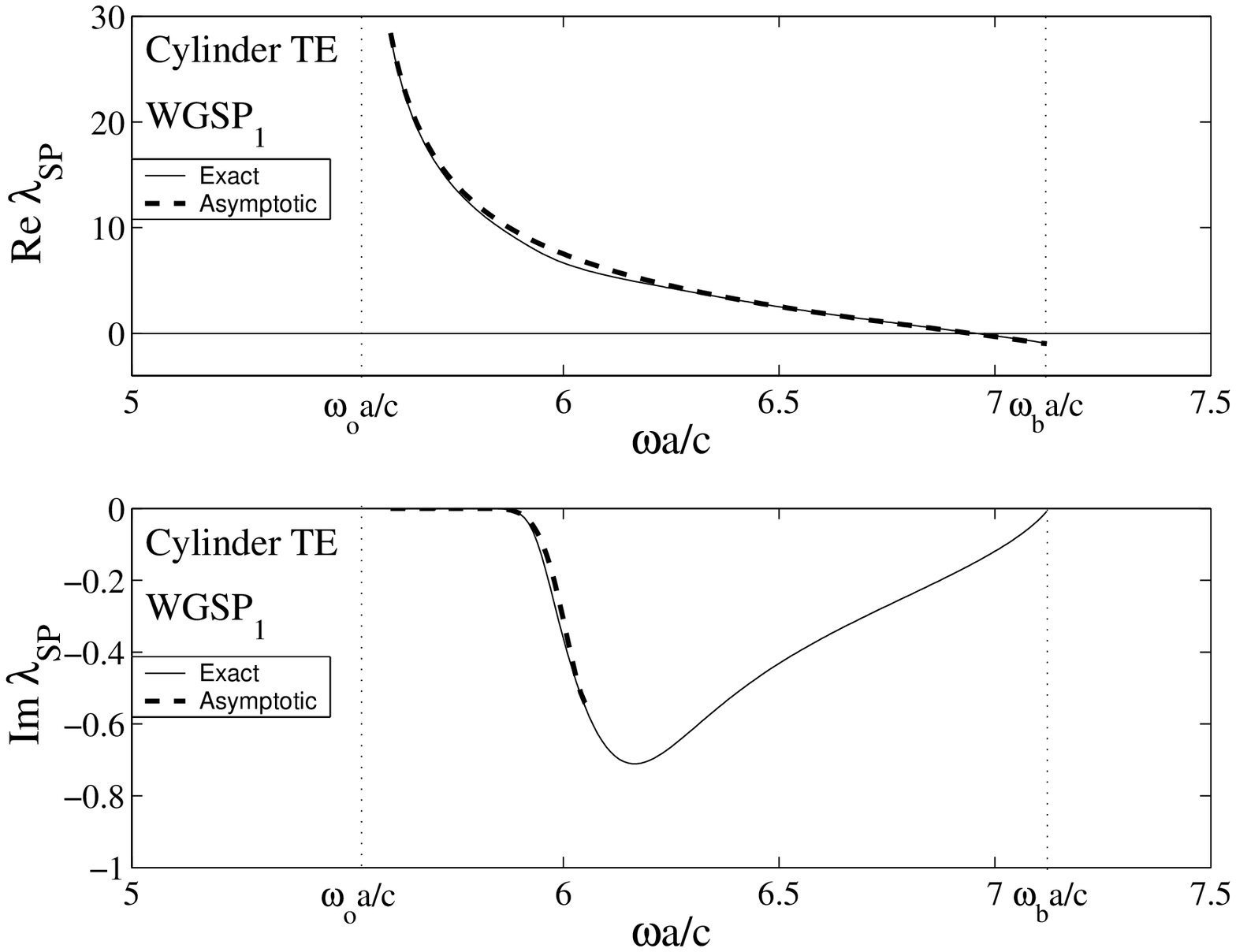}
\caption{\label{fig:RPWGSPasTE} Regge trajectory for the Regge
pole associated with ${\mathrm{WGSP}_1}$ (TE theory). Comparison
between exact and asymptotic theories.}
\end{figure}
\begin{figure}
\includegraphics[height=8cm,width=8.6cm]{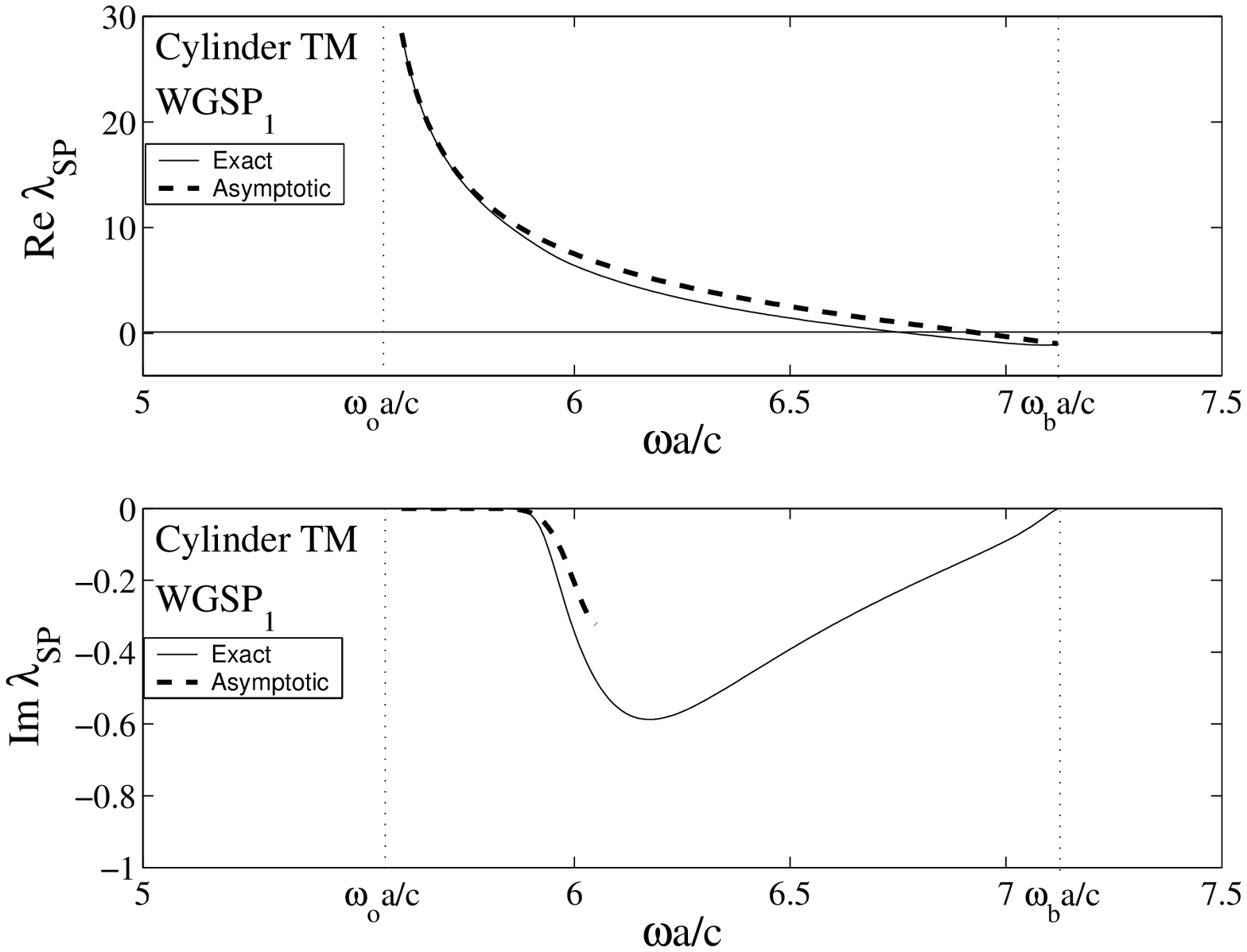}
\caption{\label{fig:RPWGSPasTM}Regge trajectory for the Regge pole
associated with ${\mathrm{WGSP}_1}$ (TE theory). Comparison
between exact and asymptotic theories.}
\end{figure}

Equations (\ref{reRPWGSPHetEas})-(\ref{imRPWGSPEas}) provide
analytic expressions for the dispersion relation and the damping
of the surface polaritons ${\mathrm{WGSP}_n}$ of the TE  and TM
theories. The corresponding wave numbers $k_{\mathrm{WGSP}_n}
(\omega)$ which are obtained from (\ref{reRPWGSPHetEas}) and
(\ref{WNSP}) are given by
\begin{equation}\label{WNWGSPn}
k_{\mathrm{WGSP}_n} (\omega) \sim |n (\omega)|  \omega /c + \left(
\frac{|n (\omega)| \omega  /c}{2a^2} \right)^{1/3} x_n.
\end{equation}
It should be noted that (\ref{1orderRPWGSP}) and therefore
(\ref{reRPWGSPHetEas}) have been obtained by following, {\it
mutatis mutandis}, the calculations done by Rayleigh a long time
ago in order to describe mathematically the whispering-gallery
phenomenon in acoustics \cite{Rayleigh1887, Rayleigh1910} (see
also Ref.~\onlinecite{StreiferKodis}). We encounter the same
phenomenon but in the context of left-handed electromagnetism and
the SP's described by the Regge poles $\lambda_{\mathrm{WGSP}_n}$
are surface waves of whispering-gallery type. Here it should be
recalled that a whispering-gallery surface wave propagates only
close to a curved interface and mainly in the internal medium in a
layer whose thickness is proportional to $k^{-2/3}$ with $k$ its
wave number. From a physical point of view, the surface polaritons
${\mathrm{WGSP}_n}$ are therefore very different from the surface
polaritons $SP_{\infty }$ which exist on both sides of the
interface. Moreover, the following important features must be also
noted:

\qquad -- These SP's have no counterparts in the plane interface
case.

\qquad -- The function $\mathrm{Re} \lambda_{\mathrm{WGSP}_n}
(\omega )$ given by (\ref{reRPSPiEas}) has a pole which is that of
$n(\omega)$ and therefore which corresponds to $\omega =
\omega_0$. Furthermore, the imaginary parts (\ref{imRPWGSPHas})
and (\ref{imRPWGSPEas}) of $\lambda_{\mathrm{WGSP}_n}$ vanish for
$\omega =\omega_0$. These results justify all our previous remarks
concerning the accumulations of resonances which converge to the
limiting frequency $\omega_0$.

\qquad -- We have numerically tested formulas
(\ref{reRPWGSPHetEas})-(\ref{imRPWGSPEas}) (see
Figs.~\ref{fig:RPWGSPasTE} and \ref{fig:RPWGSPasTM} for
$\mathrm{WGSP}_1$). They provide very good approximations for
$\mathrm {Re} \lambda_{\mathrm{WGSP}_1}(\omega)$ in the full
frequency range $]\omega_0,\omega_b[$ where the cylinder presents
left-handed behavior.  They also provide very good approximations
for $\mathrm {Im} \lambda_{\mathrm{WGSP}_1}(\omega)$ in a rather
large frequency range above the limiting frequency $\omega_0$.

\section{Conclusion and perspectives}

In the present article, we have considered the scattering of TE
and TM electromagnetic waves by a cylinder fabricated from a
left-handed material. We have mainly emphasized the resonant
aspects of the problem. We have shown that the long-lived resonant
modes can be classified into distinct families, each family being
generated by one SP and we have physically described all the SP's
orbiting around the cylinder by providing, for each one, a
numerical and a semiclassical description of its dispersion
relation and its damping.

We have also shown that the left-handed cylindrical interface can
support both TE- and TM-polarized SP's. For each polarization, we
have found that there exists a particular SP which corresponds, in
the large-radius limit, to the SP which is supported by the plane
interface and which has been theoretically described in
Refs.~\onlinecite{RuppinPLA00,Darmanyanetal03,ShadrivovEtAl04}.
But there also exists, for each polarization, an infinite family
of SP's of whispering-gallery type and these have no analogs in
the plane interface case. The existence of these SP's of
whispering-gallery type is certainly the main result of our
article. Here, in order to appreciate it, it should be recalled
that on a cylindrical metal-dielectric or semiconductor-dielectric
interface, the SP's of whispering-gallery type are absent. There
exists only one SP \cite{AnceyDFG2004}. It can be excited in the
$H$ polarization configuration -- i.e., by TE waves -- and in the
large-radius limit it corresponds to the SP which is supported by
the plane interface. From the point of view of SP physics,
left-handed cylinders are therefore much richer systems than
metallic or semiconducting cylinders and this could have important
consequences in term of practical applications.

We are now completing a theoretical and numerical study in order
to provide a better physical understanding of the resonant modes
of left-handed cylinders \cite{AnceyDFG2006}. We more particularly
focus on the repartition of the electromagnetic energy density --
both inside and outside the cylinder -- associated with all the
resonant modes. That enables us to physically distinguish on the
one hand the resonant modes generated by bulk polaritons from
those generated by SP's, and on the other hand, the resonant modes
generated by SP's of $\mathrm{SP}_\infty$ type from those
generated by SP's of whispering-gallery type.

It should be also noted that in our paper we have assumed for the
effective electric permittivity and the effective magnetic
permeability of the left-handed material the simple expressions
respectively given  by (\ref{PetP1}) and (\ref{PetP2}). In fact,
actual left-handed materials have a more complicated permittivity.
Indeed, they consist of periodic arrangements of split ring
resonators and wires and it has been recently shown that the split
ring resonators present a strong electric response in addition to
the magnetic one (see Ref.~\onlinecite{KoschnyETAL2004}). In fact,
it is possible to take into account this response by assuming that
the effective electric permittivity is given by
\cite{KoschnyETAL2004}
\begin{equation}
\epsilon (\omega) = \frac{\omega ^2 - \omega_p^2}{\omega ^2 -
{\tilde \omega}_0 ^2} \label{PetP1bis}
\end{equation}
with ${\tilde \omega}_0 \not= 0$ while the effective magnetic
permeability is still given by (\ref{PetP2}). Such a modification
does not change radically our previous analysis. {\it Mutatis
mutandis}, our results remain valid. Indeed, we have theoretically
-- i.e., semiclassically -- and numerically reexamined, in this
new context, the resonant aspects of a left-handed cylinder linked
with SP's.  We have obtained the following results:

\qquad -- The surface wave ${\mathrm{SP}_\infty}$ of the TE theory
still exists. It generates a family of resonances which converges
to the limiting frequency $\omega_s$ satisfying $\epsilon
(\omega_s)+1=0$ and which is now given by $\omega_s = \sqrt{
 ({\omega_p}^2+{{\tilde \omega}_0}^2)/ 2 }$ instead of
 (\ref{TEaccSPinf2}).  This SP is still described by
 Eqs.~(\ref{reRPSPiHas}) and (\ref{imRPSPiHas}) with now
 $\epsilon (\omega)$ given by (\ref{PetP1bis}). Of course, its
behavior depends on the relative order of the characteristic
frequencies  ${\tilde \omega}_0, \omega_0, \omega_b, \omega_p$. In
general, it presents a left-handed behavior if $\omega_s$ lies in
the frequency range where $n (\omega) <0$.

\qquad -- The surface wave ${\mathrm{SP}_\infty}$ of the TM theory
still exists. It generates a family of resonances which converges
to the limiting frequency $\omega_f$ satisfying $\mu
(\omega_f)+1=0$ and which is still given by
 (\ref{TMaccSPinf2}). This SP is still described by
 Eqs.~(\ref{reRPSPiEas}) and (\ref{imRPSPiEas}) and, in general,
 it presents a left-handed behavior if $\omega_f$ lies
in the frequency range where $n (\omega) <0$.

\qquad -- For both polarizations, there still exists the family of
SP's of whispering-gallery type described by
Eqs.~(\ref{reRPWGSPHetEas})-(\ref{imRPWGSPEas}) in the
neighborhood of the pole $\omega_0$ of $\mu (\omega)$.  They
generate an infinity of resonances which converges to the limiting
frequency $\omega_0$. In general, they present a left-handed
behavior if $\omega_0$ lies in the frequency range where $n
(\omega) <0$.

 \qquad -- But there also exists, for both polarizations, a new
family of SP's of whispering-gallery type in the neighborhood of
the pole ${\tilde \omega}_0$ of $\epsilon (\omega)$. They are also
described by (\ref{reRPWGSPHetEas})-(\ref{imRPWGSPEas}) (in this
frequency range) and they generate an infinity of resonances which
converges to the limiting frequency ${\tilde \omega}_0$. In
general, they present a left-handed behavior if ${\tilde
\omega}_0$ lies in the frequency range where $n (\omega) <0$.

Finally, it is worthwhile pointing out the possible application of
the CAM method in the context of multiple scattering by
left-handed cylinders. This could be achieved by extending to
electromagnetism of dispersive media the approaches previously
developed in the context of quantum mechanics and acoustics (see,
for example, Ref.~\onlinecite{Wirzba1999,GabrielliMercier2002}).
SP's probably induce a strong coupling between the different
cylinders and the CAM method could permit us to understand it and
to quantify its intensity.

\begin{acknowledgments}
We are grateful to Bruce Jensen for help with the English.
\end{acknowledgments}

\bibliography{SPonLHM}

\begin{thebibliography}{58}
\expandafter\ifx\csname natexlab\endcsname\relax\def\natexlab#1{#1}\fi
\expandafter\ifx\csname bibnamefont\endcsname\relax
  \def\bibnamefont#1{#1}\fi
\expandafter\ifx\csname bibfnamefont\endcsname\relax
  \def\bibfnamefont#1{#1}\fi
\expandafter\ifx\csname citenamefont\endcsname\relax
  \def\citenamefont#1{#1}\fi
\expandafter\ifx\csname url\endcsname\relax
  \def\url#1{\texttt{#1}}\fi
\expandafter\ifx\csname urlprefix\endcsname\relax\def\urlprefix{URL }\fi
\providecommand{\bibinfo}[2]{#2}
\providecommand{\eprint}[2][]{\url{#2}}

\bibitem[{\citenamefont{Veselago}(1968)}]{Veselago}
\bibinfo{author}{\bibfnamefont{V.~G.} \bibnamefont{Veselago}},
  \bibinfo{journal}{Sov.\ Phys.\ Usp.} \textbf{\bibinfo{volume}{10}},
  \bibinfo{pages}{509} (\bibinfo{year}{1968}).

\bibitem[{\citenamefont{Smith and Kroll}(2000)}]{Smith2000}
\bibinfo{author}{\bibfnamefont{D.~R.} \bibnamefont{Smith}} \bibnamefont{and}
  \bibinfo{author}{\bibfnamefont{N.}~\bibnamefont{Kroll}},
  \bibinfo{journal}{Phys.\ Rev.\ Lett.} \textbf{\bibinfo{volume}{85}},
  \bibinfo{pages}{2933} (\bibinfo{year}{2000}).

\bibitem[{\citenamefont{Pendry et~al.}(1996)\citenamefont{Pendry, Holden,
  Stewart, and Youngs}}]{Pendry96}
\bibinfo{author}{\bibfnamefont{J.~B.} \bibnamefont{Pendry}},
  \bibinfo{author}{\bibfnamefont{A.~J.} \bibnamefont{Holden}},
  \bibinfo{author}{\bibfnamefont{W.~J.} \bibnamefont{Stewart}},
  \bibnamefont{and} \bibinfo{author}{\bibfnamefont{I.}~\bibnamefont{Youngs}},
  \bibinfo{journal}{Phys.\ Rev.\ Lett.} \textbf{\bibinfo{volume}{76}},
  \bibinfo{pages}{4773} (\bibinfo{year}{1996}).

\bibitem[{\citenamefont{Pendry et~al.}(1998)\citenamefont{Pendry, Holden,
  Robbins, and Stewart}}]{Pendry98}
\bibinfo{author}{\bibfnamefont{J.~B.} \bibnamefont{Pendry}},
  \bibinfo{author}{\bibfnamefont{A.~J.} \bibnamefont{Holden}},
  \bibinfo{author}{\bibfnamefont{D.~J.} \bibnamefont{Robbins}},
  \bibnamefont{and} \bibinfo{author}{\bibfnamefont{W.~J.}
  \bibnamefont{Stewart}}, \bibinfo{journal}{J.\ Phys.: Condens. Matter}
  \textbf{\bibinfo{volume}{10}}, \bibinfo{pages}{4785} (\bibinfo{year}{1998}).

\bibitem[{\citenamefont{Pendry et~al.}(1999)\citenamefont{Pendry, Holden,
  Robbins, and Stewart}}]{Pendry99}
\bibinfo{author}{\bibfnamefont{J.~B.} \bibnamefont{Pendry}},
  \bibinfo{author}{\bibfnamefont{A.~J.} \bibnamefont{Holden}},
  \bibinfo{author}{\bibfnamefont{D.~J.} \bibnamefont{Robbins}},
  \bibnamefont{and} \bibinfo{author}{\bibfnamefont{W.~J.}
  \bibnamefont{Stewart}}, \bibinfo{journal}{IEEE Trans.\ Microwave Theory
  Tech.} \textbf{\bibinfo{volume}{47}}, \bibinfo{pages}{2075}
  (\bibinfo{year}{1999}).

\bibitem[{\citenamefont{Smith et~al.}(2000)\citenamefont{Smith, Padilla, Vier,
  Nemat-Nasser, and Schultz}}]{Smith2000a}
\bibinfo{author}{\bibfnamefont{D.~R.} \bibnamefont{Smith}},
  \bibinfo{author}{\bibfnamefont{W.~J.} \bibnamefont{Padilla}},
  \bibinfo{author}{\bibfnamefont{D.~C.} \bibnamefont{Vier}},
  \bibinfo{author}{\bibfnamefont{S.~C.} \bibnamefont{Nemat-Nasser}},
  \bibnamefont{and} \bibinfo{author}{\bibfnamefont{S.}~\bibnamefont{Schultz}},
  \bibinfo{journal}{Phys.\ Rev.\ Lett.} \textbf{\bibinfo{volume}{84}},
  \bibinfo{pages}{4184} (\bibinfo{year}{2000}).

\bibitem[{\citenamefont{Shelby et~al.}(2001{\natexlab{a}})\citenamefont{Shelby,
  Smith, Nemat-Nasser, and Schultz}}]{Smith2001a}
\bibinfo{author}{\bibfnamefont{R.~A.} \bibnamefont{Shelby}},
  \bibinfo{author}{\bibfnamefont{D.~R.} \bibnamefont{Smith}},
  \bibinfo{author}{\bibfnamefont{S.~C.} \bibnamefont{Nemat-Nasser}},
  \bibnamefont{and} \bibinfo{author}{\bibfnamefont{S.}~\bibnamefont{Schultz}},
  \bibinfo{journal}{Appl.\ Phys.\ Lett.} \textbf{\bibinfo{volume}{78}},
  \bibinfo{pages}{4} (\bibinfo{year}{2001}{\natexlab{a}}).

\bibitem[{\citenamefont{Shelby et~al.}(2001{\natexlab{b}})\citenamefont{Shelby,
  Smith, and Schultz}}]{Smith2001b}
\bibinfo{author}{\bibfnamefont{R.~A.} \bibnamefont{Shelby}},
  \bibinfo{author}{\bibfnamefont{D.~R.} \bibnamefont{Smith}}, \bibnamefont{and}
  \bibinfo{author}{\bibfnamefont{S.}~\bibnamefont{Schultz}},
  \bibinfo{journal}{Science} \textbf{\bibinfo{volume}{292}},
  \bibinfo{pages}{77} (\bibinfo{year}{2001}{\natexlab{b}}).

\bibitem[{\citenamefont{Pendry and Smith}(2004)}]{PendrySmith2004}
\bibinfo{author}{\bibfnamefont{J.~B.} \bibnamefont{Pendry}} \bibnamefont{and}
  \bibinfo{author}{\bibfnamefont{D.~R.} \bibnamefont{Smith}},
  \bibinfo{journal}{Phys.\ Today} \textbf{\bibinfo{volume}{57}},
  \bibinfo{pages}{37} (\bibinfo{year}{2004}).

\bibitem[{\citenamefont{Berrier et~al.}(2004)\citenamefont{Berrier, Mulot,
  Swillo, Qiu, Thylen, Talneau, and Anand}}]{BerrierETAL2004}
\bibinfo{author}{\bibfnamefont{A.}~\bibnamefont{Berrier}},
  \bibinfo{author}{\bibfnamefont{M.}~\bibnamefont{Mulot}},
  \bibinfo{author}{\bibfnamefont{M.}~\bibnamefont{Swillo}},
  \bibinfo{author}{\bibfnamefont{M.}~\bibnamefont{Qiu}},
  \bibinfo{author}{\bibfnamefont{L.}~\bibnamefont{Thylen}},
  \bibinfo{author}{\bibfnamefont{A.}~\bibnamefont{Talneau}}, \bibnamefont{and}
  \bibinfo{author}{\bibfnamefont{S.}~\bibnamefont{Anand}},
  \bibinfo{journal}{Phys.\ Rev.\ Lett.} \textbf{\bibinfo{volume}{93}},
  \bibinfo{pages}{073902} (\bibinfo{year}{2004}).

\bibitem[{\citenamefont{Kuzmiak and Maradudin}(2002)}]{KuzmiakMaradudin02}
\bibinfo{author}{\bibfnamefont{V.}~\bibnamefont{Kuzmiak}} \bibnamefont{and}
  \bibinfo{author}{\bibfnamefont{A.~A.} \bibnamefont{Maradudin}},
  \bibinfo{journal}{Phys.\ Rev.\ B} \textbf{\bibinfo{volume}{66}},
  \bibinfo{pages}{045116} (\bibinfo{year}{2002}).

\bibitem[{\citenamefont{Newton}(1982)}]{New82}
\bibinfo{author}{\bibfnamefont{R.~G.} \bibnamefont{Newton}},
  \emph{\bibinfo{title}{Scattering Theory of Waves and Particles}}
  (\bibinfo{publisher}{Springer-Verlag, New York}, \bibinfo{year}{1982}),
  \bibinfo{edition}{2nd} ed.

\bibitem[{\citenamefont{Nussenzveig}(1992)}]{Nus92}
\bibinfo{author}{\bibfnamefont{H.~M.} \bibnamefont{Nussenzveig}},
  \emph{\bibinfo{title}{Diffraction Effects in Semiclassical Scattering}}
  (\bibinfo{publisher}{Cambridge University Press, Cambridge},
  \bibinfo{year}{1992}).

\bibitem[{\citenamefont{Mills and Burstein}(1974)}]{MillsB74}
\bibinfo{author}{\bibfnamefont{D.~L.} \bibnamefont{Mills}} \bibnamefont{and}
  \bibinfo{author}{\bibfnamefont{E.}~\bibnamefont{Burstein}},
  \bibinfo{journal}{Rep.\ Prog.\ Phys.} \textbf{\bibinfo{volume}{37}},
  \bibinfo{pages}{817} (\bibinfo{year}{1974}).

\bibitem[{\citenamefont{Agranovich and Mills}(1982)}]{Agranovich82}
\bibinfo{author}{\bibfnamefont{V.~M.} \bibnamefont{Agranovich}}
  \bibnamefont{and} \bibinfo{author}{\bibfnamefont{D.~L.} \bibnamefont{Mills}},
  \emph{\bibinfo{title}{Surface Polaritons: Electromagnetic Waves at Surfaces
  and Intersurfaces}} (\bibinfo{publisher}{North-Holland, Amsterdam},
  \bibinfo{year}{1982}).

\bibitem[{\citenamefont{Raether}(1988)}]{Raether88}
\bibinfo{author}{\bibfnamefont{H.}~\bibnamefont{Raether}},
  \emph{\bibinfo{title}{Surface Plasmons \rm{Vol. 111 of Springer Tracts in
  Modern Physics}}} (\bibinfo{publisher}{Springer-Verlag, Berlin},
  \bibinfo{year}{1988}).

\bibitem[{\citenamefont{Sernelius}(2001)}]{Sernelius01}
\bibinfo{author}{\bibfnamefont{B.~E.} \bibnamefont{Sernelius}},
  \emph{\bibinfo{title}{Surface Modes in Physics}}
  (\bibinfo{publisher}{Wiley-VCH Verlag GmbH, Weinheim}, \bibinfo{year}{2001}).

\bibitem[{\citenamefont{McGurn and Maradudin}(1993)}]{McGurnMaradudin93}
\bibinfo{author}{\bibfnamefont{A.~R.} \bibnamefont{McGurn}} \bibnamefont{and}
  \bibinfo{author}{\bibfnamefont{A.~A.} \bibnamefont{Maradudin}},
  \bibinfo{journal}{Phys.\ Rev.\ B} \textbf{\bibinfo{volume}{48}},
  \bibinfo{pages}{17576} (\bibinfo{year}{1993}).

\bibitem[{\citenamefont{Kuzmiak et~al.}(1994)\citenamefont{Kuzmiak, Maradudin,
  and Pincemin}}]{KusmiakMP1994}
\bibinfo{author}{\bibfnamefont{V.}~\bibnamefont{Kuzmiak}},
  \bibinfo{author}{\bibfnamefont{A.~A.} \bibnamefont{Maradudin}},
  \bibnamefont{and} \bibinfo{author}{\bibfnamefont{F.}~\bibnamefont{Pincemin}},
  \bibinfo{journal}{Phys.\ Rev.\ B} \textbf{\bibinfo{volume}{50}},
  \bibinfo{pages}{16835} (\bibinfo{year}{1994}).

\bibitem[{\citenamefont{Sigalas et~al.}(1995)\citenamefont{Sigalas, Chan, Ho,
  and Soukoulis}}]{SigalasCHS1995}
\bibinfo{author}{\bibfnamefont{M.~M.} \bibnamefont{Sigalas}},
  \bibinfo{author}{\bibfnamefont{C.~T.} \bibnamefont{Chan}},
  \bibinfo{author}{\bibfnamefont{K.~M.} \bibnamefont{Ho}}, \bibnamefont{and}
  \bibinfo{author}{\bibfnamefont{C.~M.} \bibnamefont{Soukoulis}},
  \bibinfo{journal}{Phys.\ Rev.\ B} \textbf{\bibinfo{volume}{52}},
  \bibinfo{pages}{11744} (\bibinfo{year}{1995}).

\bibitem[{\citenamefont{Zhang et~al.}(1996)\citenamefont{Zhang, Hu, Lei, Xu,
  and Ming}}]{ZhangHLXM1996}
\bibinfo{author}{\bibfnamefont{W.}~\bibnamefont{Zhang}},
  \bibinfo{author}{\bibfnamefont{A.}~\bibnamefont{Hu}},
  \bibinfo{author}{\bibfnamefont{X.}~\bibnamefont{Lei}},
  \bibinfo{author}{\bibfnamefont{N.}~\bibnamefont{Xu}}, \bibnamefont{and}
  \bibinfo{author}{\bibfnamefont{N.}~\bibnamefont{Ming}},
  \bibinfo{journal}{Phys.\ Rev.\ B} \textbf{\bibinfo{volume}{54}},
  \bibinfo{pages}{10280} (\bibinfo{year}{1996}).

\bibitem[{\citenamefont{Kuzmiak et~al.}(1997)\citenamefont{Kuzmiak, Maradudin,
  and McGurn}}]{KusmiakMG1997}
\bibinfo{author}{\bibfnamefont{V.}~\bibnamefont{Kuzmiak}},
  \bibinfo{author}{\bibfnamefont{A.~A.} \bibnamefont{Maradudin}},
  \bibnamefont{and} \bibinfo{author}{\bibfnamefont{A.~R.}
  \bibnamefont{McGurn}}, \bibinfo{journal}{Phys.\ Rev.\ B}
  \textbf{\bibinfo{volume}{55}}, \bibinfo{pages}{4298} (\bibinfo{year}{1997}).

\bibitem[{\citenamefont{Kuzmiak and Maradudin}(1997)}]{KusmiakM1997}
\bibinfo{author}{\bibfnamefont{V.}~\bibnamefont{Kuzmiak}} \bibnamefont{and}
  \bibinfo{author}{\bibfnamefont{A.~A.} \bibnamefont{Maradudin}},
  \bibinfo{journal}{Phys.\ Rev.\ B} \textbf{\bibinfo{volume}{55}},
  \bibinfo{pages}{7427} (\bibinfo{year}{1997}).

\bibitem[{\citenamefont{van~der Lem and Moroz}(2000)}]{Moroz2000}
\bibinfo{author}{\bibfnamefont{H.}~\bibnamefont{van~der Lem}} \bibnamefont{and}
  \bibinfo{author}{\bibfnamefont{A.}~\bibnamefont{Moroz}},
  \bibinfo{journal}{J.\ Opt.\ A:\ Pure Appl.\ Opt.}
  \textbf{\bibinfo{volume}{2}}, \bibinfo{pages}{395} (\bibinfo{year}{2000}).

\bibitem[{\citenamefont{Sakoda}(2001)}]{Sakoda2001}
\bibinfo{author}{\bibfnamefont{K.}~\bibnamefont{Sakoda}},
  \emph{\bibinfo{title}{Optical Properties of Photonic Crystals}}
  (\bibinfo{publisher}{Springer-Verlag, Berlin}, \bibinfo{year}{2001}).

\bibitem[{\citenamefont{Ito and Sakoda}(2001)}]{Sakoda2001b}
\bibinfo{author}{\bibfnamefont{T.}~\bibnamefont{Ito}} \bibnamefont{and}
  \bibinfo{author}{\bibfnamefont{K.}~\bibnamefont{Sakoda}},
  \bibinfo{journal}{Phys.\ Rev.\ B} \textbf{\bibinfo{volume}{64}},
  \bibinfo{pages}{045117} (\bibinfo{year}{2001}).

\bibitem[{\citenamefont{Moreno et~al.}(2002)\citenamefont{Moreno, Erni, and
  Hafner}}]{MorenoEH2002}
\bibinfo{author}{\bibfnamefont{E.}~\bibnamefont{Moreno}},
  \bibinfo{author}{\bibfnamefont{D.}~\bibnamefont{Erni}}, \bibnamefont{and}
  \bibinfo{author}{\bibfnamefont{C.}~\bibnamefont{Hafner}},
  \bibinfo{journal}{Phys.\ Rev.\ B} \textbf{\bibinfo{volume}{65}},
  \bibinfo{pages}{155120} (\bibinfo{year}{2002}).

\bibitem[{\citenamefont{Ochiai and Sanchez-Dehesa}(2002)}]{OchiaiSD2002}
\bibinfo{author}{\bibfnamefont{T.}~\bibnamefont{Ochiai}} \bibnamefont{and}
  \bibinfo{author}{\bibfnamefont{J.}~\bibnamefont{Sanchez-Dehesa}},
  \bibinfo{journal}{Phys.\ Rev.\ B} \textbf{\bibinfo{volume}{65}},
  \bibinfo{pages}{245111} (\bibinfo{year}{2002}).

\bibitem[{\citenamefont{Ebbesen et~al.}(1998)\citenamefont{Ebbesen, Lezec,
  Ghaemi, Thio, and Wolff}}]{Ebbesen98}
\bibinfo{author}{\bibfnamefont{T.~W.} \bibnamefont{Ebbesen}},
  \bibinfo{author}{\bibfnamefont{H.~J.} \bibnamefont{Lezec}},
  \bibinfo{author}{\bibfnamefont{H.~F.} \bibnamefont{Ghaemi}},
  \bibinfo{author}{\bibfnamefont{T.}~\bibnamefont{Thio}}, \bibnamefont{and}
  \bibinfo{author}{\bibfnamefont{P.~A.} \bibnamefont{Wolff}},
  \bibinfo{journal}{Nature (London)} \textbf{\bibinfo{volume}{391}},
  \bibinfo{pages}{667} (\bibinfo{year}{1998}).

\bibitem[{\citenamefont{Barnes et~al.}(2004)\citenamefont{Barnes, Murray,
  Dintinger, Devaux, and Ebbesen}}]{Ebbesen2004}
\bibinfo{author}{\bibfnamefont{W.~L.} \bibnamefont{Barnes}},
  \bibinfo{author}{\bibfnamefont{W.~A.} \bibnamefont{Murray}},
  \bibinfo{author}{\bibfnamefont{J.}~\bibnamefont{Dintinger}},
  \bibinfo{author}{\bibfnamefont{E.}~\bibnamefont{Devaux}}, \bibnamefont{and}
  \bibinfo{author}{\bibfnamefont{T.~W.} \bibnamefont{Ebbesen}},
  \bibinfo{journal}{Phys.\ Rev.\ Lett.} \textbf{\bibinfo{volume}{92}},
  \bibinfo{pages}{107401} (\bibinfo{year}{2004}).

\bibitem[{\citenamefont{Pendry}(2000)}]{Pendry00}
\bibinfo{author}{\bibfnamefont{J.~B.} \bibnamefont{Pendry}},
  \bibinfo{journal}{Phys.\ Rev.\ Lett.} \textbf{\bibinfo{volume}{85}},
  \bibinfo{pages}{3966} (\bibinfo{year}{2000}).

\bibitem[{\citenamefont{Feise et~al.}(2001)\citenamefont{Feise, Bevelacqua, and
  Schneider}}]{Feise01}
\bibinfo{author}{\bibfnamefont{M.~W.} \bibnamefont{Feise}},
  \bibinfo{author}{\bibfnamefont{P.~J.} \bibnamefont{Bevelacqua}},
  \bibnamefont{and} \bibinfo{author}{\bibfnamefont{J.~B.}
  \bibnamefont{Schneider}}, \bibinfo{journal}{Phys.\ Rev.\ B}
  \textbf{\bibinfo{volume}{66}}, \bibinfo{pages}{035113}
  (\bibinfo{year}{2001}).

\bibitem[{\citenamefont{Haldane}(unpublished)}]{Haldane02}
\bibinfo{author}{\bibfnamefont{F.~D.~M.} \bibnamefont{Haldane}},
  \emph{\bibinfo{title}{{\rm cond-mat/0206420}}} (\bibinfo{year}{unpublished}).

\bibitem[{\citenamefont{Rao and Ong}(2003)}]{Rao03}
\bibinfo{author}{\bibfnamefont{X.~S.} \bibnamefont{Rao}} \bibnamefont{and}
  \bibinfo{author}{\bibfnamefont{C.~K.} \bibnamefont{Ong}},
  \bibinfo{journal}{Phys.\ Rev.\ B} \textbf{\bibinfo{volume}{68}},
  \bibinfo{pages}{113103} (\bibinfo{year}{2003}).

\bibitem[{\citenamefont{Shadrivov et~al.}(2003)\citenamefont{Shadrivov, Zharov,
  and Kivshar}}]{ShadrivovEtAl03}
\bibinfo{author}{\bibfnamefont{I.~V.} \bibnamefont{Shadrivov}},
  \bibinfo{author}{\bibfnamefont{A.~A.} \bibnamefont{Zharov}},
  \bibnamefont{and} \bibinfo{author}{\bibfnamefont{Y.~S.}
  \bibnamefont{Kivshar}}, \bibinfo{journal}{Appl.\ Phys.\ Lett.}
  \textbf{\bibinfo{volume}{83}}, \bibinfo{pages}{2713} (\bibinfo{year}{2003}).

\bibitem[{\citenamefont{Ruppin}(2000{\natexlab{a}})}]{RuppinPLA00}
\bibinfo{author}{\bibfnamefont{R.}~\bibnamefont{Ruppin}},
  \bibinfo{journal}{Phys.\ Lett.\ A} \textbf{\bibinfo{volume}{277}},
  \bibinfo{pages}{61} (\bibinfo{year}{2000}{\natexlab{a}}).

\bibitem[{\citenamefont{Darmanyan et~al.}(2003)\citenamefont{Darmanyan,
  Nevi\`ere, and Zakhidov}}]{Darmanyanetal03}
\bibinfo{author}{\bibfnamefont{S.~A.} \bibnamefont{Darmanyan}},
  \bibinfo{author}{\bibfnamefont{M.}~\bibnamefont{Nevi\`ere}},
  \bibnamefont{and} \bibinfo{author}{\bibfnamefont{A.~A.}
  \bibnamefont{Zakhidov}}, \bibinfo{journal}{Opt.\ Commun.}
  \textbf{\bibinfo{volume}{225}}, \bibinfo{pages}{233} (\bibinfo{year}{2003}).

\bibitem[{\citenamefont{Shadrivov et~al.}(2004)\citenamefont{Shadrivov,
  Sukhorukov, Kivshar, Zharov, Boardman, and Egan}}]{ShadrivovEtAl04}
\bibinfo{author}{\bibfnamefont{I.~V.} \bibnamefont{Shadrivov}},
  \bibinfo{author}{\bibfnamefont{A.~A.} \bibnamefont{Sukhorukov}},
  \bibinfo{author}{\bibfnamefont{Y.~S.} \bibnamefont{Kivshar}},
  \bibinfo{author}{\bibfnamefont{A.~A.} \bibnamefont{Zharov}},
  \bibinfo{author}{\bibfnamefont{A.~D.} \bibnamefont{Boardman}},
  \bibnamefont{and} \bibinfo{author}{\bibfnamefont{P.}~\bibnamefont{Egan}},
  \bibinfo{journal}{Phys.\ Rev.\ E} \textbf{\bibinfo{volume}{69}},
  \bibinfo{pages}{016617} (\bibinfo{year}{2004}).

\bibitem[{\citenamefont{Ruppin}(2000{\natexlab{b}})}]{RuppinSSC00}
\bibinfo{author}{\bibfnamefont{R.}~\bibnamefont{Ruppin}},
  \bibinfo{journal}{Solid State Comm.} \textbf{\bibinfo{volume}{116}},
  \bibinfo{pages}{411} (\bibinfo{year}{2000}{\natexlab{b}}).

\bibitem[{\citenamefont{Ancey et~al.}(2004)\citenamefont{Ancey, D\'ecanini,
  Folacci, and Gabrielli}}]{AnceyDFG2004}
\bibinfo{author}{\bibfnamefont{S.}~\bibnamefont{Ancey}},
  \bibinfo{author}{\bibfnamefont{Y.}~\bibnamefont{D\'ecanini}},
  \bibinfo{author}{\bibfnamefont{A.}~\bibnamefont{Folacci}}, \bibnamefont{and}
  \bibinfo{author}{\bibfnamefont{P.}~\bibnamefont{Gabrielli}},
  \bibinfo{journal}{Phys.\ Rev.\ B} \textbf{\bibinfo{volume}{70}},
  \bibinfo{pages}{245406} (\bibinfo{year}{2004}).

\bibitem[{\citenamefont{Dingle}(1973)}]{Dingle73}
\bibinfo{author}{\bibfnamefont{R.~D.} \bibnamefont{Dingle}},
  \emph{\bibinfo{title}{Asymptotic Expansions: Their Derivation and
  Interpretation}} (\bibinfo{publisher}{Academic Press, London},
  \bibinfo{year}{1973}).

\bibitem[{\citenamefont{Berry}(1989)}]{Berry89}
\bibinfo{author}{\bibfnamefont{M.~V.} \bibnamefont{Berry}},
  \bibinfo{journal}{Proc.\ R.\ Soc.\ London} \textbf{\bibinfo{volume}{A 422}},
  \bibinfo{pages}{7} (\bibinfo{year}{1989}).

\bibitem[{\citenamefont{Berry and Howls}(1990)}]{BerryHowls90}
\bibinfo{author}{\bibfnamefont{M.~V.} \bibnamefont{Berry}} \bibnamefont{and}
  \bibinfo{author}{\bibfnamefont{C.~J.} \bibnamefont{Howls}},
  \bibinfo{journal}{Proc.\ R.\ Soc.\ London} \textbf{\bibinfo{volume}{A 430}},
  \bibinfo{pages}{653} (\bibinfo{year}{1990}).

\bibitem[{\citenamefont{Segur et~al.}(1991)\citenamefont{Segur, Tanveer, and
  Levine}}]{SegurTL91}
\bibinfo{author}{\bibfnamefont{H.}~\bibnamefont{Segur}},
  \bibinfo{author}{\bibfnamefont{S.}~\bibnamefont{Tanveer}}, \bibnamefont{and}
  \bibinfo{author}{\bibfnamefont{H.}~\bibnamefont{Levine}},
  \emph{\bibinfo{title}{Asymptotics Beyond all Orders}}
  (\bibinfo{publisher}{Plenum, New York}, \bibinfo{year}{1991}).

\bibitem[{\citenamefont{Mott and Massey}(1965)}]{Mott65}
\bibinfo{author}{\bibfnamefont{N.~F.} \bibnamefont{Mott}} \bibnamefont{and}
  \bibinfo{author}{\bibfnamefont{H.~S.~W.} \bibnamefont{Massey}},
  \emph{\bibinfo{title}{The Theory of Atomic Collisions}}
  (\bibinfo{publisher}{Oxford University Press, Oxford}, \bibinfo{year}{1965}).

\bibitem[{\citenamefont{Abramowitz and Stegun}(1965)}]{AS65}
\bibinfo{author}{\bibfnamefont{M.}~\bibnamefont{Abramowitz}} \bibnamefont{and}
  \bibinfo{author}{\bibfnamefont{I.~A.} \bibnamefont{Stegun}},
  \emph{\bibinfo{title}{Handbook of Mathematical Functions}}
  (\bibinfo{publisher}{Dover, New-York}, \bibinfo{year}{1965}).

\bibitem[{\citenamefont{Watson}(1918)}]{Watson18}
\bibinfo{author}{\bibfnamefont{G.~N.} \bibnamefont{Watson}},
  \bibinfo{journal}{Proc.\ Roy.\ Soc.\ London A}
  \textbf{\bibinfo{volume}{100}}, \bibinfo{pages}{83} (\bibinfo{year}{1918}).

\bibitem[{\citenamefont{Nussenzveig}(1965)}]{Nuss65}
\bibinfo{author}{\bibfnamefont{H.~M.} \bibnamefont{Nussenzveig}},
  \bibinfo{journal}{Ann.\ Phys.\ (N.Y.)} \textbf{\bibinfo{volume}{34}},
  \bibinfo{pages}{23} (\bibinfo{year}{1965}).

\bibitem[{\citenamefont{Watson}(1995)}]{WatsonBessel}
\bibinfo{author}{\bibfnamefont{G.~N.} \bibnamefont{Watson}},
  \emph{\bibinfo{title}{Theory of Bessel Functions}}
  (\bibinfo{publisher}{Cambridge University Press, Cambridge},
  \bibinfo{year}{1995}), \bibinfo{edition}{2nd} ed.

\bibitem[{\citenamefont{Stokes}(1847)}]{StokesPh1}
\bibinfo{author}{\bibfnamefont{G.~G.} \bibnamefont{Stokes}},
  \bibinfo{journal}{Trans.\ Cambridge Philos.\ Soc.}
  \textbf{\bibinfo{volume}{9}}, \bibinfo{pages}{379} (\bibinfo{year}{1847}).

\bibitem[{\citenamefont{Stokes}(1864)}]{StokesPh2}
\bibinfo{author}{\bibfnamefont{G.~G.} \bibnamefont{Stokes}},
  \bibinfo{journal}{Trans.\ Cambridge Philos.\ Soc.}
  \textbf{\bibinfo{volume}{10}}, \bibinfo{pages}{106} (\bibinfo{year}{1864}).

\bibitem[{\citenamefont{Rayleigh}(1976)}]{Rayleigh1887}
\bibinfo{author}{\bibfnamefont{J.~W.~S.} \bibnamefont{Rayleigh}},
  \emph{\bibinfo{title}{The Theory of Sound \rm{reprinted by Dover}}}
  (\bibinfo{publisher}{Dover, New York}, \bibinfo{year}{1976}).

\bibitem[{\citenamefont{Rayleigh}(1910)}]{Rayleigh1910}
\bibinfo{author}{\bibfnamefont{J.~W.~S.} \bibnamefont{Rayleigh}},
  \bibinfo{journal}{Phil.\ Mag.} \textbf{\bibinfo{volume}{20}},
  \bibinfo{pages}{1001} (\bibinfo{year}{1910}).

\bibitem[{\citenamefont{Streifer and Kodis}(1964)}]{StreiferKodis}
\bibinfo{author}{\bibfnamefont{W.}~\bibnamefont{Streifer}} \bibnamefont{and}
  \bibinfo{author}{\bibfnamefont{R.~D.} \bibnamefont{Kodis}},
  \bibinfo{journal}{Q.\ Appl.\ Math.} \textbf{\bibinfo{volume}{21}},
  \bibinfo{pages}{285} (\bibinfo{year}{1964}).

\bibitem[{\citenamefont{Ancey et~al.}(In preparation)\citenamefont{Ancey,
  D\'ecanini, Folacci, and Gabrielli}}]{AnceyDFG2006}
\bibinfo{author}{\bibfnamefont{S.}~\bibnamefont{Ancey}},
  \bibinfo{author}{\bibfnamefont{Y.}~\bibnamefont{D\'ecanini}},
  \bibinfo{author}{\bibfnamefont{A.}~\bibnamefont{Folacci}}, \bibnamefont{and}
  \bibinfo{author}{\bibfnamefont{P.}~\bibnamefont{Gabrielli}}
  (\bibinfo{year}{In preparation}).

\bibitem[{\citenamefont{Koschny et~al.}(2004)\citenamefont{Koschny, Kafesaki,
  Economou, and Soukoulis}}]{KoschnyETAL2004}
\bibinfo{author}{\bibfnamefont{T.}~\bibnamefont{Koschny}},
  \bibinfo{author}{\bibfnamefont{M.}~\bibnamefont{Kafesaki}},
  \bibinfo{author}{\bibfnamefont{E.~N.} \bibnamefont{Economou}},
  \bibnamefont{and} \bibinfo{author}{\bibfnamefont{C.~M.}
  \bibnamefont{Soukoulis}}, \bibinfo{journal}{Phys.\ Rev.\ Lett.}
  \textbf{\bibinfo{volume}{93}}, \bibinfo{pages}{107402}
  (\bibinfo{year}{2004}).

\bibitem[{\citenamefont{Wirzba}(1999)}]{Wirzba1999}
\bibinfo{author}{\bibfnamefont{A.}~\bibnamefont{Wirzba}},
  \bibinfo{journal}{Phys.\ Rep.} \textbf{\bibinfo{volume}{309}},
  \bibinfo{pages}{1} (\bibinfo{year}{1999}).

\bibitem[{\citenamefont{Gabrielli and
  Mercier-Finidori}(2002)}]{GabrielliMercier2002}
\bibinfo{author}{\bibfnamefont{P.}~\bibnamefont{Gabrielli}} \bibnamefont{and}
  \bibinfo{author}{\bibfnamefont{M.}~\bibnamefont{Mercier-Finidori}},
  \bibinfo{journal}{Phys.\ Rev.\ E} \textbf{\bibinfo{volume}{66}},
  \bibinfo{pages}{046629} (\bibinfo{year}{2002}).

\end{thebibliography}

\end{document}